\documentclass[iop,revtex4,numberedappendix]{emulateapj}
\newcommand{\bvec}[1]{\textbf{#1}}

\newcommand{\mytilde}{\raise.19ex\hbox{$\scriptstyle\sim$}}
\newcommand{\elgordo}{ACT-CL~J0102$-$4915}
\newcommand{\kms}{$\mbox{km}~\mbox{s}^{-1}$}
\newcommand{\hubblel}{h_{70}^{-2}}
\newcommand{\hubblem}{h_{70}^{-1}}
\newcommand{\hubbleml}{h_{70}}
\newcommand{\solarm}{$\hubblem 10^{14}M_{\sun}$}

\shorttitle{HST/ACS Weak-lensing Study of ACT-CL J0102$-$4915}
\shortauthors{Jee et al.}

\begin{document}

\title{WEIGHING ``EL GORDO" WITH A PRECISION SCALE: 
 \textbf{\emph{HUBBLE SPACE TELESCOPE}} WEAK-LENSING ANALYSIS OF THE MERGING
GALAXY CLUSTER  
ACT-CL~J0102$-$4915 AT $Z=0.87$}


\author{M.~JAMES~JEE\altaffilmark{1},  
JOHN~P.~HUGHES\altaffilmark{2},  
FELIPE~MENANTEAU\altaffilmark{2,3,4}, \\ 
CRIST{\'O}BAL SIF{\'O}N\altaffilmark{5},  
RACHEL MANDELBAUM\altaffilmark{6},  
L. FELIPE BARRIENTOS\altaffilmark{7},
LEOPOLDO INFANTE\altaffilmark{7}, AND 
KAREN Y. NG\altaffilmark{1} }

\altaffiltext{1}{Department of Physics, University of California, Davis, One Shields Avenue, 
Davis, CA 95616, 
USA}
\altaffiltext{2}{Department of Physics \& Astronomy, Rutgers University, 136 Frelinghysen 
Rd., Piscataway, 
NJ 08854, USA}
\altaffiltext{3}{National Center for Supercomputing Applications, University of Illinois at Urbana-Champaign, 1205 W. Clark St., Urbana, IL 61801, USA}
\altaffiltext{4}{Department of Astronomy, University of Illinois at Urbana-Champaign, 1002 W. Green St, Urbana, IL 61801, USA}

\altaffiltext{5}{Leiden Observatory, Leiden University, PO Box 9513, NL-2300 RA Leiden, 
Netherlands}
\altaffiltext{6}{Department of Physics, Carnegie Mellon University, Pittsburgh, PA 15213, 
USA}
\altaffiltext{7}{ Departamento de Astronom{\'i}a y Astrof{\'i}sica, Facultad de
F{\'i}sica, Ponticia Universidad Cat{\'o}lica de Chile, Casilla 306,
Santiago 22, Chile}

\begin{abstract}
We present a {\it Hubble Space Telescope} weak-lensing study of the
merging galaxy cluster ``El Gordo'' (\elgordo)~at $z=0.87$ discovered
by the Atacama Cosmology Telescope (ACT) collaboration as the
strongest Sunyaev-Zel'dovich decrement in its $\mytilde1000$ deg$^2$
survey.  Our weak-lensing analysis confirms that \elgordo\ is indeed
an extreme system consisting of two massive ($\gtrsim10^{15}M_{\sun}$
each) subclusters with a projected separation of $
\mytilde0.7\hubblem$ Mpc.  This binary mass structure revealed by our
lensing study is consistent with the cluster galaxy distribution and
the dynamical study carried out with 89 spectroscopic members.  We
estimate the mass of \elgordo\ by simultaneously fitting two
axisymmetric Navarro-Frenk-White (NFW) profiles allowing their centers
to vary. We use only a single parameter for the NFW mass profile by
enforcing the mass-concentration relation from numerical simulations.
Our Markov-Chain-Monte-Carlo (MCMC) analysis shows that the masses of
the northwestern (NW) and the southeastern (SE) components are
$M_{200c}=(1.38\pm0.22)\times10^{15} \hubblem M_{\sun}$ and
$(0.78\pm0.20)\times10^{15} \hubblem M_{\sun}$, respectively,
where the quoted errors include only $1\sigma$ statistical
uncertainties determined by the finite number of source galaxies.
These mass estimates are subject to additional uncertainties
(20--30\%) due to the possible presence of triaxiality,
correlated/uncorrelated large scale structure, and departure of the
cluster profile from the NFW model.  
The lensing-based velocity
dispersions are $1133_{-61}^{+58}~ \mbox{km}~\mbox{s}^{-1}$ and
$1064_{-66} ^{+62}~\mbox{km}~\mbox{s}^{-1}$ for the NW and SE
components, respectively, which are consistent with their
spectroscopic measurements ($1290\pm134~\mbox{km}~\mbox{s}^{-1}$ and
$1089\pm200~\mbox{km}~ \mbox{s}^{-1}$, respectively).
The centroids of both components
are tightly constrained ($\mytilde4\arcsec$) and close to the optical
luminosity centers.  The X-ray and mass peaks are spatially offset by 
$\mytilde8\arcsec$ ($\mytilde100\hubblem$ kpc), which is significant
at the $\mytilde2\sigma$ confidence level.  The mass
peak, however, does not lead the gas peak in the direction
expected {\it if} we are viewing the cluster soon after first core
passage during a high speed merger.  Under the
assumption that the merger is happening in the plane of the sky,
extrapolation of the two NFW halos to a radius
$r_{200a}=2.4\hubblem$~Mpc yields a combined mass of
$M_{200a}=(3.13\pm0.56)\times10^{15}\hubblem M_{\sun}$. This
extrapolated total mass is consistent with our two-component-based
dynamical analysis and previous X-ray measurements, projecting
\elgordo\ to be the most massive cluster at $z>0.6$ known to date.
\end{abstract}


\section{INTRODUCTION} \label{section_introduction}
One of the fundamental questions in modern cosmology is how do the
large scale structures of the universe form and evolve.  Galaxy
clusters are the largest structures of the universe detached from the
Hubble flow and gravitationally bound.  Advancement in observational
techniques over the past few decades has enabled us to map the
distributions of the three different cluster constituents, namely,
galaxies, hot plasma, and dark matter in such detail that the
comparison of their distributions reveals on-going activities.  These
studies strongly suggest that almost all galaxy clusters possess hints
of current merging activities, which is consistent with our
theoretical understanding that galaxy clusters constantly grow by
accreting new galaxies/groups/clusters along their neighboring
filaments.

Within the hierarchical structure formation paradigm, merging is the
primary mechanism for structures to grow, and thus detailed studies of
cluster mergers provide key information to advance our understanding
of the large-scale structure evolution of the universe. Apart from the
cosmological context, merging clusters also serve as useful
astrophysical laboratories, where physical properties of galaxies,
intracluster gas, and dark matter can be probed.  Although still in
their infancy, numerical studies of merging clusters or cluster
substructures will soon play pivotal roles in our understanding of
properties of dark matter (e.g., Randall et al.\ 2008, Rocha et
al.\ 2013).

Cluster collisions are not rare. However, enlarging the sample of
prominent colliding clusters at close separation is not easy because
of the limited observational time window (a few Gyrs)  in which to catch the merger.
In addition, high-resolution
X-ray imaging is required to reliably confirm the stage of the merger
either through the temperature structure or offsets between galaxies
and plasma. To date, only a handful of clusters are known to possess
merging features convincingly indicative of a recent core pass-through
(e.g., Markevitch et al.\ 2002, Markevitch et al.\ 2005, Dawson et
al.\ 2012, Merten et al.\ 2011).

The galaxy cluster \elgordo\ at $z=0.87$, recently discovered as the
most significant Sunyaev-Zel'dovich (SZ) decrement by the Atacama
Cosmology Telescope (ACT) collaboration (Marriage et al.\ 2011), is a
new example of this rare class of cluster.  The cluster is
particularly interesting because several lines of evidence indicate
that the system might be a high-redshift analog of the ``Bullet
Cluster'' (1E0657-56) at $z=0.3$ (Markevitch et al.\ 2002; Clowe et
al.\ 2006).  Menanteau et al.\ (2012; hereafter M12) suggest that the
cool peak of the X-ray emission, whose Fe abundance is substantially
enhanced with respect to the rest of the cluster, may correspond to
the ``bullet" of the Bullet Cluster. Also, M12 find that the plasma
temperature structure is reminiscent of shock heating. Moreover, their
radio data analysis shows the presence of an intense double radio
relic, which has now been confirmed by higher resolution, 
multi-frequency radio imaging (Lindner et al. 2013).

Now the conspicuously missing information, which prevents us from a
deeper understanding of the system, is the underlying dark matter
distribution. Therefore, in this paper we present a high-resolution
weak-lensing analysis of the cluster with {\it Hubble Space Telescope}
imaging.  If the cluster is indeed a post-collision system of two
massive clusters as suggested by M12, we expect to observe two
corresponding mass clumps as in the Bullet Cluster. Also, it will be
interesting to examine if we can detect a similar offset between dark
matter and plasma, which has been used as direct evidence for the
presence of collisionless dark matter in the Bullet Cluster.  Of
course, the mass will be one of the key ``ingredients'' necessary to
set up the initial conditions for follow-up numerical studies.

Another critical discussion that this weak-lensing study can
facilitate is our understanding of the cosmological implication of
\elgordo.  Discovery of even a single sufficiently massive cluster at
high redshift can provide a non-negligible challenge to the currently
accepted and well tested $\Lambda$CDM structure formation paradigm if
the probability of finding such a cluster within the survey is
extremely small (e.g., Brodwin et al.\ 2012; Stanford et al.\ 2012;
Jee et al.\ 2009; Foley et al.\ 2011; Planck Collaboration et
al.\ 2011).  As noted above, \elgordo\ is the most significant SZ
decrement in the ACT survey. which covers nearly 1000 square degrees
(Hasselfield et al.\ 2013).  In addition, both the South Pole
Telescope (SPT) and the {\it Planck} data confirm that the cluster is an
extreme case (Williamson et al.\ 2011, Planck Collaboration et
al.\ 2013a).  It has been argued, based on the strength of the SZ
signal, the dynamical velocity dispersion
($\sigma_v=1321\pm106~\mbox{km}~\mbox{s}^{-1}$), integrated X-ray
temperature $T_X=14.5\pm1.0~$keV, and X-ray luminosity
$L_X\sim2\times10^{45}~\mbox{erg}~\mbox{s}^{-1}$ (in the 0.5--2.0 keV
band), that \elgordo\ might be the most massive cluster known to date
at $z>0.6$ (M12).  However, for such an extreme merger system
conversion of these mass proxy measurements to actual mass estimates
may not be completely reliable.  Thus, our weak gravitational lensing
analysis of \elgordo, which does not depend on the detailed physics of
the cluster baryons, provides an independent comparison to these
previous non-lensing-based mass estimates.  Recently Zitrin et
al.\ (2013) presented a strong gravitational lensing analysis of
\elgordo\ which revealed its binary mass distribution and presented a
total mass estimate of $M_{200} \sim 2.3\times 10^{15}\, M_\odot$.

The format of this paper is as follows. In
\textsection\ref{section_obs}, we describe our {\it HST} data.  Both
theoretical background and technical issues on weak-lensing are
presented in \textsection \ref{section_analysis}.  We show our mass
reconstruction and provide weak-lensing mass estimates in \textsection
\ref{section_result}.  \textsection\ref{section_discussion} presents
our interpretation of the weak-lensing results in the context of the
cluster merger and the cosmology. We conclude in
\textsection\ref{section_conclusion}.  Unless otherwise indicated, we
assume a cosmology, where $\Omega_M=0.3$, $ \Omega_{\Lambda}=0.7$, and
$H_0=70h_{70}~ \mbox{km}~\mbox{s}^{-1}~\mbox{Mpc}^{-1}$ We use the AB
magnitude system throughout and quote all uncertainties at the
1-$\sigma$ confidence level.

\section{OBSERVATIONS} \label{section_obs}

The galaxy cluster \elgordo\ has been observed with the {\it Hubble
  Space Telescope} under programs PROP 12755 (PI: J.~Hughes) and PROP 12477 (PI: W.~High). The
weak-lensing analysis presented in this paper is based on our
combined analysis of the data  obtained under these two programs. 

\begin{figure}
\includegraphics[width=8.5cm]{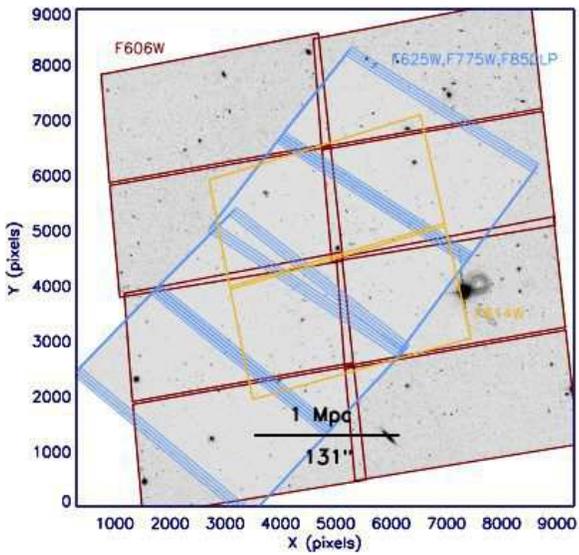}
\caption{Layout of the ACS pointings for the observation of \elgordo. The F606W imaging (PROP 12477) covers  the $\mytilde6\arcmin\times6\arcmin$ area in a $2\times2$ mosaic pattern. The F625W, F775W, and F850LP filters (PROP 12755) image the $\mytilde6\arcmin\times3\arcmin$ strip containing the two cluster galaxy overdensity regions. Only a single pointing is used
for F814W (PROP 12477).
\label{fig_pointing}}
\end{figure}

\elgordo\ was observed on 2012 September and October with the Wide Field Channel
(WFC) of ACS.  
In PROP 12755, two ACS pointings in F625W, F775W, and F850LP  were used to image a $\mytilde6\arcmin \times3\arcmin$ strip in the NW-SE orientation to cover the two dominant substructures of
\elgordo\ traced by the cluster galaxy distribution (M12).  
PROP 12477 observed the cluster in the $2\times2$ mosaic pattern with F606W and
in a single pointing with F814W.
The total integration per pointing is 1,920 s, 2,344 s, 2,512 s, 1,916 s, and 2,516 s for F606W, F625W,
F775W, F814W, and F850LP, respectively. Figure~\ref{fig_pointing} illustrates the different pointings of the two {\it HST} programs.

We measure galaxy shapes in F606W, F625W, F775W, and F814W and combine the results wherever multiple 
shapes are available. Although measuring shapes in F850LP is
possible, it is not optimal. This filter's point spread function (PSF)
is larger and it has a peculiar horizontal spike due to the
anti-halation layer (Jee et al.\ 2007a), resulting in significantly
larger ellipticity errors.

Our data reduction starts with the {\tt FLC} images provided by the
STScI pipeline, which removes bias striping noise from post-Servicing
Mission 4 (SM4) images and corrects for charge transfer inefficiency
(CTI) using the latest pixel-based method (Anderson \& Berdin 2010;
Ubeda \& Anderson 2012); CTI occurs because of charge traps in the
CCD, and we provide somewhat detailed discussions in
\textsection\ref{section_cti1} and Appendix~\ref{section_cti} of the
influence of CTI on our study. In summary, our tests show that the
current level of the STScI pixel-based correction for the CTI problem
is sufficient for our weak-lensing analysis.

We compute and refine WCS offsets between different pointings by
cross-correlating common astronomical sources. We estimate that the
mean image registration error is $\mytilde0.01$~pixels (i.e., the
average alignment error divided by the square root of the number of
common objects), which would induce an ellipticity bias of
$\mytilde10^{-3}$ for the smallest galaxies that we use for lensing
analysis. This accuracy surpasses the current weak-lensing
requirement.  We perform cosmic ray removal, sky subtraction, and
image combination using the MultiDrizzle (Koekemoer et al.\ 2002)
software with the ``Lanczos3" drizzling kernel and an output pixel
scale of $0\farcs05$. This combination of drizzling kernel and pixel
scale has been shown to provide de-facto the sharpest
point-spread-function (PSF) with minimal noise correlations in our
previous weak-lensing studies (e.g., Jee et al.\ 2007a; 2009; 2011).

Note that some weak-lensing studies (e.g., COSMOS; Koekemoer et
al.\ 2007) prefer the ``Gauss'' drizzling kernel with an output pixel
scale of $0\farcs03$ arguing that the choice reduces the effects of
aliasing although the resulting image shows more correlated noise
between pixels.  Our experiment shows that both schemes (i.e.,
$0\farcs03$ with the Gauss kernel vs.~$0\farcs05$ with the Lanczos3
kernel) produce very similar results in the weak-lensing analysis of
\elgordo.  Readers are referred to
Appendix~\ref{section_drizzle_comparison} for the description of our
comparison.

We create a detection image by weight-averaging all five filter
images. Objects are detected with SExtractor (Bertin \& Arnouts 1996)
run in a dual-image mode. We identify sources by looking for at least
10 connected pixels above 1.5 times the sky rms.  The total number of
detected sources is $\mytilde12,228$ within the $\mytilde6\arcmin
\times6\arcmin$ area.  Along the boundaries of the ACS observation
footprints, some sources are clipped and contaminated by cosmic rays.
We use SExtractor's {\tt FLAGS} to discard the clipped objects. For
cosmic rays along the boundaries, we visually scan the images and
manually identify corrupted objects.
After applying our source selection criteria
(\textsection\ref{section_redshift}), the number density of sources is
$\mytilde101$ arcmin$^{-2}$. Smaller weights are applied to fainter
sources in our weak-lensing analysis, and thus the effective number
density is slightly smaller ($\mytilde11$\%) than this value.

\section{ANALYSIS }\label{section_analysis}

\subsection{Theoretical Background}
The transformation of a galaxy image by weak gravitational lensing is
described by the following equation:
\begin{equation}
\textbf{A}=(1-\kappa) \left (\begin{array} {c c} 1 - g _1 & -g _2 \\
                      -g_2 & 1+g _1
          \end{array}  \right ), \label{eqn_lens_trans}
\end{equation}
\noindent
where $\kappa$ is the projected mass density in units of the critical
surface density $ \Sigma_c$ and $g$ is the reduced shear $g =
\gamma/(1-\kappa)$.  The critical surface density is given by
\begin{equation}
\Sigma_c = \frac{c^2}{4 \pi G D_l \beta }, \label{eqn_sigma_c}
\end{equation}
\noindent
where $c$ is the speed of light, $G$ is the gravitational constant,
and $D_l$ is the angular diameter distance to the lens.  In
equation~\ref{eqn_sigma_c}, $\beta$ is the distance ratio
$D_{ls}/D_s$, where $D_{ls}$ and $D_s$ are the angular diameter
distances between the lens and the source, and between the observer
and the source, respectively. Because source galaxies are at different
redshifts, it is necessary to define an effective redshift of the
source plane to estimate $\beta$ (\textsection\ref{section_redshift}).

According to Equation~\ref{eqn_lens_trans}, a circular image is
transformed into an ellipse, whose ellipticity is defined as
\begin{equation}
g=(g_1^2+g_2^2)^{1/2}=(1-r)/(1+r),
\end{equation}
 where $r$ is the aspect ratio of
the semi-minor axis to the semi-major axis. The position angle of the
ellipse is $(1/2)~ \mbox{tan}^{-1}(g_2/ g_1)$. In practice, measurement
of the weak-lensing signal is much more complicated because 1) galaxy
shapes possess intrinsic ellipticity, 2) imperfect charge transfers
elongate object ellipticity in the readout direction, 3) the PSFs
bias/dilute the signal, 4) sources are at different redshifts, 5)
galaxy morphology affects the response to shear (often referred to as
shear responsivity), etc.  In this paper, all these subtleties are
addressed in detail.

\subsection{CTI Correction} \label{section_cti1}
In a CCD, electrons are transferred pixel-to-pixel to the readout
register.  However, because of defects in the silicon, some fraction
of electrons remain trapped and are released after a characteristic
time delay $\tau$.  This unfortunate fractional charge transfer lapse
happens in every pixel-to-pixel transfer, and can be visually
identified as long trails along the readout direction.  Although
almost every CCD is subject to this CTI, in particular this is a
non-trivial issue for the {\it HST} detectors exposed to constant space
radiation leading to considerable charge trap creation over time.  In
addition, the Wide Field Channel (WFC) detector of ACS consists of two
large-format CCDs ($4096\times2048$), which causes significant CTI
effects for the pixels farthest from the readout register compared to
the Wide Field Planetary Camera 2 (WFPC2) comprised of relatively
small ($800\times800$) CCDs.

Our {\it HST}/ACS data of \elgordo\ were taken more than ten years
after the instrument was installed in 2002 March. Therefore, the
cumulative damage on the CCDs is severe, and the resulting CTI is one
of our initial concerns for the use of the ACS instrument in
weak-lensing studies. However, we find that the most recent
pixel-based method by Ubeda \& Anderson (2012) corrects for the CTI
robustly to the extent that the residual shear systematic error is
less than $\delta \gamma < 0.01$.  Compared to the level of the
statistical noise and the large lensing signal, we conclude that this
level of systematic error is negligible. Our detailed discussion on
this issue is presented in Appendix~\ref{section_cti}.

\subsection{PSF Modeling} \label{section_PSF}
One of the most critical instrumental signatures for weak-lensing
analysis is the PSF. An anisotropic PSF distorts the images of
galaxies and induces coherent ellipticity alignments, mimicking
gravitational lensing. In addition, PSF blurring dilutes the apparent
ellipticity. The removal of these two PSF effects in the shear
measurement is discussed in
\textsection\ref{section_shear_measurement}. Here, we focus on
modeling of the PSF for ACS.

In general, the number of high S/N stars within an ACS field is too
small to describe the complicated spatial variation of the ACS PSF
(Jee et al.\ 2007a). Because of the so-called ``focus breathing," the
{\it HST} PSF is also time-dependent. Potentially, these obstacles
prevent us from modeling the PSF accurately in the region, where no
nearby stars are usable to infer the PSF. However, fortunately, the
{\it HST} PSF pattern is repetitive (Jee et al.\ 2007a), and this
allows us to construct a PSF library from dense stellar field
observations and apply the library to the weak-lensing field.

Finding a good PSF template that closely matches one's target frame is
important.  With simulations using {\it HST}/ACS globular cluster
data, Jee et al.\ (2007a) demonstrated that the success rate is over
90\% when the second moments of $\mytilde10$ (out of $\mytilde1000$)
bright stars (S/N $>$ 20) are used, where the matching is considered a
success if the resulting residual (data-model) PSF ellipticity
correlation computed from all ($\mytilde1000$) bright stars is less
than $10^{-5}$.  This is a conservative criterion given the small size
of the {\it HST} PSF.

The current mosaic image of \elgordo\ consists of multiple pointings, and
each pointing is comprised of short ($\sim500$~s) exposures. Because
our weak-lensing shapes are measured on the mosaic image (not on the
individual exposure data), it is important to combine our PSF models
for all individual exposures in such a way that the stacked PSF at the
galaxy location is robustly represented. Therefore, we carefully trace
the image registration procedures including the choice of drizzling
kernel and pixel scale, the shift and rotation, and the weight
distribution.  We achieve an rms scatter of the residual PSF
ellipticity of $\mytilde0.008$. This small level of scatter is more than
sufficient for the current cluster weak-lensing analysis.  The PSF
library of Jee et al.\ (2007a) is publicly
available\footnote{http://acs.pha.jhu.edu/$\mytilde$mkjee/acs\_psf/}.

\subsection{Shear Measurement} \label{section_shear_measurement}

We measure galaxy ellipticity by fitting an elliptical Gaussian to
pixellated images. The resulting semi-major and -minor axes, $a$ and
$b$, and orientation $\phi$ are used to define the following $e_1$ and
$e_2$ components:
\begin{eqnarray}
e_1&=&e \cos{2 \phi},  \\
e_2&=&e \sin{2 \phi},  
\end{eqnarray}
where
\begin{equation}
e=\frac{a-b} {a+b}.
\end{equation}
Because observed galaxy images are smeared by PSFs, we convolve the
elliptical Gaussian by model PSFs prior to fitting.  This so-called
``forward modeling" removes both anisotropic and smearing effects of
the PSF by performing this indirect deconvolution. Other techniques
include moment-based methods such as Kaiser, Squires, and Broadhurst
(1995; hereafter KSB), which performs quasi-deconvolution using
moments of both stars and galaxies to remove these PSF effects.

Although the anisotropic PSF effect is taken care of by the above
forward modeling, our shear measurement with the above method is still
subject to a multiplicative bias, which means that the raw ellipticity
should be multiplied by a calibration factor (higher than unity) to
match gravitational shear.

This is because
1) the exact ellipticity transformation rule for a given shear
$\gamma$ depends on the galaxy surface brightness profile
(morphology-dependent shear responsivity),
2) galaxy surface brightness profiles are not Gaussian (model
incompleteness),
3) ellipticity measurement is a non-linear mapping,
4) galaxies possess their own intrinsic ellipticities associated with
their formation environment (intrinsic alignment).
Issues 1), 2), and 3) can be addressed both analytically (Bernstein \&
Jarvis 2002) and numerically (through image simulations).  In Jee et
al.\ (2013), we find that the simple analytic recipe of Bernstein \&
Jarvis (2002) agrees reasonably well with our image simulation result
for bright galaxies. However, the difference is large for low S/N
galaxies. This S/N-dependent bias (so-called ``noise bias'') is an
important issue for future cosmic shear surveys (Melchior \& Viola
2012, Refregier et al.\ 2012).  In this paper, we address this shear
calibration through image simulations and determine the calibration
factor.  This type of shear calibration also takes care of the impact
of systematics due to neighboring objects statistically because we
ensure the number density of simulated objects to match observed
values.  Our calibration factor varies from $\mytilde1.05$ to $\mytilde1.16$
depending on the S/N value of the object. Although in principle, one can
use S/N-dependent calibration, we find that for the current weak-lensing
analysis it is sufficient to apply a global factor $\mytilde1.11$ (our
ellipticities should be multiplied by 1.11 to give shear) to the entire
source population; the difference in the final result  is 
only at the sub-percent level. However, we note that in future cosmic shear surveys it
is critical to use S/N-dependent calibration because the S/N-dependence implies also
redshift-dependence.
The
intrinsic alignment 4) is an important topic again in cosmic shear
studies. However, this can safely be ignored in the current cluster
lensing, where the strength of the lensing signal is a few
orders-of-magnitude higher than intrinsic alignment signals (e.g., Hao
et al.\ 2011, Schneider et al.\ 2013).

We measure object ellipticities in F606W, F625W, F775W, and F814W. 
Where available,  ellipticities from different colors are combined; we found no
statistically significant difference in galaxy shapes among different filters.
We use inverse-variances
derived from ellipticity errors as weights. Combining shapes from different filters allows us
to use fainter galaxies for our weak-lensing than using shapes from a single filter. For example,
when we use shapes only from F775W, the number density of sources is $\mytilde93$ arcmin$^{-2}$ 
within the $6\arcmin\times3\arcmin$ strip (after applying our source selection criteria described in
\textsection\ref{section_redshift}). 
Adding the shapes from F606W, F625W, and F814W increased
the source number density to $\mytilde101$  arcmin$^{-2}$. 

\begin{figure}
\includegraphics[width=8cm]{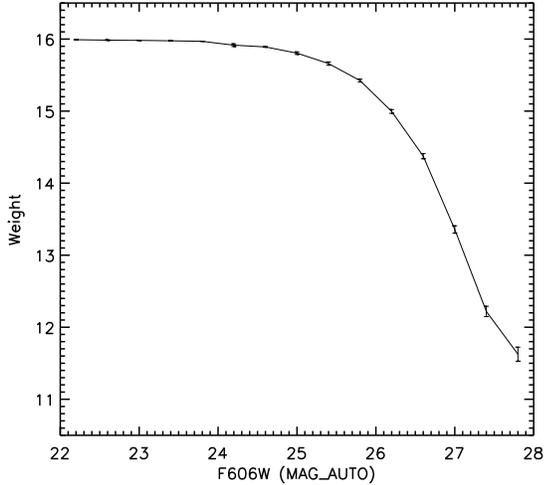}
\caption{Magnitude vs. ellipticity weight. We display our ellipticity weight (equation~\ref{eqn_ell_weight}) as a function of the F606W magnitude.
\label{fig_weight}}
\end{figure}

These raw numbers should not be confused with effective number densities. Effective number
densities are the results that we obtain after taking into account the fact that fainter sources provide noisier measurements and thus are de-weighted. We define our weight as
\begin{equation}
\mu_i = \frac{1}{\sigma^{2}_{SN} + (\delta e_i)^2}, \label{eqn_ell_weight}
\end{equation}
\noindent
where $\sigma_{SN}$ is the shape noise ($\mytilde 0.25$) and $\delta e_i$ is the measurement error for the $i^{th}$ galaxy.
Then, the corresponding $n_{eff}$ is 
\begin{equation}
n_{\rm eff} = \frac{1}{A} \sum \frac{\sigma_{SN}^2} {\sigma_{SN}^2 + (\delta e_i)^2},  \label{eqn_neff}
\end{equation}
\noindent
where
$A$ is the field area.
According to the above weighting scheme, $n_{eff}$ is $\mytilde90$ sources arcmin$^{-2}$, $\mytilde11$\% smaller than the raw number density $\mytilde101$  arcmin$^{-2}$. We display in Figure~\ref{fig_weight} our relation between
source magnitude and weight.

\subsection{Source Selection and Redshift Estimation} \label{section_redshift}

We combine imaging data from two programs, and this somewhat complicates source selection and their
redshift estimation. The $6\arcmin\times3\arcmin$ stripe area (hereafter Region A) observed by PROP 12755 is imaged in F625W, F775W, and F850LP. Thus, it is possible to employ the traditional red-sequence technique to identify and remove cluster galaxies in this region based on the ACS data. However, about 48\% of our target area is observed only with F606W (hereafter Region B). For this region, we can remove only bright cluster galaxies utilizing the photometric redshift catalog of M12. Fortunately, since Region B is far from the cluster center,  the expected number of cluster galaxies is small. Furthermore, because the lensing signal is much weaker than in Region A, less stringent contamination rate (and source redshift) estimation is required for our weak-lensing analysis of the cluster.

\subsubsection{Region A}

We rely on the traditional method, which discards cluster members identified through
their 4000~\AA~break feature.  Although it is a powerful method to select ``red" cluster galaxies,
the procedure includes a significant fraction of non-background galaxies into the source
catalog. We quantify the dilution of the lensing
signal due to this contamination statistically, utilizing a
high-fidelity photometric redshift catalog obtained from control fields.

\begin{figure}
\includegraphics[width=8cm]{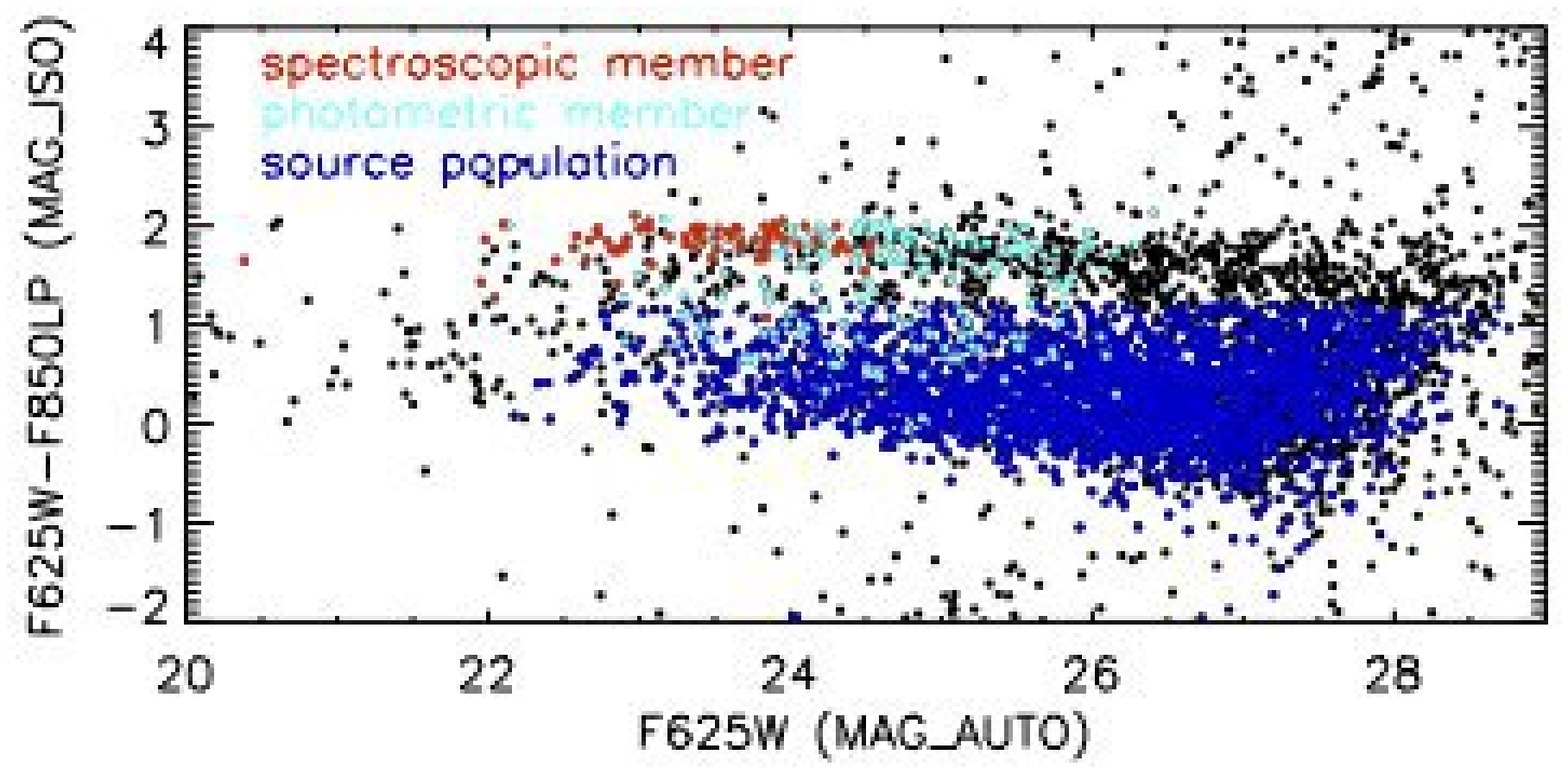}
\includegraphics[width=8cm]{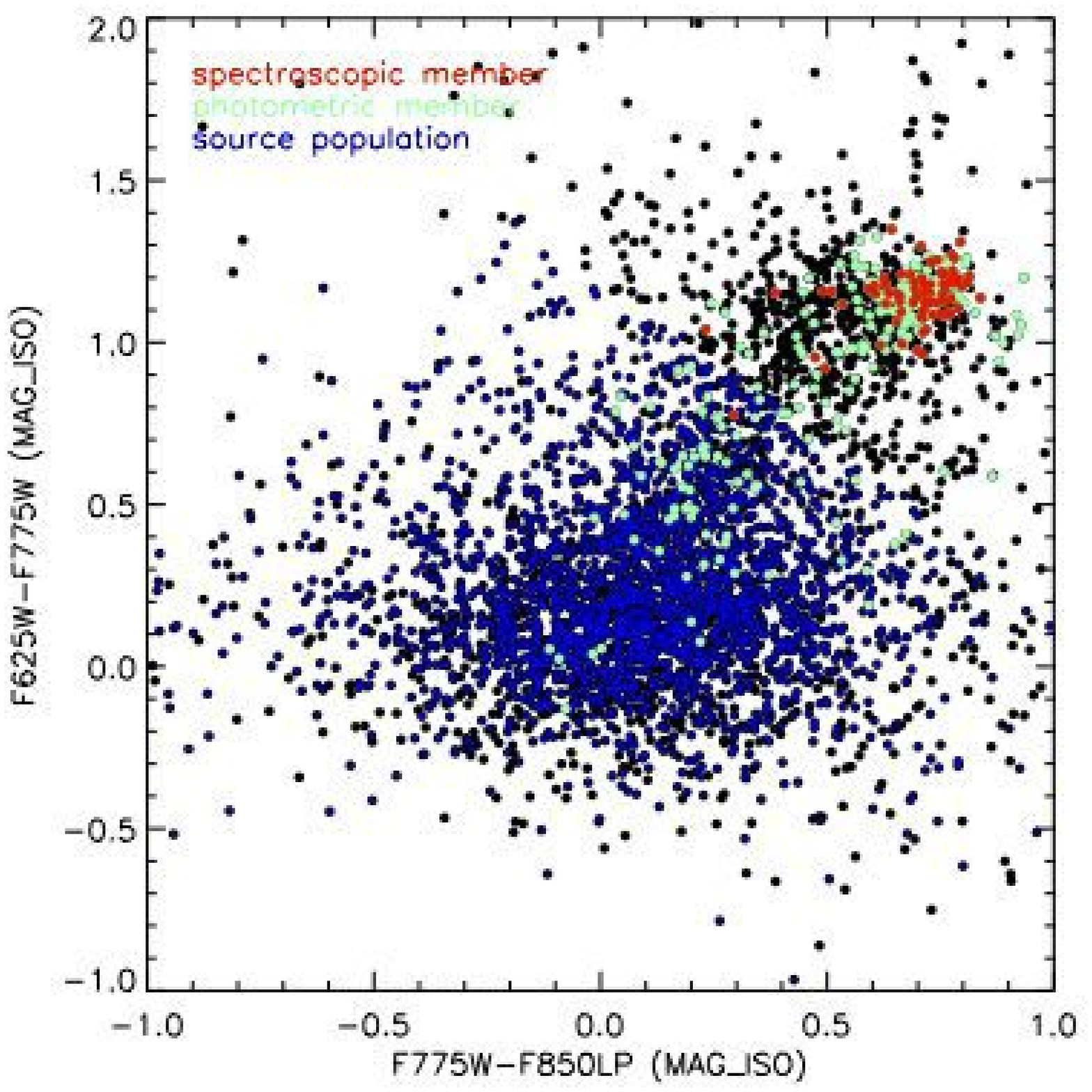}

\caption{Source galaxy selection. Top:  color-magnitude relation.
The locus of the red-sequence is clearly visible down to F625W$\mytilde26$.
This color-magnitude relation is used to select our sources (blue). 
Bottom: color-color relation.
We considered using a color-color diagram for source selection. However, compared
to the color-magnitude diagram (top), this scheme does not provide an advantage over the
red-sequence method for the
current dataset. The
red-sequence is already well-separated with the color magnitude diagram. The blue cluster member
candidates are severely blended with our source population.
\label{fig_cmd}}
\end{figure}

\begin{figure}
\includegraphics[width=8cm]{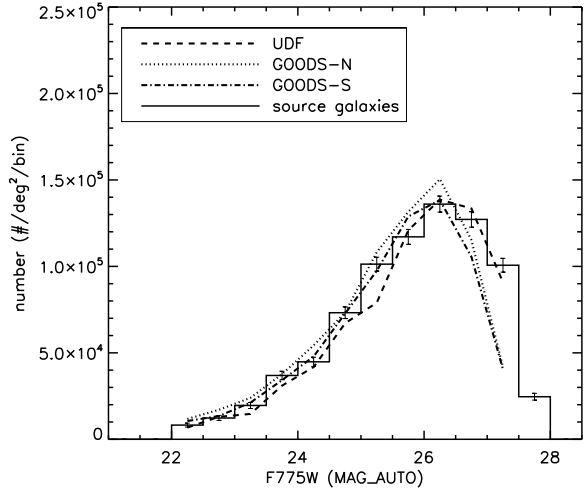}
\caption{Magnitude distribution of source galaxies. We use the
  publicly available {\it HST} images of the UDF and GOODS data to
  examine a potential excess in the magnitude distribution of our
  source galaxies with respect to these control field galaxies. We add
  noise to the UDF and GOODS images to match their depth to our ACS
  image of \elgordo. We find that the magnitude distribution of our
  source galaxies is similar to those from GOODS and UDF without any
  hint of an excess.
\label{fig_mag_dist}}
\end{figure}

The redshifted 4000~\AA~break at $z=0.87$ ($\mytilde7480$~\AA) is
well-bracketed by F625W and F850LP.  In the top panel of Figure~\ref{fig_cmd}, the
color-magnitude diagram clearly shows the red-sequence cluster
galaxies of \elgordo\ (the F625W$-$F850LP color of the bright end is
$\mytilde2$).  We define our source galaxies as the population bluer
than this red-sequence in the magnitude range $22< F775W<28$. 
In addition, we use the photometric redshift catalog of M12 to identify and remove
blue cluster members.

The blue points in Figure~ \ref{fig_cmd} represent the
source population selected in this way.  We find that only one
spectroscopic member (out of 89) is included in this selection.  
We note, however, that we preferentially selected red galaxies for our spectroscopic follow-up.
The resulting S/N threshold of the source at the faint end is about 5.

One may suggest source selection using all of the three filters (bottom panel of
Figure~\ref{fig_cmd}). However, we find that
this method does not provide advantage over the above color-magnitude relation.
The red-sequence is already well-separated with the color magnitude diagram. The blue cluster
candidates are severely blended with our source population in both panels.

Our source magnitude range is broad ($22< F775W<28$), and the choice of the upper and lower limits can be a
subject for debate. Obviously, we desire to avoid the cases where the sources near the bright end are mostly foreground galaxies or where the sources near the faint end are too noisy to be used for weak-lensing studies. Our test shows that the sources at both ends possess significant lensing signals (Appendix~\ref{section_mag_test}).

We applied a S/N cut by imposing that sources should have ellipticity measurement errors less than 0.25.
To avoid the pixellation artifact mentioned in Jee et
al.\ (2013), we require  the semi-minor axis $b$ (pre-seeing) 
to be greater than 0.4 pixels. Along with this constraint, we ensure that the half-light radius ($r_h$)
should be greater than 1.2 pixels (larger than the PSF). 

For lenses at high redshift, care must be taken in estimating source
redshifts because 
1) many cluster members are blue and thus their 4000~\AA ~feature is
weak,
2) the lensing signal is sensitive to source redshifts, and
3) the width of the source redshift distribution is non-negligible.

We address 1) by comparing the magnitude distribution of the source
population with those from other control fields. If a significant
fraction of blue cluster members is included in our source catalog
from the selection procedure just described, we expect to observe an
excess with respect to some control fields.  We use the publicly
available {\it HST} images of the Ultra Deep Field (UDF; Beckwith et
al.\ 2006 ) and the Great Observatories Origins Deep Survey (GOODS;
Giavalisco et al.\ 2004). Because these images are significantly
deeper than the current \elgordo\ images, it is necessary to add noise
to the control field images in order to match the depth. In addition,
we need to apply the same selection criteria such as magnitude range,
size, color, etc.  We display the comparison in
Figure~\ref{fig_mag_dist}. We find that the magnitude distribution of
our source galaxies is similar to those from GOODS and UDF without any
evidence for an excess from blue cluster member contamination.

Although no statistically significant excess is detected in the above comparison,
the top panel of Figure~\ref{fig_cmd} shows that our source selection
includes some blue cluster member candidates defined by the M12 photo-$z$ catalog.
For sources with F775W$<26.5$, the fraction is about $\mytilde5$\%.
Although negligibly small, this contamination rate is included in our source redshift estimation.

After confirming that the contamination fraction of blue cluster
members in our source catalog is negligibly small, we address 2) and 3)
using the UDF and GOODS photometric redshift catalogs of Coe et
al.\ (2006) and Dahlen et al.\ (2010), respectively.

The UDF photometric redshifts are derived from the ultra-deep ACS and
NICMOS images. For example, the limiting AB magnitude (S/N=10 in 0.2
arcsec$^{2}$) is $\mytilde29$ in F435W, F606W, F775W, and F850LP. This
catalog has been widely applied to our previous weak-lensing studies
of high-redshift clusters (e.g., Jee et al.\ 2011).  Because the UDF
field consists of a single ACS pointing ($3\arcmin\times3\arcmin$), it
does not allow us to address cosmic variance, which becomes a limiting factor in
high-redshift cluster mass determination. In Jee et al.\ (2009), it
was realized that sample variance was an important issue for deriving
the mass of the then highest-redshift cluster XMMU J2235.3-2557 at
$z=1.4$. Jee et al.\ (2009) estimated that the uncertainty on the
effective redshift of background galaxies used for their weak-lensing
study was $\simeq0.06$, based on examining two
additional redshift catalogs (Hubble Deep Field North: HDF-N, and
Ultra Deep Field Parallel Field: UDF-P).

Here, we expand the study of Jee et al.\ (2009) utilizing the
photometric redshift catalog from the GOODS data.  GOODS consists of
two separate fields, each of which covers an area of 10$\times$16
arcmin$^{2}$. We combine the public release of the GOODS photometric
catalog version 2.0\footnote{http://www.stsci.edu/science/goods/} with
the Dahlen et al.\ (2010) photo-$z$ catalog; this procedure is
necessary because the photo-$z$ catalog does not contain
photometry. The combined catalog contains 37,238 and 32,508 objects
for the northern and southern fields, respectively.  The number
density exceeds $\mytilde200$ arcmin$^{-2}$. This is much higher than
our source density ($\mytilde100$ arcmin$^{-2}$), although we believe
that a substantial fraction of the faint GOODS galaxies may not have
secure photometric redshifts.  These faint sources are likely to
default to a prior, for which Dahlen et al.\ (2010) use a luminosity
function.  Note that Coe et al.\ (2010) used the HDF-N prior for their
UDF photo-$z$ estimation.

As shown in Equation~\ref{eqn_sigma_c}, the lensing signal (surface
mass density) is scaled by the angular diameter ratio
$D_{ls}/D_s$. Because sources at lower than the cluster redshift are
not lensed, we compute $\beta$ as follows:
\begin{equation}
\beta  = \left < \mbox{max} \left (0,\frac{D_{ls}}{D_s} \right ) \right >.
\end{equation}
\noindent
Thus, the corresponding redshift is referred to as {\it effective}
(rather than {\it mean}) source redshift.  As mentioned above, we
apply our source selection criteria to the UDF and GOODS galaxies. In
addition, it is important to correct for the difference in depth and
the magnitude-dependent weighting scheme used in shear estimation.

After taking into account these subtleties, we obtain $\beta=0.257$
from the UDF photometric redshift catalog of Coe et al.\ (2006)
corresponding to an effective source plane redshift $z_{\rm eff}=1.32$.
Without the last two corrections, $\beta$ would be higher by
$\mytilde4$\%.  Substituting $\beta=0.257$ into
equation~\ref{eqn_sigma_c} gives $\Sigma_c \simeq 4050~M_{\sun}
\mbox{pc}^{-2}$. The resulting non-background contamination rate is
$\mytilde31$\%.

We obtain similar values from the GOODS photometric redshift catalog
of Dahlen et al.\ (2010).  The northern and southern fields yield
$\beta=0.268$ ($z_{\rm eff}$=1.35) and $0.262$ ($z_{\rm eff}$=1.33),
respectively. To estimate the Poisson scatter, we define 7 and 8
non-overlapping ACS pointings (i.e., size of the UDF) in the northern
and southern fields, respectively, and measure $\beta$ from each
pointing. The standard deviations in $\beta$ are 0.008 and 0.005 for
GOODS-N and -S, respectively. The $\beta$ value from UDF is consistent
with the mean value in GOODS-S and about 1 $\sigma$ lower than the
mean value in GOODS-N.  If we adopt the larger difference (between
GOODS-N and UDF) as the representative scatter in $\beta$, the
difference of $\Delta \beta=0.01$ ($\Delta z_{\rm eff}\simeq0.03$)
gives rise to a $\mytilde4$\% shift in mass.  The small difference in
$\beta$ between the two GOODS fields that are widely separated on the
sky is encouraging, although one has yet to establish the scatter
arising from cosmic variance based on more surveys available in
the future. Hereafter, we adopt the UDF photo-$z$ result to scale our
lensing signal.

As mentioned above, the width of the distribution also affects the
scaling of the lensing signal.  We obtain $ \left < \beta^2 \right
>=0.119\pm0.005$. Thus, the observed reduced shear $g^{\prime}$ is
related to the true reduced shear $g$ via $g^{\prime} = [1 + (\left
  <\beta^2 \right >/ \left <\beta \right >^2-1) \kappa ]g \simeq
(1+0.79 \kappa) g$ (Seitz \& Schneider 1997).  This correction is
important in the region where $\kappa$ is high, and we address the
effect when we fit a model to the observed cluster shear profile for
our mass determination. Omission of this correction alone leads to
overestimation of the cluster mass by as much as $\mytilde15$\%.

\subsubsection{Region B}
From the photo-$z$ catalog of M12, we select 410 objects with $0.7<z_{phot}<0.96$.
Among these 410 objects,  75 sources are located in Region B. After removing these cluster
member candidates, we apply the $22<F606W<28$ magnitude cut. 
As is done for the sources in Region A, 
we further trim the source catalog by ensuring that the sources are
sufficiently large ($b>0.4$ and $r_h>1.2$) and their ellipticity has
a reasonable measurement uncertainty $\delta e<0.25$.

We attempted to estimate a cluster galaxy contamination rate in these sources by
comparing their magnitude distribution with those from the UDF and GOODS fields
as is done for the sources in Region A. 
However, in contrast to our initial expectation,  we cannot find any excess in Region B.
We suspect that this may be because the  contamination from  ``red" cluster members is insignificant
in Region B or the source density in the field happens to be intrinsically low. 

The total number of sources in Region B is 3,255 ($\mytilde102$~arcmin$^{-2}$).
This source density is similar to that in Region A. This may seem surprising because in Region A
we combine the shapes from four different filters, which give more galaxies with smaller shape
measurement errors. However, in Region A there are more cluster galaxies, which we discard. This
offsets the increase due to the combination of different filters.

Using the UDF photo-$z$ catalog of Coe et al. (2006), we obtain $\beta=0.276$ corresponding
to the effective redshift $z_{eff}=1.37$. This value is slightly higher than the estimate $z_{\rm eff}=1.32$ in Region A.
When we use the GOODS photometric redshift catalog of Dahlen et al.\ (2010), the north and south fields
give $\beta=0.268$ ($z_{eff}=1.34$) and $0.263$ ($z_{eff}=1.33$), respectively. For our subsequent
analysis, the UDF result is adopted. However, we note that these different $\beta$ values causes only
insignificant changes in our final weak-lensing analysis. For example, the cluster masses are affected at most by $\mytilde4$\%, which is much smaller than the contribution from other sources.
In addition, we stress that the cluster lensing signal is dominated by the sources in Region A, although the sources in Region B helps us to reduce statistical uncertainties by allowing us to measure more robust tangential shears at large radii from complete circles.

\section{RESULTS}  \label{section_result}

\subsection{Two-dimensional Mass Reconstruction} \label{section_massmap}

\begin{figure*}
\includegraphics[width=9.5cm]{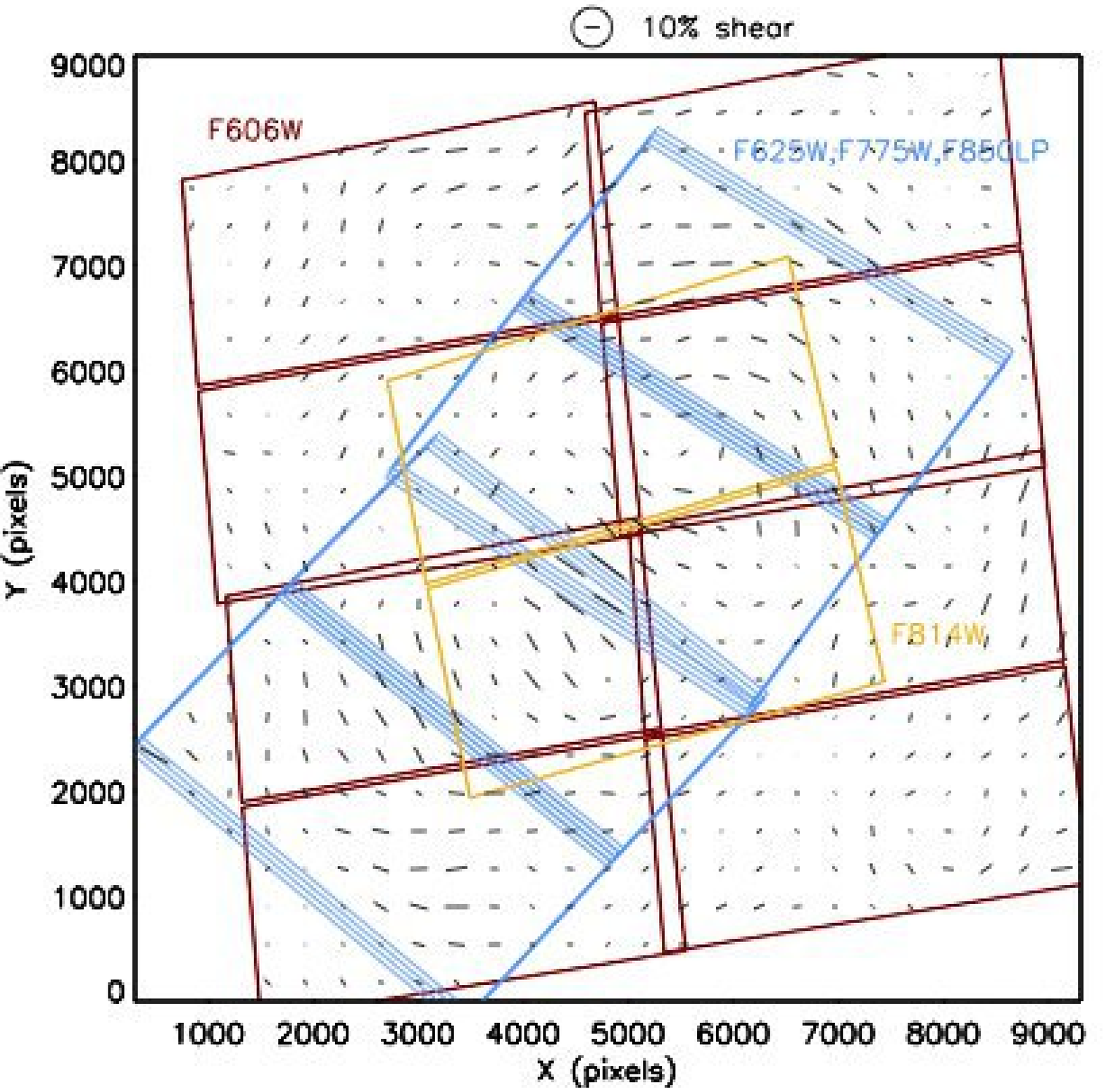}
\includegraphics[width=9.5cm]{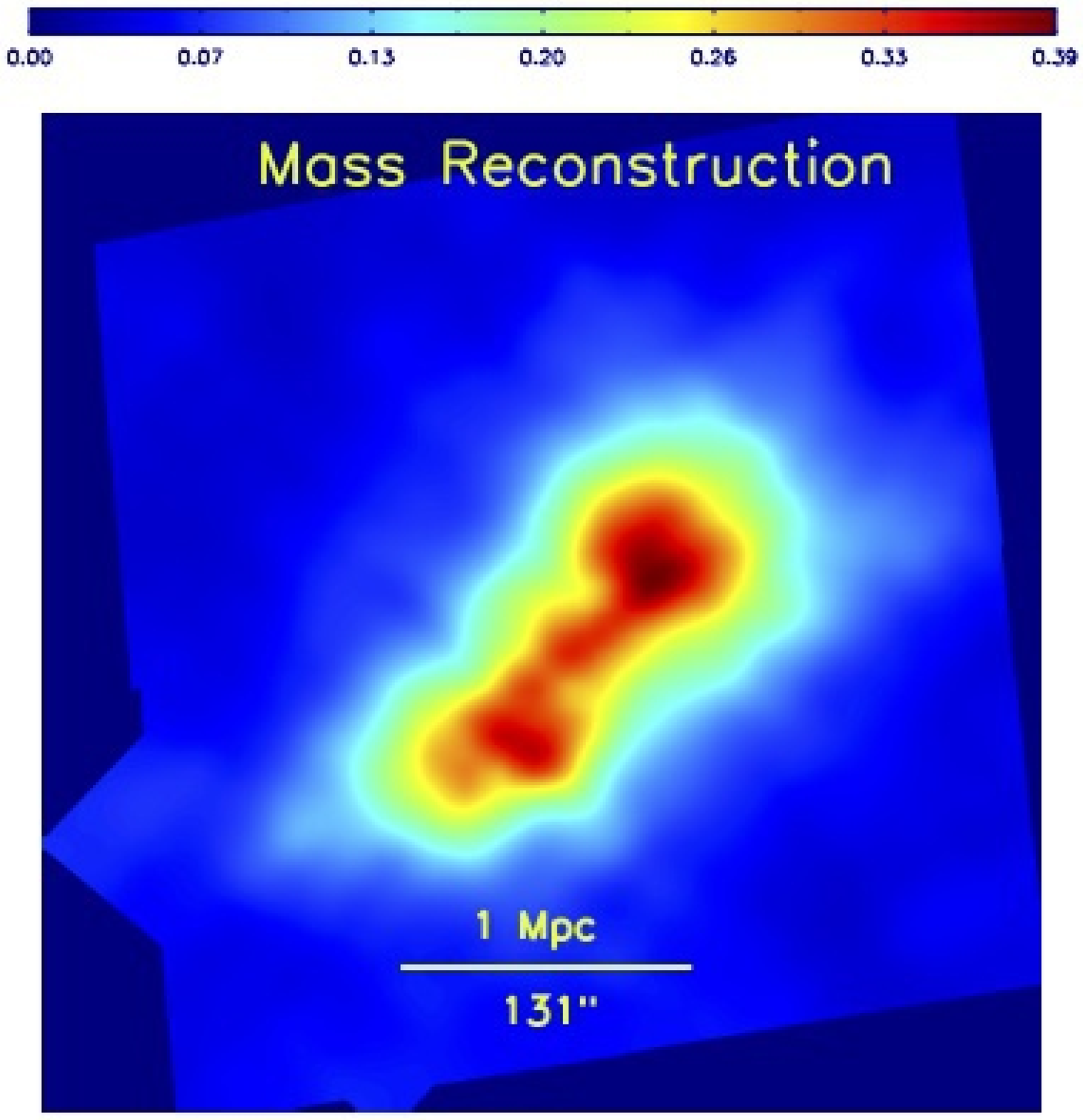}
\includegraphics[width=9.5cm]{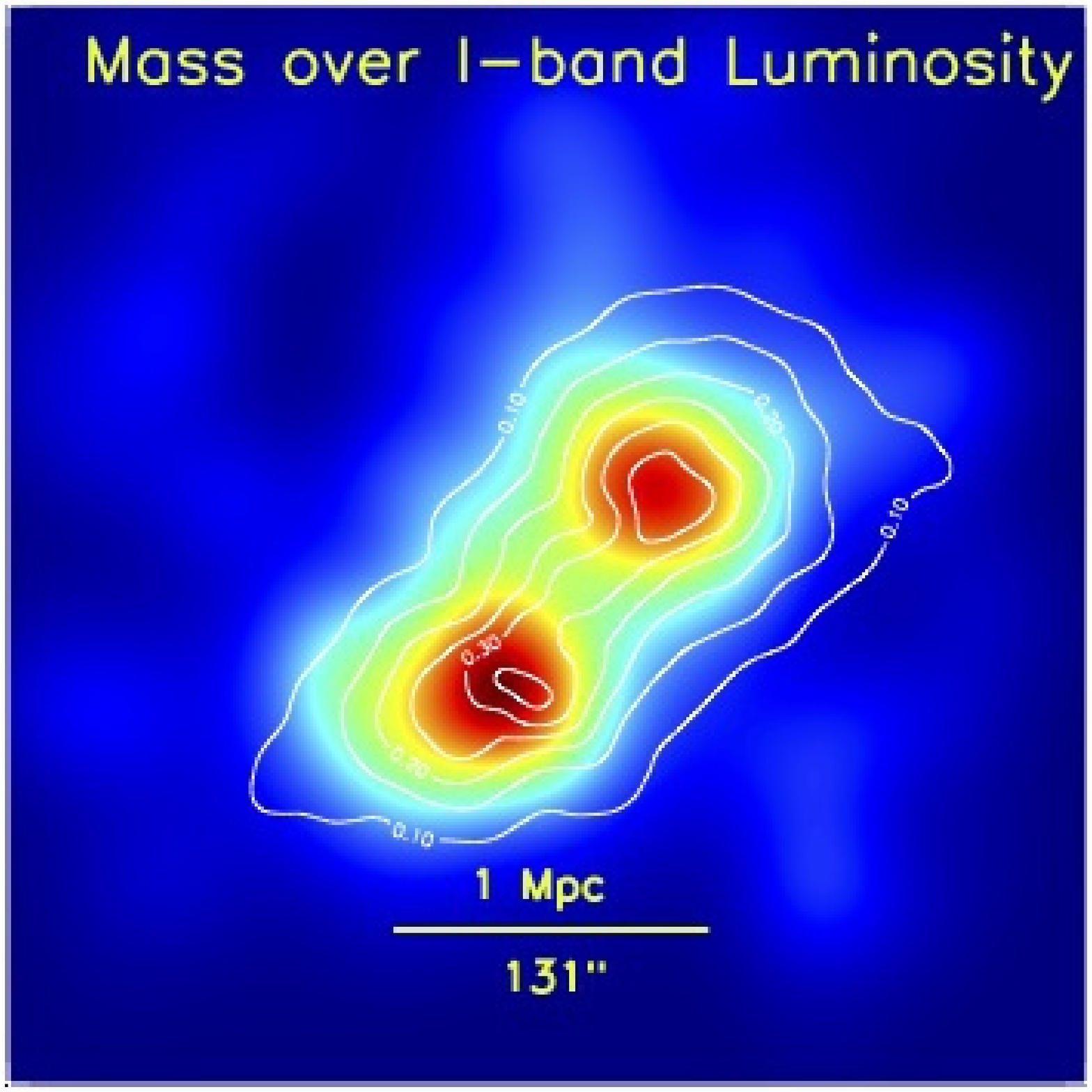}
\includegraphics[width=9.5cm]{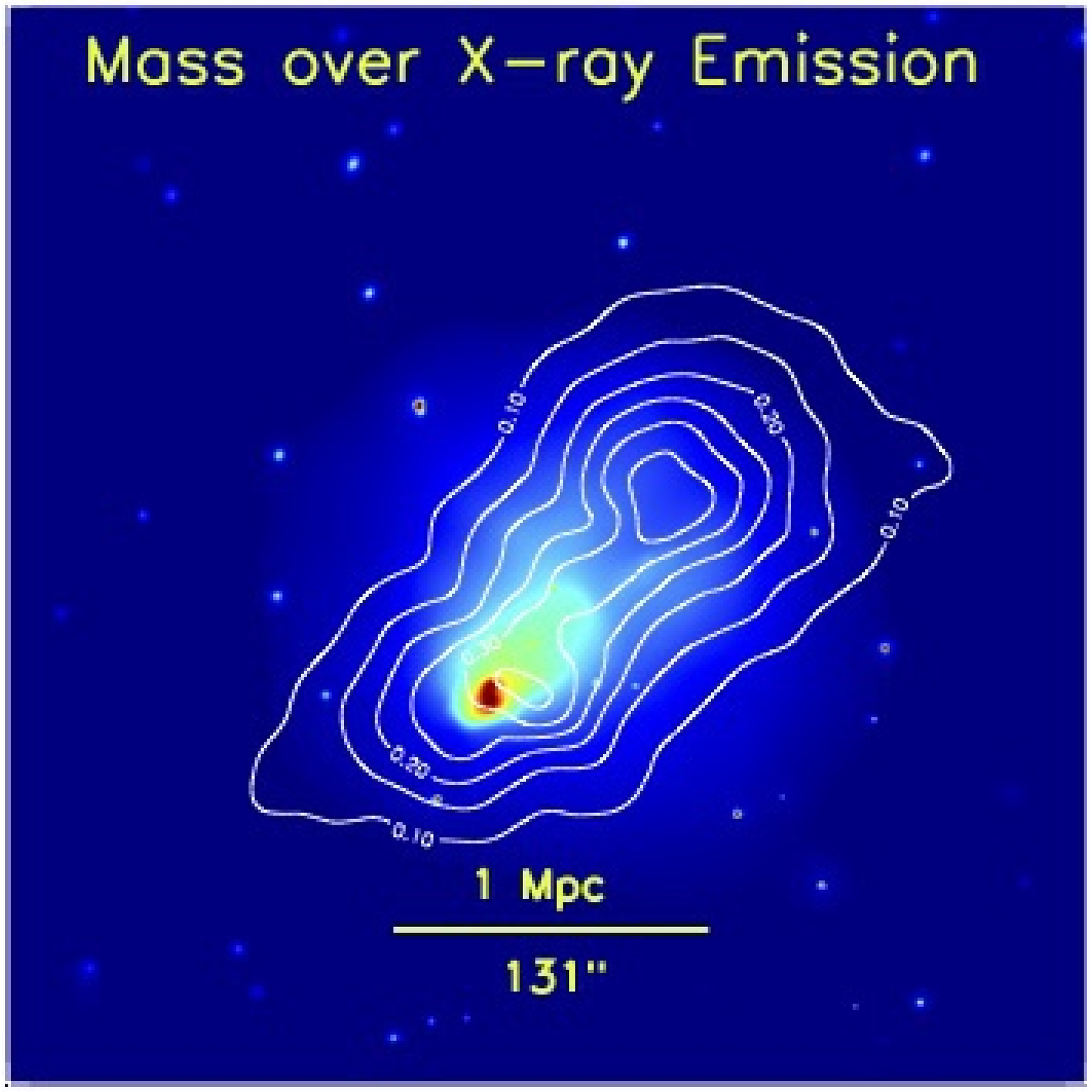}
\vspace{-1cm}

\caption{Two dimensional mass reconstruction of \elgordo. The
  ``whisker'' plot in the upper-left panel shows the smoothed ellipticity
  variation of background galaxies.  The orientation and length of the
  sticks represent the position angle and magnitude of the
  ellipticity, respectively. The stick inside the circle above the
  plot illustrates the size of a 10\% shear whereas the diameter of
  the circle shows the size (FWHM=$30\arcsec$) of the Gaussian
  smoothing kernel used here. 
  The upper-right panel displays the
  resulting two-dimensional mass reconstruction. 
  We performed the mass-sheet degeneracy ($\kappa \rightarrow
  1-\lambda + \lambda \kappa $) transformation in such a way that
  $\kappa$ becomes zero near the map boundaries.
  We overlay the mass contours on
 the smoothed optical luminosity and X-ray emission in the lower-left and -right panels, respectively.
\label{fig_whisker_n_mass}}
\end{figure*}

Weak-lensing distorts shapes of background galaxies only slightly, and
thus the detection of the signal requires averaging over a number of
galaxy shapes. The left panel of Figure~ \ref{fig_whisker_n_mass}
displays a so-called ``whisker" plot, where each whisker represents
both the direction and magnitude of the local average ellipticity. In
principle, it is possible to convert this shear field into a mass
distribution through the following relation:
\begin{equation}
\kappa (\bvec{x}) = \frac{1}{\pi} \int D^*(\bvec{x}-\bvec{x}^\prime) \gamma (\bvec{x}^\prime) 
d^2 \bvec{x} 
\label{k_of_gamma},
\end{equation}
where $D(\bvec{x} ) = - 1/ (x_1 - i x_2 )^2$ is the convolution
kernel.  However, the process is sensitive to noise, as is often the
case with inversion problems.  Many algorithms to overcome the
pitfalls of this direct inversion have been suggested. In this study,
we use the maximum entropy reconstruction method of Jee et
al.\ (2007b), which revised the earlier algorithm of Seitz et
al.\ (1998).  The method implements not only the non-linear relation between
galaxy ellipticity and shear, but also the S/N-dependent smoothing of the lensing signal utilizing the entropy of the convergence
values.
We verify that very similar results are obtained with
different mass reconstruction algorithms (e.g., Kaiser \& Squires
1993; Fischer \& Tyson 1997; Lombardi \& Bertin 1999) except that the
maximum entropy method yields the fewest spurious features near the
field edges.

\begin{figure}
\includegraphics[width=8.5cm]{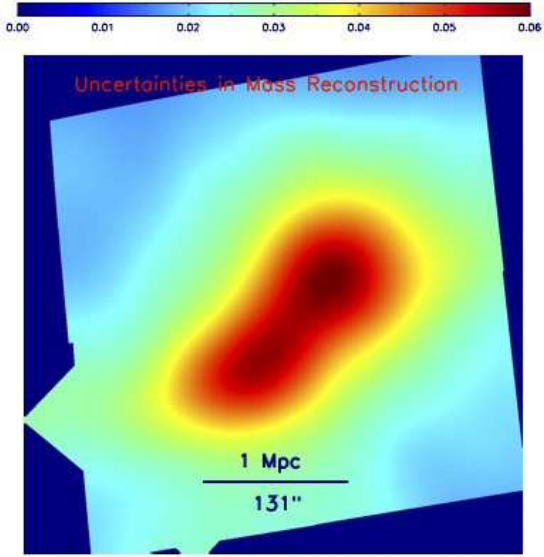}
\caption{Convergence ($\kappa$) error distribution derived from
our Hessian matrix. The rms value ranges
from $\delta \kappa\sim0.02$ to $\mytilde0.06$. Since our entropy regularization dampens
possible fluctuations near the field boundaries, the rms values are much lower in those regions.
This trend is reversed in mass reconstructions performed without any regularization because
fewer galaxies are available near the boundaries.
\label{fig_rmsmap}}
\end{figure}

The uncertainties in our mass reconstruction are estimated from a Hessian matrix.
The Hessian matrix is derived by taking the second derivatives of the target function
(i.e., function to minimize) with respect to parameters (i.e., convergence $\kappa$).
Under the assumption that the error distribution is Gaussian, we can adopt the resulting matrix
elements as inverse of the covariances.
Readers are referred to Bridle et al. (1998) for further details. We note that bootstrapping
is not a feasible solution because our maximum entropy method is slow
(several hours to converge). The resulting convergence
rms map is displayed in Figure~\ref{fig_rmsmap}.  The rms value ranges
from $\delta \kappa\sim0.02$ to $\mytilde0.06$. Since our entropy regularization dampens
possible fluctuations near the field boundaries, the rms values are much lower in those regions.
This trend is reversed in mass reconstructions performed without any regularization because
fewer galaxies are available near the boundaries.

\begin{figure*}
\hspace{-0.8cm}
\includegraphics[width=20cm]{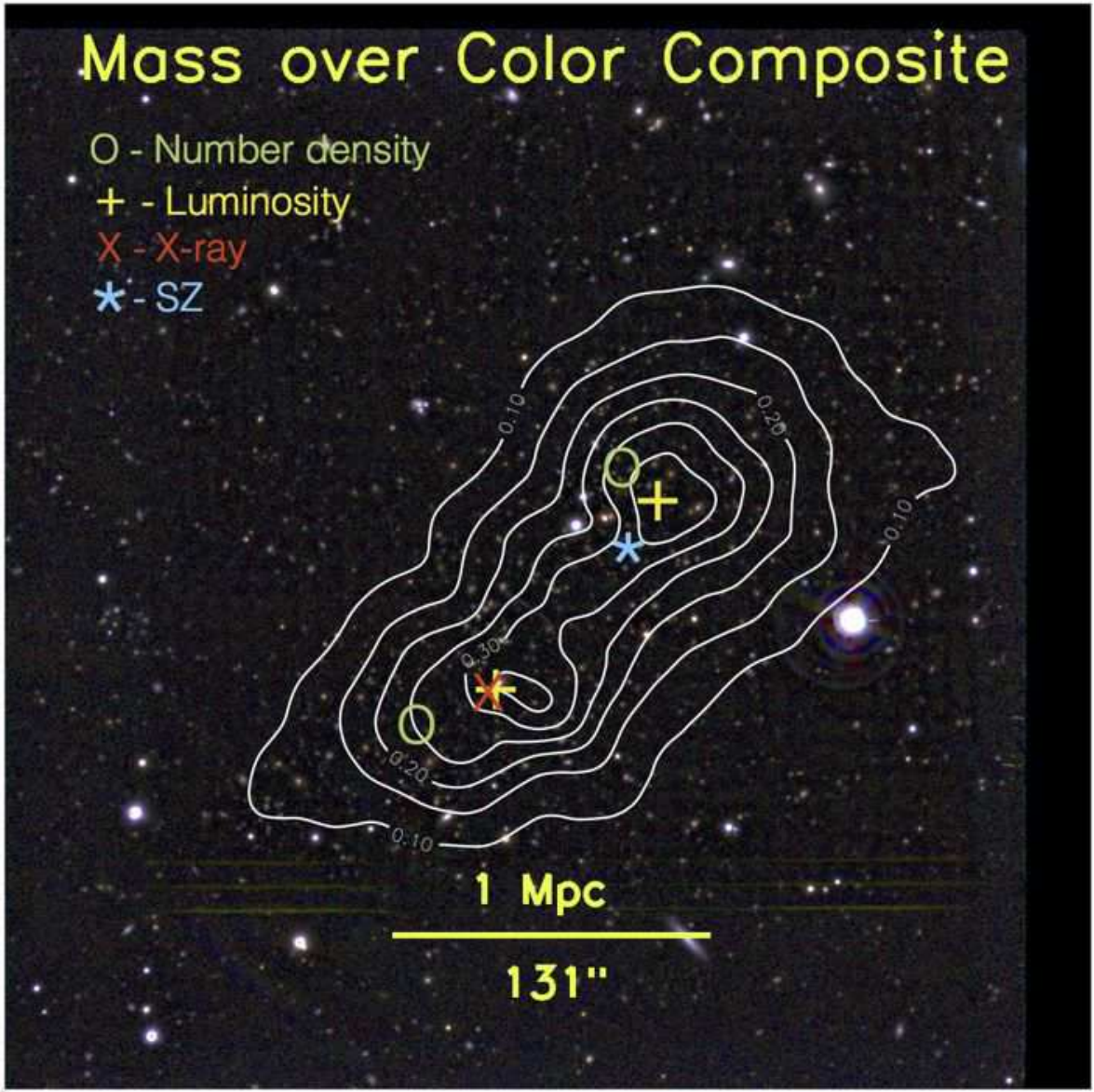}
\vspace{-2.0cm}
\caption{Mass contours overlaid on the VLT-SOAR color-composite image (using
  the $z$, $i$, and $r$ filters for red, green, and blue,
  respectively). Various measures of the cluster centroids are
  shown as illustrated.  The mass contours are depicted using solid
  white lines. North is up and east is left. 
\label{fig_massmap_contour}}
\end{figure*}

\begin{figure*}
\hspace{-0.8cm}
\includegraphics[width=9.2cm]{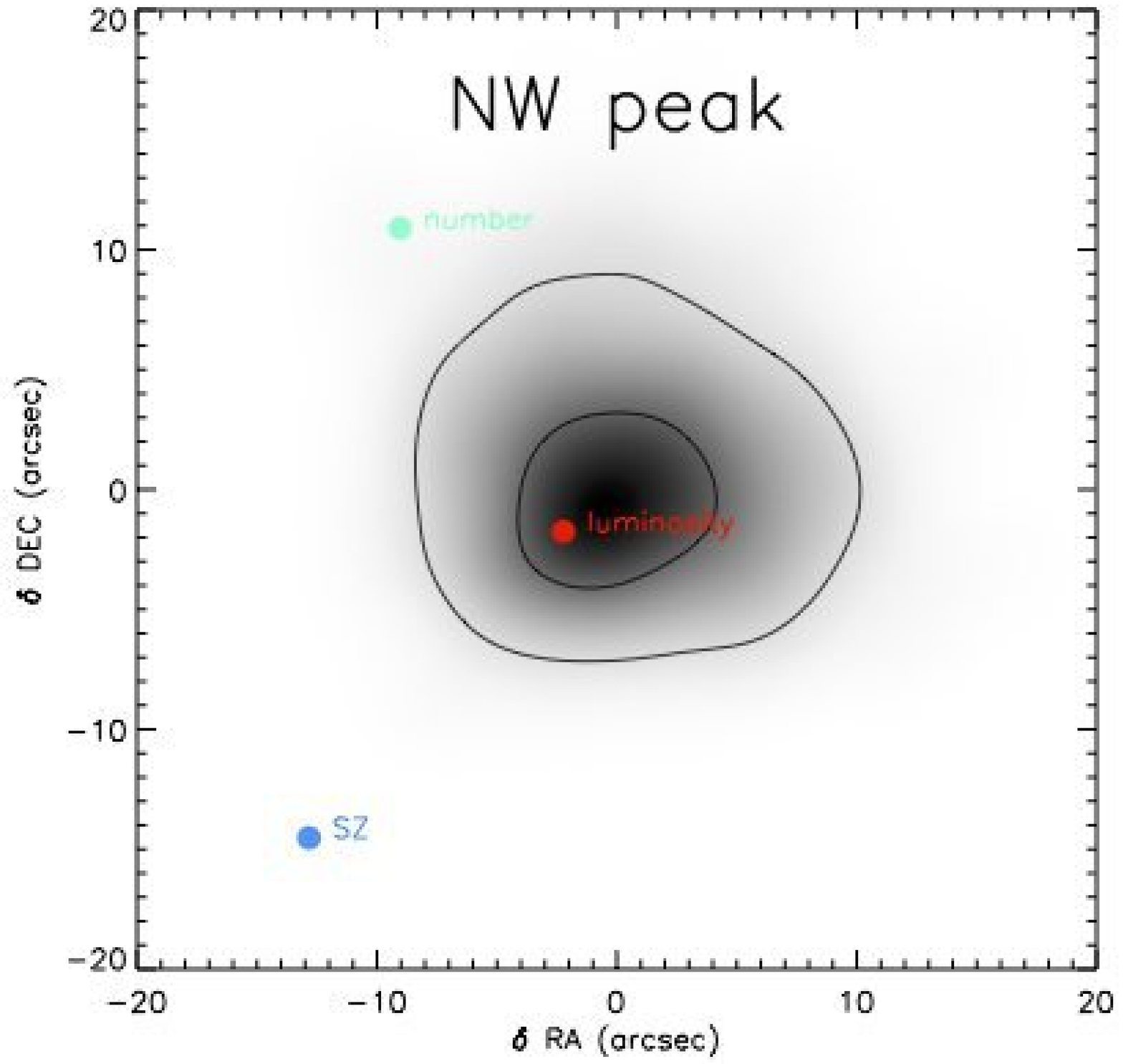}
\includegraphics[width=9.2cm]{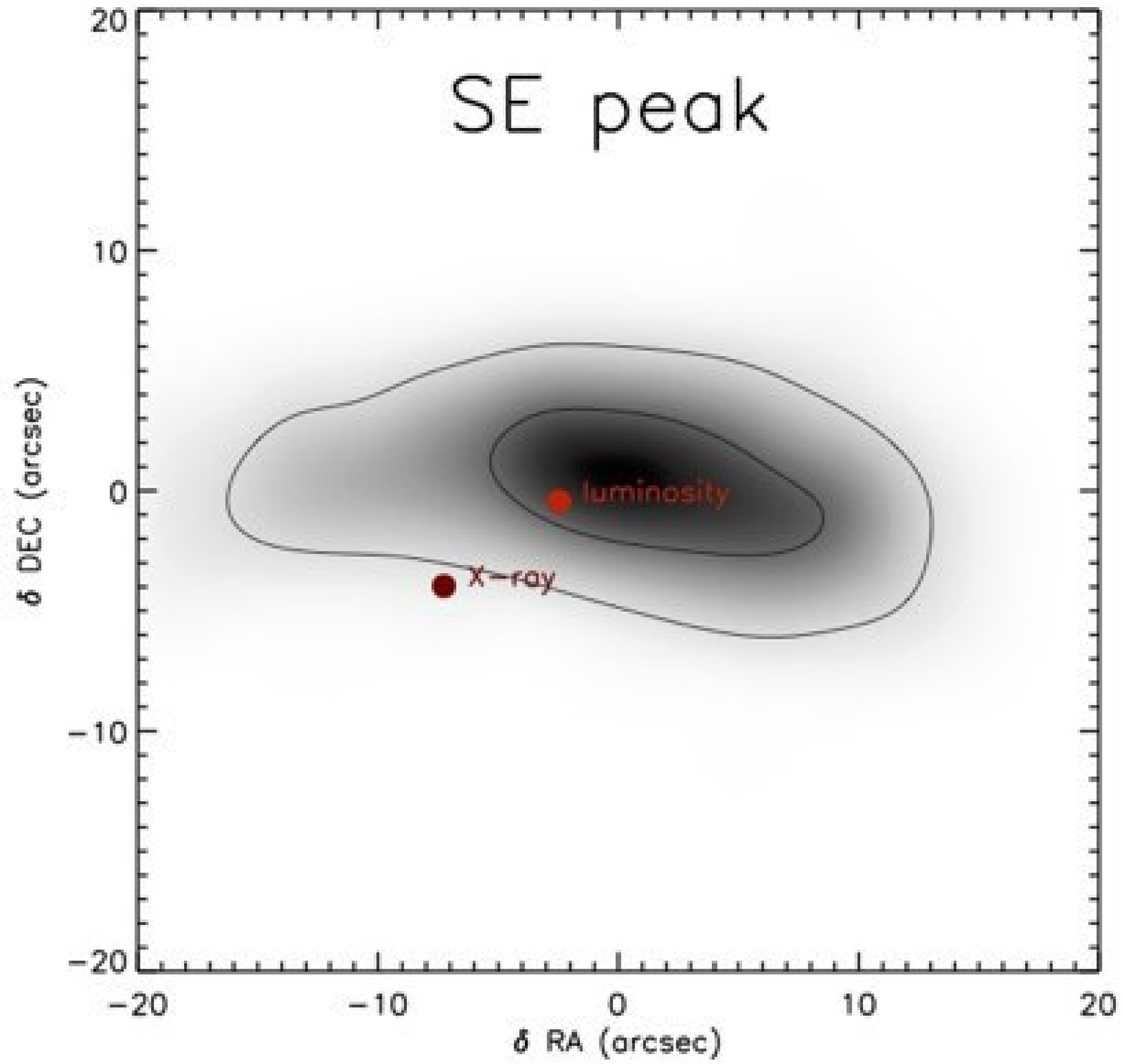}
\caption{Bootstrapping test of centroid distribution. We use the {\tt FIATMAP} code
to carry out the experiment. The inner- and outer contours represent 1$\sigma$ and 2$\sigma$
limits. The centroid distribution of the SE peak is elongated east-west, which is similar to
the profile of the convergence $\kappa$.
 The (0,0) positions correspond
  to positions of R.A.=01:02:50.601, Decl.=$-$49:15:04.48 for the NW
  component and R.A.=01:02:56.312, Delc.=$-$49:16:23.15 for the SE
  one.
\label{fig_bootstrap_centroid}}
\end{figure*}

Our weak-lensing analysis reveals that \elgordo\ consists of two
massive subclusters separated by $\mytilde700\hubblem$ kpc, closely resembling
the cluster galaxy light distribution (see 
Figure~\ref{fig_whisker_n_mass}).  The shear peaks are located at
R.A.=01:02:50.60, Decl.=$-$49:15:04.5 for the NW component and
R.A.=01:02:56.31, Delc.=$-$49:16:23.2 for the SE one.  This bimodal
distribution can also be inferred by the whisker plot showing the
tangential alignments of the sticks around these two mass clumps. The
X-ray image, however, does not show any significant gas overdensity
for the northwestern (NW) mass clump (the lower-right panel of Figure~
\ref{fig_whisker_n_mass}).  Figure~\ref{fig_massmap_contour}
summarizes the comparison of the mass centroids with the centroids
of X-ray emission, number density, luminosity, and SZ decrement.

\subsection{Significance and Locations of the Weak-Lensing Mass Peaks}

In order to investigate the statistical significance of the difference
between the two mass peaks and other centroids, we measure mass peak
centroids from the 1000 bootstrap runs.
We use the {\tt FIATMAP} code (Fischer \& Tyson 1997) to carry out this
experiment.
 The centroid is
determined by iteratively computing first moments from the convergence
map. We use a FWHM=20$\arcsec$ Gaussian for weighting
$\kappa$. Figure~\ref{fig_bootstrap_centroid} displays the results for
both mass peaks, where the (0,0) position is referenced to the 
location of each component's shear peak. The
peak of the NW mass clump coincides with the NW galaxy luminosity
peak.  Interestingly, the centroid of the SZ decrement is close to
this luminosity peak, although it is unclear whether or not the SZ
centroid is physically representative of the location of the NW
cluster potential well.  The southeastern (SE) mass peak is close
to the corresponding SE galaxy luminosity peak.
The X-ray emission is strongest near this SE mass peak. The distance
between the X-ray and SE mass peaks is $\mytilde8\arcsec$.  Based on
the above bootstrapping analysis, the significance of the offset is at
the $\mytilde2\sigma$ level.  In contrast to the galaxy light
distribution, the galaxy number density peaks show large offsets with
respect to the mass centroids. The NW number density centroid is
offset from the corresponding mass peak by $\mytilde15\arcsec$
($\gtrsim 2\sigma$), and the SE number density centroid is separated
from the SE mass peak by $\mytilde400\hubblem$ kpc ($\mytilde50\arcsec$),
which is outside the plotting region in
Figure~\ref{fig_bootstrap_centroid}.

We measure the significance of the mass peaks utilizing the rms map
(Figure~\ref{fig_rmsmap}). The background level is determined within
the mass reconstruction field.
Because the lensing signal is still strong at the field boundary, this
estimation of the significance should be regarded as conservative.  
As we lift the mass-sheet degeneracy by enforcing
the convergence values near the field boundaries to approach zero,
we effectively take into account the nonlinear $g=\gamma/(1-\kappa)$ relation
between shear and convergence.
The significance of the NW mass clump ($\mytilde9\sigma$ within a
$r=50\arcsec$ aperture) is slightly higher than that of the SE mass
clump ($\mytilde 6\sigma$) and is consistent with the mass ratio of
the two subclusters (see
\textsection\ref{section_mass_determination}).

\subsection{Mass Determination of \elgordo} \label{section_mass_determination}

\begin{figure}
\begin{center}
\includegraphics[width=8.2cm]{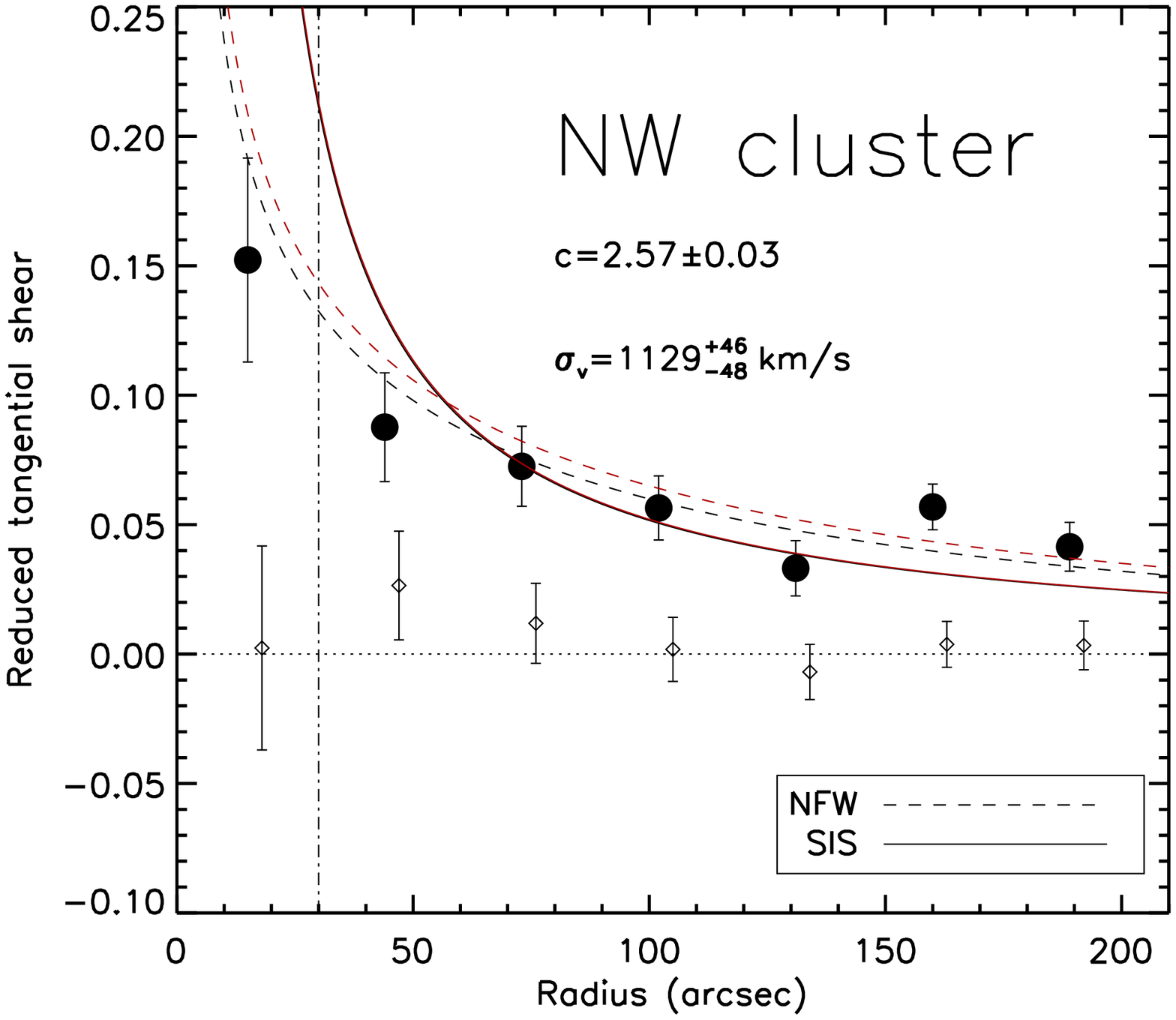}
\includegraphics[width=8.2cm]{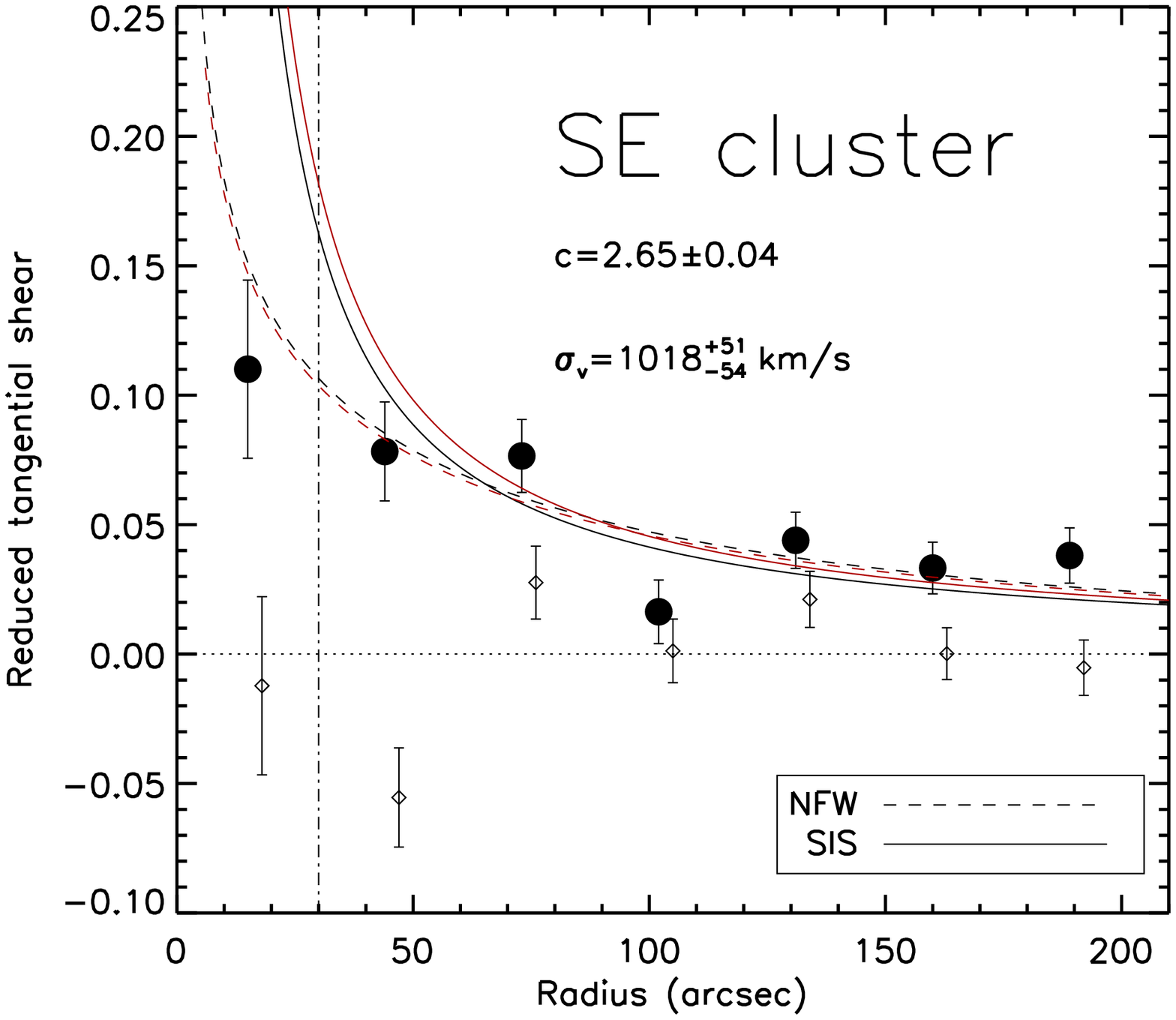}
\includegraphics[width=8.2cm]{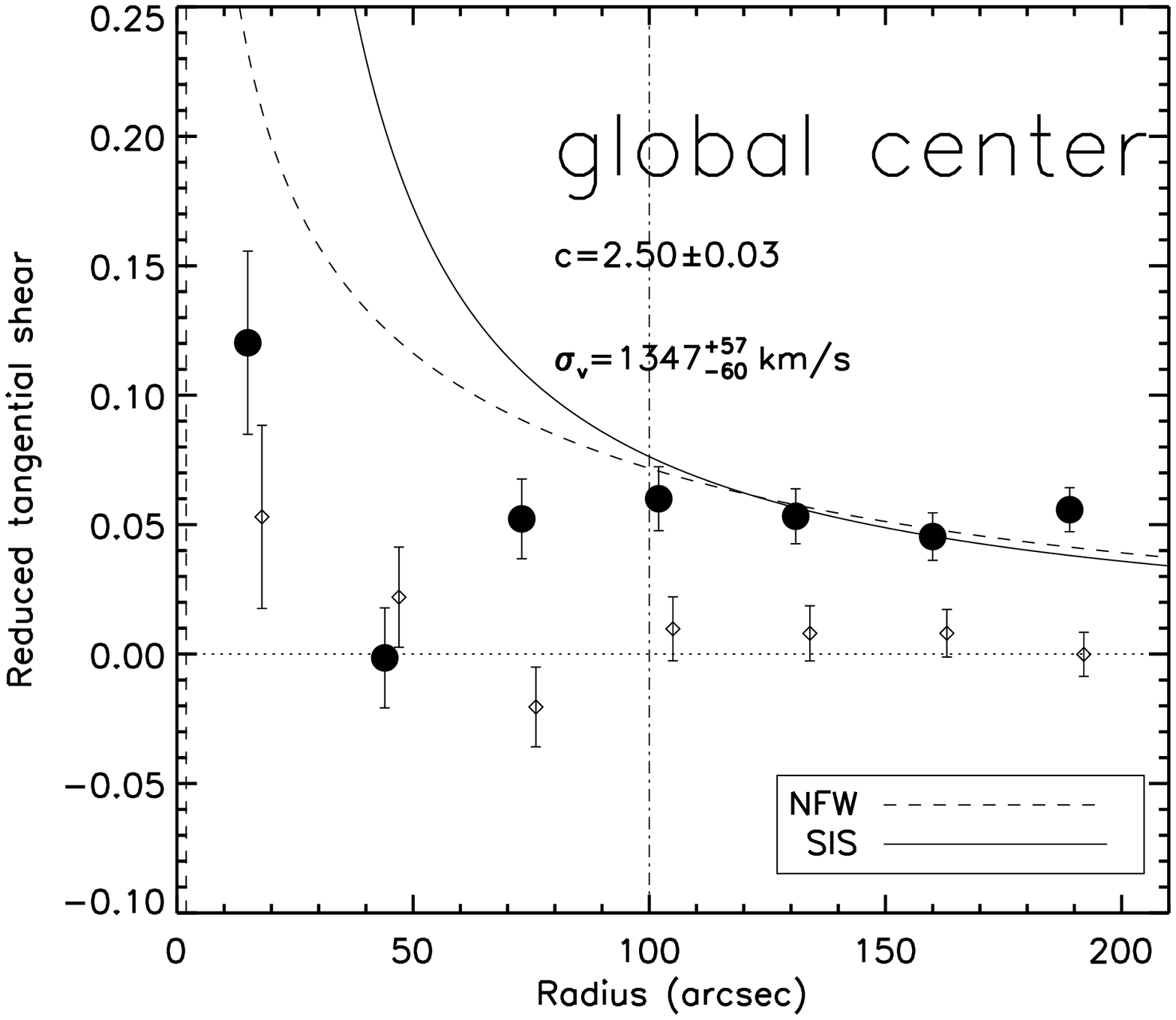}
\end{center}
\caption{Reduced tangential shear profile. Filled circles represent
  the tangential shear while open diamonds show the 45-deg rotation
  test results. The vertical dot-dashed line shows the cut-off radius, inside
  which the signal is not used to constrain the models.  The solid and
  dashed lines represent the best-fit SIS and NFW models,
  respectively.  The black lines are the results when we use this 1D
  tangential shear profiles whereas the red lines show the results
  from our simultaneous 2D fits (see text).  The displayed
  concentration  and predicted velocity dispersions
  are the results from the 1D fits.
\label{fig_tangential_shear}}
\end{figure}

One of the most cosmologically interesting properties of \elgordo\ is
its virial mass.  The cluster redshift $z=0.87$ approximately
corresponds to half the age of the universe, and according to the
standard $\Lambda$CDM prediction, the $z=0.87$ mass function at the
high end is expected to be several orders of magnitude lower than the
amplitude at $z=0$. Therefore, the cosmological leverage of
\elgordo\ is significant {\it if} the mass of the cluster is indeed as
high as indicated by its X-ray, SZ, and dynamical data.  M12 showed
that the combined mass estimate of \elgordo\ is near the 95\% exclusion
curve of Mortonson et al.\ (2011) when the full ACT+SPT 2800 deg$^2$
survey area is considered, although M12 caution that, because of the
large mass uncertainty, it would be premature to regard the cluster as
a challenge to the current $\Lambda$CDM paradigm.

The mass distribution of \elgordo\ is bimodal, and thus we must take
care in determining the virial mass of the system. Because the current
ACS data covers a small area, it is not feasible to use a
model-independent method such as aperture mass densitometry, which
requires shear measurement over a considerably larger area.
Therefore, in this study we use parametric models and compare the
expected shears with the observed galaxy ellipticity.

Conventionally, this parametric approach is implemented by first
measuring a one-dimensional (1D) azimuthally averaged tangential shear
profile around the center of each mass component and then fitting an
independent model to each profile. Although our presentation below
includes the results from this approach, in the study of \elgordo\ we
regard this method as biased for the following reasons.

The main weakness of this method is that this 1D analysis assumes that
there is only a single halo whereas the shapes of the source galaxies
are affected by other substructures. Since \elgordo\ is comprised of at least
two massive halos, this weak-lensing signal interference
will bias the cluster mass non-negligibly, although
we cannot make a general statement on the direction of the shift.
Another potential issue is the sensitivity of the profile shape to the choice of
the halo center.  In general, choosing centers determined by
mass reconstruction can bias the mass estimate high because
the decision is based on the noisy realization.
In this study, we assume that the luminosity peaks
(Figure~\ref{fig_massmap_contour}) are also the centers of the dark
matter halos, which turns out to be in good agreement with the results
from the two-dimensional (2D) analysis discussed below. Fortunately, in
the case of \elgordo, the centroid of each component is tightly constrained
by the lensing.

Our second approach is to fit two model halos simultaneously to the
ensemble of individual galaxy ellipticities.  This overcomes the
pitfalls of the first method. Now we free the centroid of each
cluster, which explores the impacts of centroid bias in cluster mass
estimation.  The uncertainties of the masses obtained from this method
are estimated, therefore, after marginalizing over the centroids of
the two halos using a Markov-Chain-Monte-Carlo (MCMC) analysis.

Figure~\ref{fig_tangential_shear} shows the 1D tangential shear
profile and the model fit results. Also displayed are the
45-$\deg$ rotation test results (diamond), which provide a useful
diagnostic for residual systematics. This so-called B-mode signal does not indicate any
significant residual systematics. Note that we corrected for the  small difference in $\beta$ when
the sources in Region A and B are combined.

Prior to fitting, we need to exclude the shears at small radii for
each halo because in this regime
1) current theory does not converge on the behavior of the halo
profile,
2) the chance of cluster galaxy contamination is high,
3) the signal shape is sensitive to the choice of center,  and
4) the weak-lensing assumption is not valid.
Currently, no simulation-based study on the optimal choice of the
cutoff radius is present.  In this paper, we adopt $r=30\arcsec$,
which is greater than the expected Einstein radius 
($\mytilde10\arcsec$) of each halo for the effective source plane
redshift $z_{\rm eff}\sim1.3$ (see \textsection\ref{section_redshift}
for our source redshift estimation).

Despite the above caveats in the 1D analysis, the amplitudes of the
tangential shears in Figure~ \ref{fig_tangential_shear} indicate that
the two halos of \elgordo\ are indeed massive. The solid and dashed
lines show our best-fit results using the singular isothermal sphere
(SIS) and the Navarro-Frenk-White (NFW; Navarro et al.\ 1997)
models. We assume the Duffy et al.\ (2008) mass-concentration relation
in our NFW fits.  The black lines are the results when we use the 1D
tangential shear profiles whereas the red lines show the results from
the 2D fits described below.  Conversion of the SIS fit result to
velocity dispersion is straightforward.  From the 1D fits, the
predicted velocity dispersions for the NW and SE subclusters are
$\sigma_v=1129_{-48} ^{+46}~\mbox{km} ~\mbox{s}^{-1}$ and
$1018_{-54}^{+51}~\mbox{km}~\mbox{s}^{-1}$, respectively. These values
are consistent with the dynamical (spectroscopic) measurements of
$1290\pm134~\mbox{km}~\mbox{s}^{-1}$ and
$1089\pm200~\mbox{km}~\mbox{s}^{-1}$ for the NW and SE components,
respectively (M12).  The bottom panel in
Figure~\ref{fig_tangential_shear} displays the results when we measure
the tangential shear around the global center of the two
components. We exclude shear values at $r<100\arcsec$ in this case. We
do not claim that this cutoff radius $r=100\arcsec$ is a legitimate
choice because the interference of the lensing signal from the two
subclusters should extend beyond this value; clearly, the shape of the
radial profile at $r<100\arcsec$ cannot be modeled by the profile of a
single halo.  For this choice of center, our 1D SIS fit predicts
$\sigma_v=1347_{-60}^{+58}~\mbox{km}~\mbox{s}^{-1}$, which is also
consistent with the direct measurement $1321\pm106~\mbox{km}~\mbox{s}
^{-1}$ (M12).

From the NFW fits, we estimate $M_{200c}
=(1.17\pm0.17)\times10^{15}\hubblem M_{\sun}$ and
$(0.79\pm0.14)\times10^{15}\hubblem M_{\sun}$ for the NW and SE
mass clumps, respectively.  Note that $M_{200c}$ refers to a mass
within a sphere inside which the mean density equals 200 times the
{\it critical} density of the universe at the cluster redshift. The circles
traced by the
corresponding $r_{200c}$ values ($\mytilde1.6\hubblem$ Mpc and
$\mytilde1.4\hubblem$ Mpc for the NW and SE, respectively) are greater
than the size of our ACS field. Thus,
a more relevant mass might be $M_{500c}$, whose defining radius is a
factor of two smaller.  Table 1 summarizes our 1-D mass estimation
results including these $M_{500c}$ values.

\begin{deluxetable*}{ccccccc}
\tablewidth{0pt}
\tabletypesize{\scriptsize}
\tablecaption{Mass Estimates of \elgordo\ based on 1D fit. \label{tab_mass1}}
\tablehead{  \colhead{Clusters} & \colhead{$\sigma_v$}  & \colhead{$c_{200c}$}  & \colhead{$R_{500c}$} & \colhead{$M_{500c}$}  & \colhead{$R_{200c}$} & \colhead{$M_{200c}$} \\
 & \colhead{(\kms)}  & \colhead{}  & \colhead{($\hubblem$ Mpc)} & \colhead{(\solarm)}  & \colhead{($\hubblem$ Mpc)} & \colhead{(\solarm)}  }
\startdata
NW                                     & $1129_{-48}^{+46}$  &  $2.57\pm0.03$  &  $0.60\pm0.03$ &  $4.00\pm0.54$  & $1.57\pm0.08$  & $11.7\pm1.7$ \\
SE                                       & $1018_{-54}^{+51}$  &  $2.65\pm0.04$  &  $0.53\pm0.03$ & $2.76\pm0.49$  &  $1.38\pm0.08$ & $7.9\pm1.4$ \\
global\tablenotemark{a}     &    $1347_{-60}^{+58}$   &  $2.50\pm0.03$ & $0.66\pm0.03$  &  $5.30\pm0.08$  & $1.74\pm0.09$ & $15.9\pm2.4$ \\
NW+SE\tablenotemark{b}     &      		...	                   &	...		                  &  $1.31\pm0.06$ & $16.8\pm3.2$ & $2.02\pm0.16$ &  $25.4\pm4.9$  
\enddata
\tablecomments{$^a$ We assume that there is a single halo centered on
  the middle point of the NW and SE subclusters.  The result is
  obtained from the outer ($>100\arcsec$) part of the tangential
  shears.
 $^b$ We sum the masses of the NW and SE subclusters obtained from the
  individual tangential shear fitting results.}
\end{deluxetable*}

Our full 2D analysis (simultaneous fitting of two halos with freed
centroids) produces cluster mass values that are slightly different from
the above 1D fitting results, although with overlapping error bars;
the predicted amplitudes are compared in
Figure~\ref{fig_tangential_shear}.  Because it is obvious that the
interference of the lensing signal between the two halos is
non-negligible, we are confident that this second method should
produce more unbiased results.  We estimate the NW and SE components'
masses to be $M_{200c} =(1.38\pm0.22)\times10^{15} \hubblem
M_{\sun}$ and $(0.78\pm0.20)\times10^{15} \hubblem M_{\sun}$,
respectively.  These values are obtained with the same ($r<30\arcsec$)
cluster-core-exclusion radius as used in our 1D fitting.   The predicted
velocity dispersions are $\sigma_v=1133_{-61}^{+58}~\mbox{km}~\mbox{s}^{-1}$
and $1064_{-66}^{+62}~\mbox{km}~\mbox{s}^{-1}$ for the NW and SE
subclusters, which are also in good agreement with the spectroscopic
measurements.  We summarize our 2D fitting
results in Table 2.

As mentioned above, the difference between our 2D and 1D results is small. Considering
the potential sources of bias in the 1D analysis discussed above, we note that this small
difference is interesting, although it is premature to draw any general conclusion from this single
case. Because most weak-lensing studies in the literature do not use this 2D simultaneous fitting to
constrain cluster masses (despite the fact that most clusters possess non-negligible substructures), 
it will be an important subject of future studies to examine the size and direction of the bias from a large sample.

\begin{deluxetable*}{ccccccc}
\tablewidth{0pt}
\tabletypesize{\scriptsize}
\tablecaption{Mass Estimates of \elgordo\ based on simultaneous 2D fit. \label{tab_mass2}}
\tablehead{  \colhead{Clusters} & \colhead{$\sigma_v$}  & \colhead{$c_{200c}$}  & \colhead{$R_{500c}$} & \colhead{$M_{500c}$} & \colhead{$R_{200c}$} & \colhead{$M_{200c}$} \\
 & \colhead{(\kms)}  & \colhead{}  & \colhead{($\hubblem$ Mpc)} & \colhead{(\solarm)}  & \colhead{($\hubblem$ Mpc)} & \colhead{(\solarm)}  }
\startdata
NW             &  $1133_{-61}^{+58}$  &  2.54$\pm$0.04  &  $0.63\pm0.03$ & $4.58\pm0.71$   & $1.65\pm0.10$ &   $13.8\pm2.2$ \\

SE              & $1064_{-66}^{+62}$  &  2.66$\pm$0.06  &   $0.53\pm0.03$ & $3.39\pm0.56$  &     $1.37\pm0.11$  &  $7.8\pm2.0$ \\
NW+SE     &      		...	                   &	...		       &               $1.34\pm0.07$ & $18.0\pm3.4$ &        $2.09\pm0.19$ &  $27.6\pm5.1$  
\enddata
\end{deluxetable*}

\begin{figure*}
\begin{center}
\includegraphics[width=8.5cm]{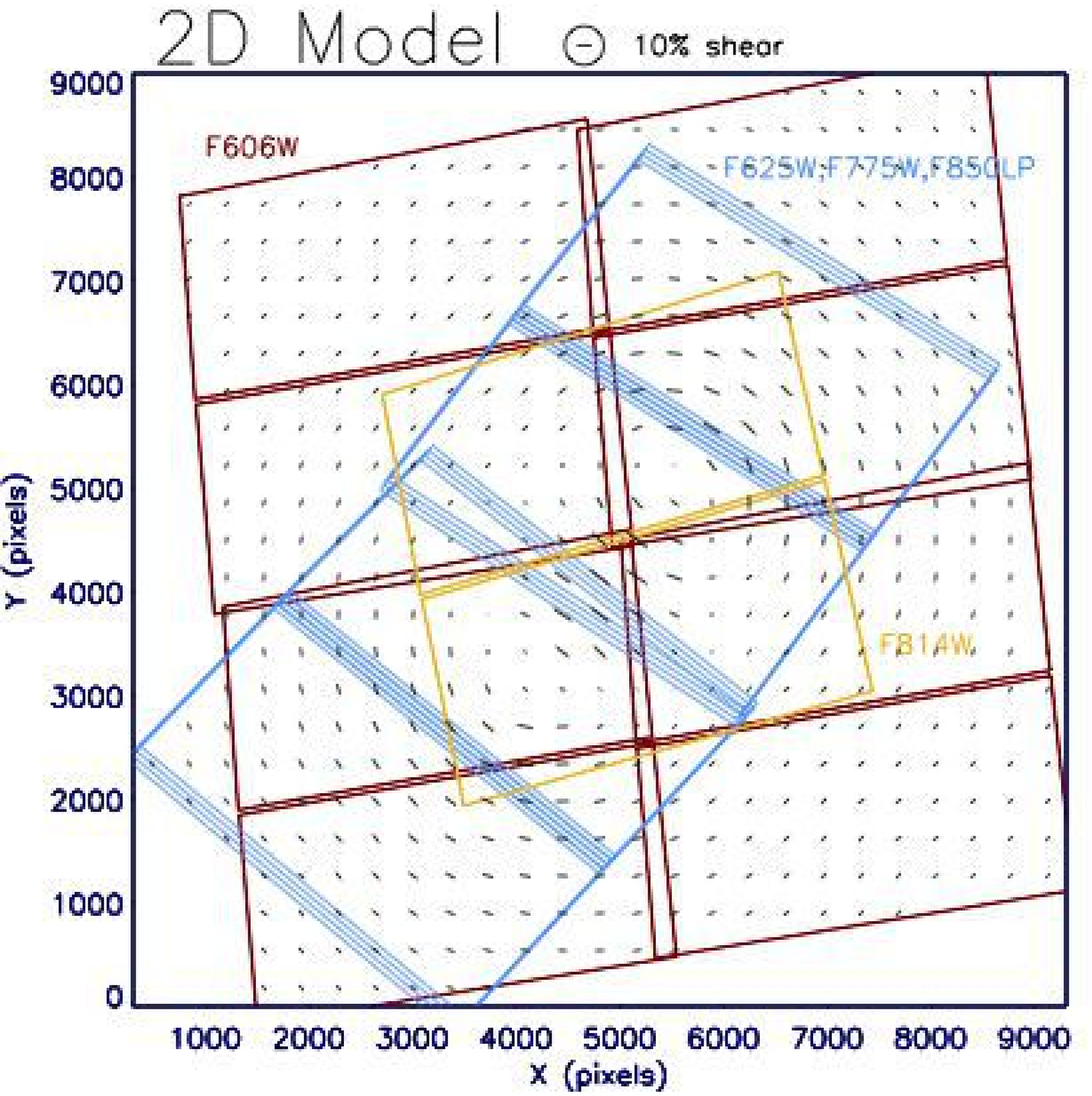}
\includegraphics[width=8.5cm]{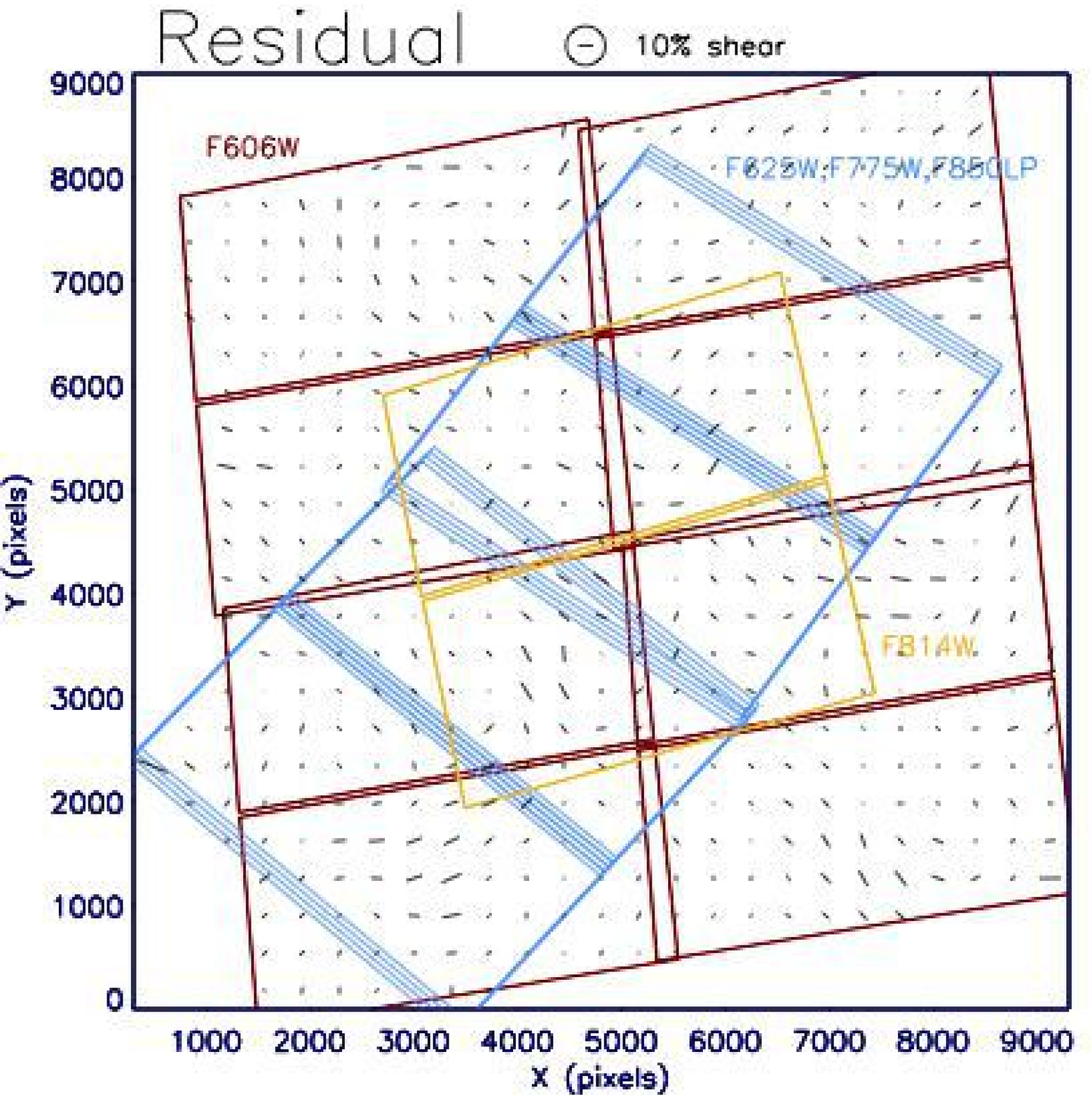}
\includegraphics[width=8.5cm]{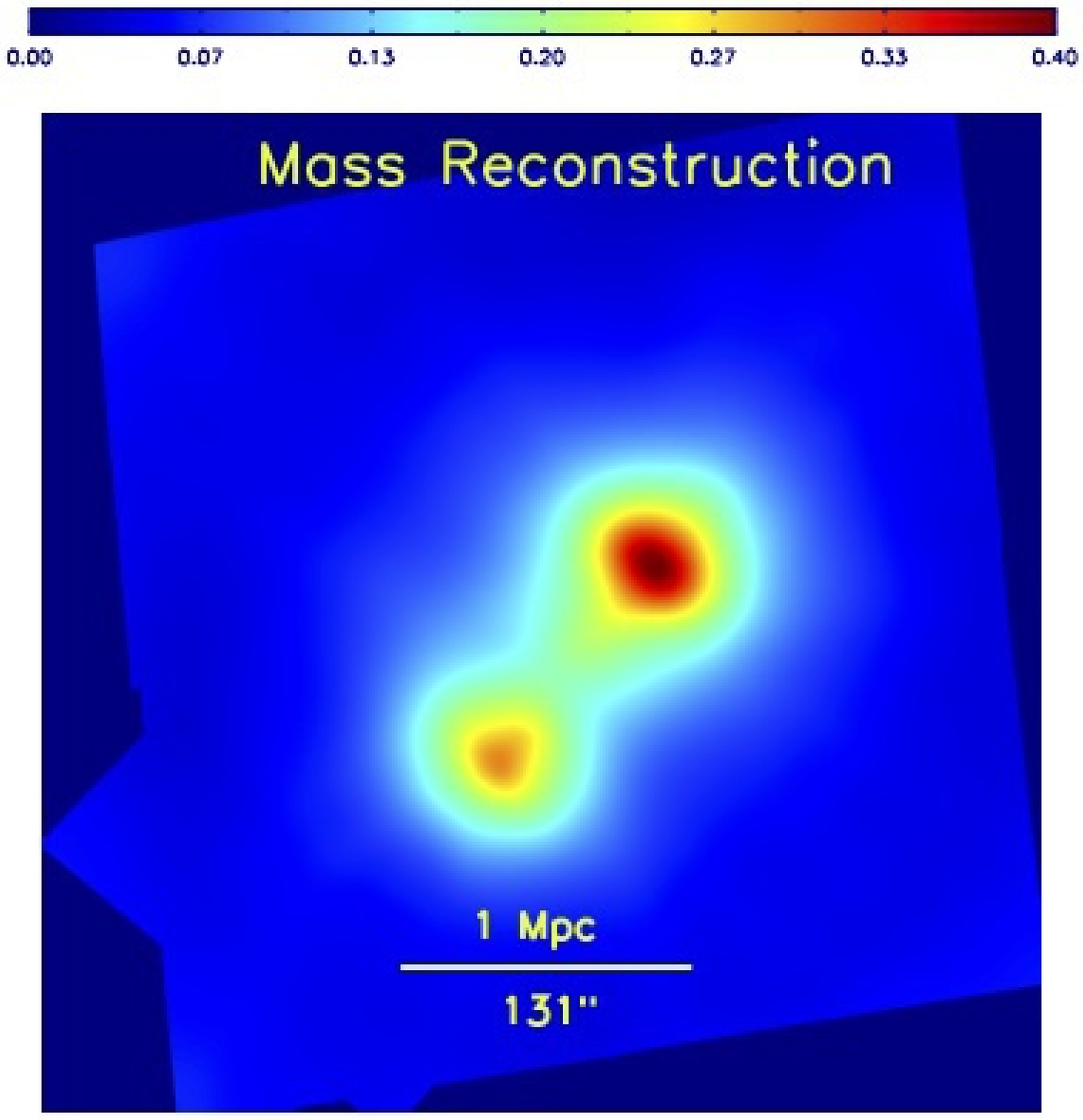}
\includegraphics[width=8.5cm]{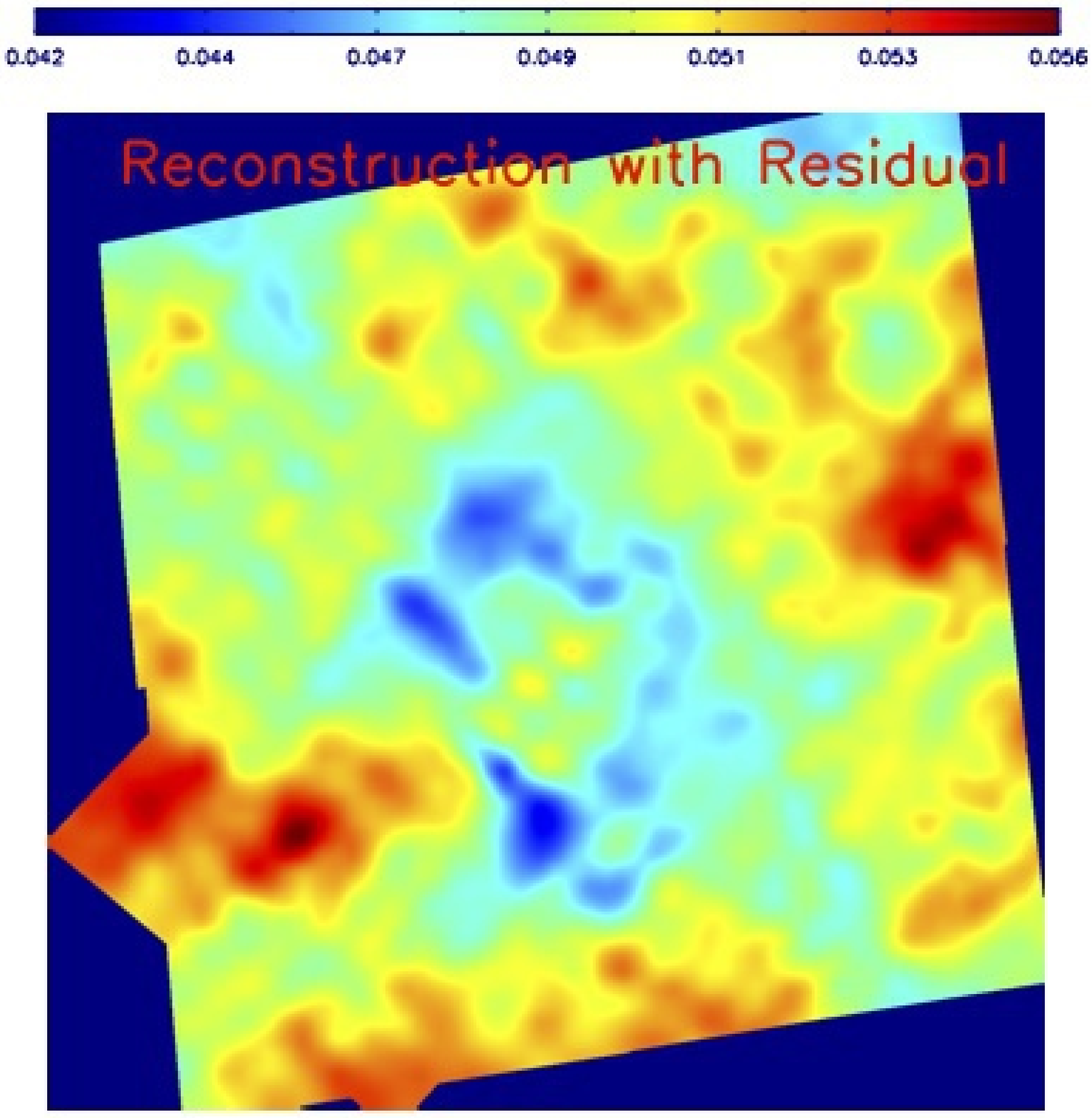}

\end{center}
\caption{Two-dimensional mass modeling of \elgordo. In the upper-left panel,
  we use our best-fit 2D NFW-fit results to display the predicted
  (noiseless) ellipticity of background galaxies and the resulting
  convergence. The upper-right panel displays the residuals, which can be 
  attributed to the
  (correlated or uncorrelated) large scale structure, departures
  from the axisymmetric NFW profile, and/or simply shape noise. 
  The corresponding convergence maps are shown in the bottom panels.
  Because of the mass-sheet ambiguity, the absolute convergence values
  are meaningless. Instead, one should pay attention to the range of the variation. 
  The pick-to-valley variation is $\mytilde0.01$.  The
  most significant residual is found in the south-eastern region. This feature is
  also seen in our original mass reconstruction and coincides with a
  galaxy overdensity.
\label{fig_2D_sanity_check}}
\end{figure*}

Although we consider that our simultaneous 2D fitting procedure is the
better method for cluster mass estimation given the small field size,
the accuracy of the mass estimate depends on the degree that the real
cluster mass distribution resembles our parametric description of
it. To assess the validity of our assumption, we create a mock galaxy
catalog based on our best-fit model and compare the result with the
source catalog obtained from our {\it HST} images. The predicted
ellipticity pattern is displayed in the upper-left panel of
Figure~\ref{fig_2D_sanity_check}.  Also shown is the resulting
convergence field (lower-left panel).  The overall patterns in the ellipticity and
convergence distributions are similar to those seen in our data
(Figure~\ref{fig_whisker_n_mass}).  In particular, the ratio of the
significance between the two mass peaks and their individual centroids
are in good agreement.  Detailed comparison is possible when their
residuals (model-data) are plotted (right panels in Figure~
\ref{fig_2D_sanity_check}).  The residual convergence (lower-right) is created from
the residual ellipticities (not by subtracting two convergence
fields) using the same maximum entropy method.  The residuals can be attributed to correlated/uncorrelated
large scale structure, departure from the axisymmetric NFW profile,
and/or shape noise.  Because of the mass-sheet ambiguity, the absolute
convergence values are meaningless. Instead, one should pay attention to the range of the variation. 
 The pick-to-valley variation is $\mytilde0.01$, which is  
 smaller than our mass reconstruction rms values (Figure~\ref{fig_rmsmap}).
 The   most significant residual is found in the south-eastern region. This feature 
  coinciding with a galaxy overdensity is
  also seen in our original mass reconstruction when a smaller smoothing scale is used.
Nevertheless, there is no associated X-ray emission.  
It is not clear at the moment whether or not the structure is associated
with \elgordo. However, if we assume that this structure is at the
cluster redshift, its mass from within an aperture of $r\sim50\arcsec$
($\sim380\hubblem$ kpc) is $\lesssim1\times10^{14} \hubblem M_{\sun}$.
Therefore, we do not regard this extra substructure as a serious
concern in our modeling of \elgordo\ with two massive NFW halos.

\begin{figure}
\begin{center}
\includegraphics[width=8cm]{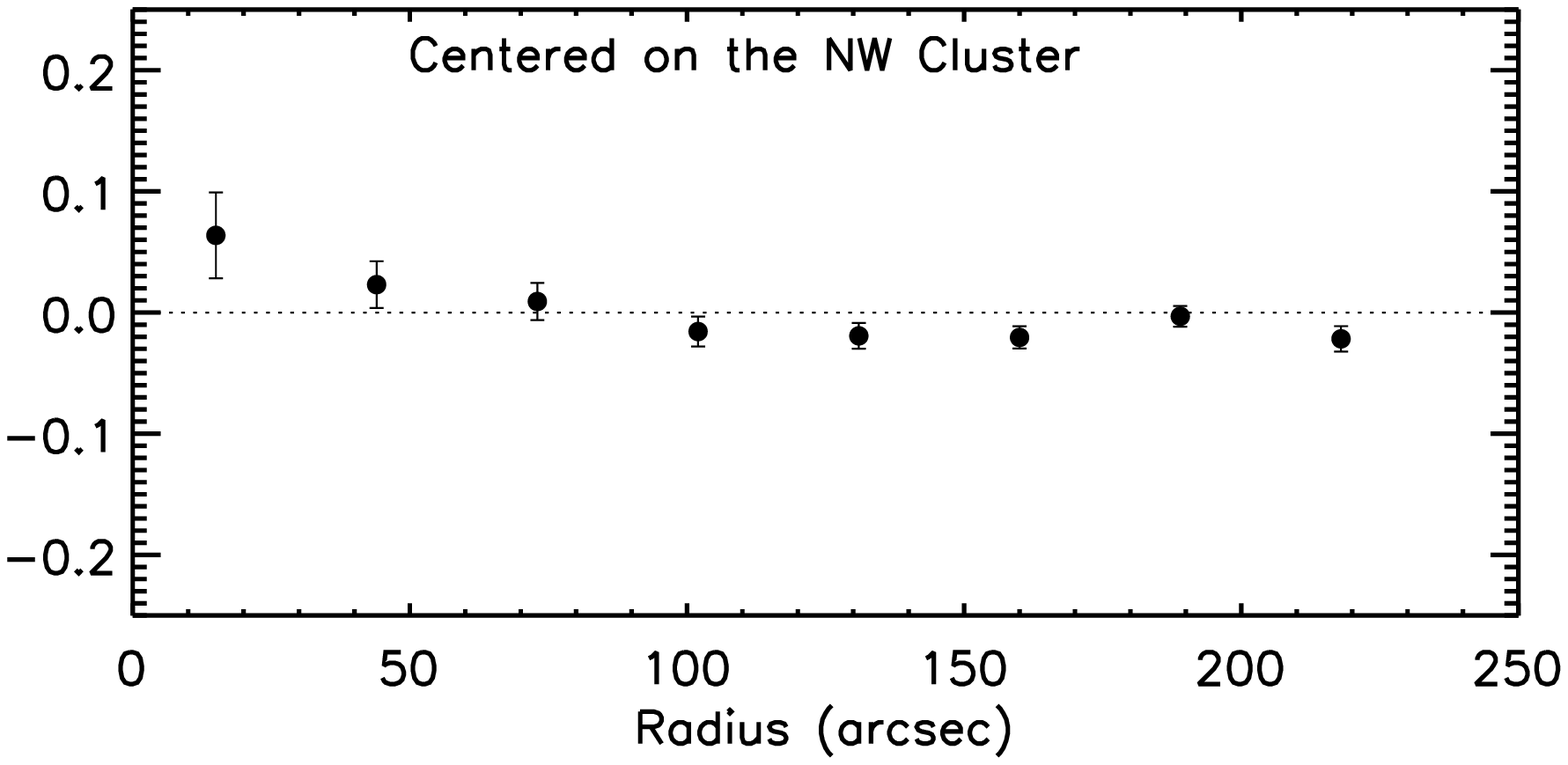}
\includegraphics[width=8cm]{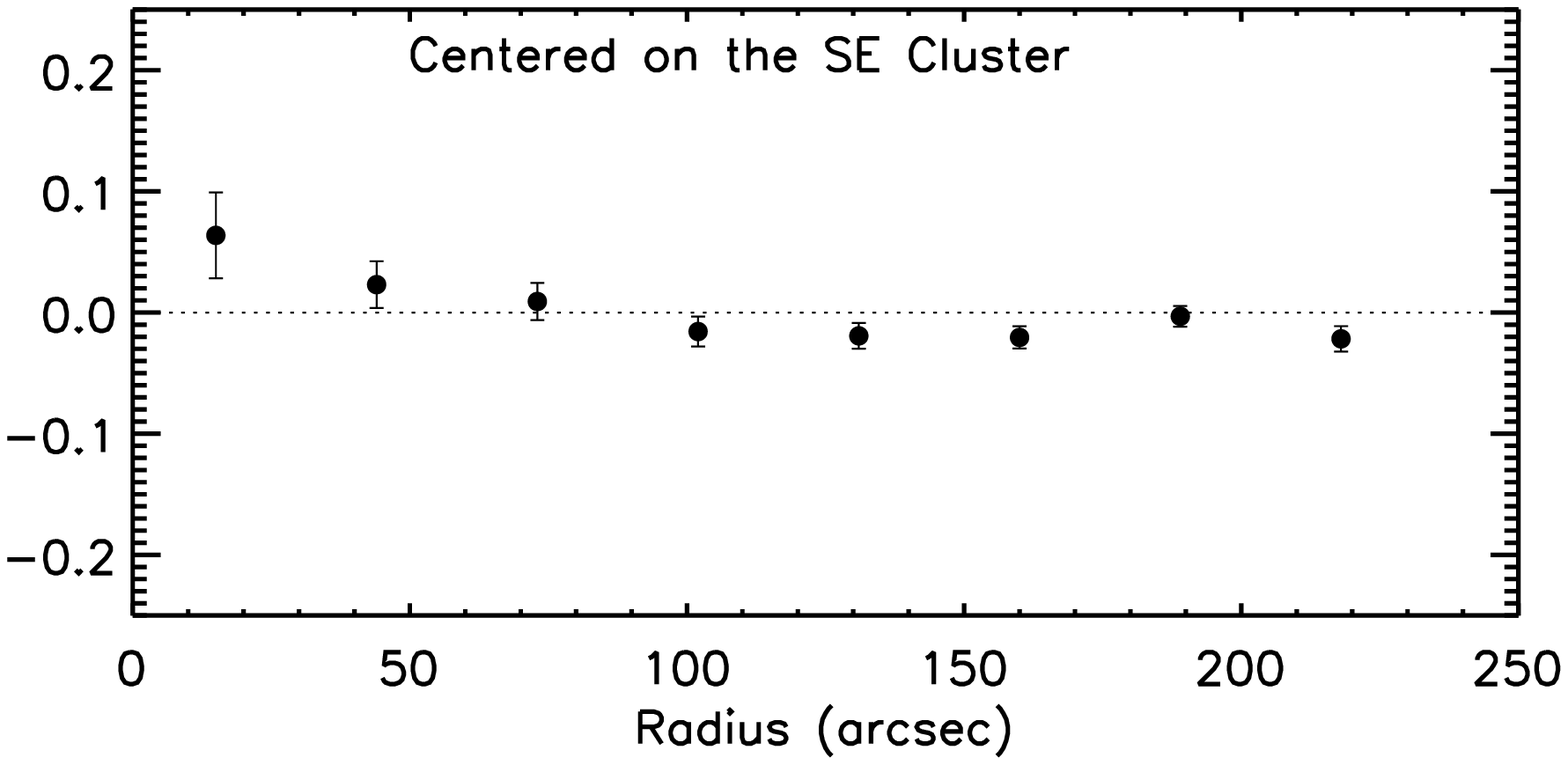}
\includegraphics[width=8cm]{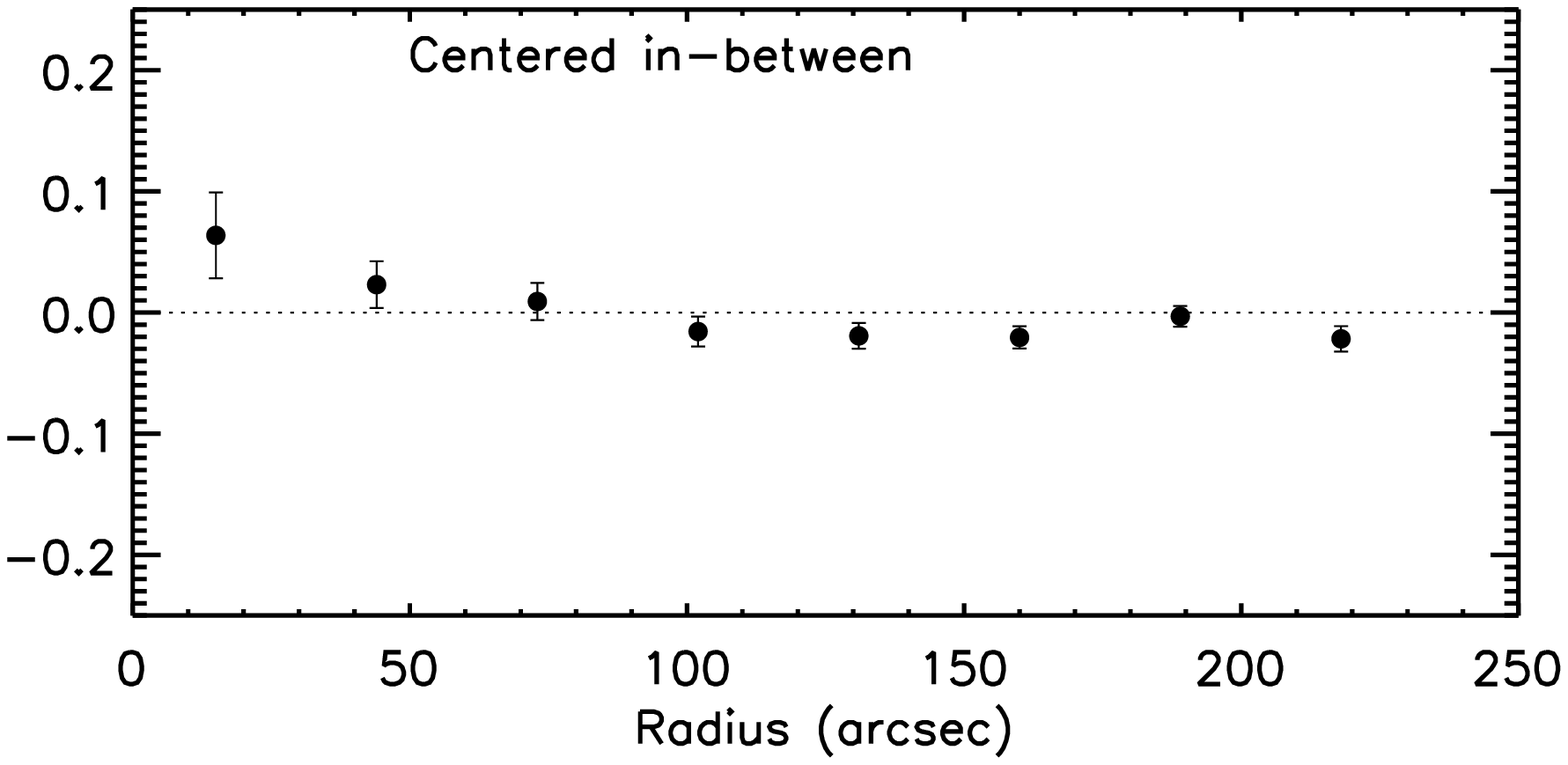}
\end{center}
\caption{Residual tangential shear profile. Shear values are estimated
  at the location of every source galaxy based on our best-fit 2D
  model, and these predicted shears are subtracted from the measured
  ellipticity.  Then, the tangential shears are computed as done in
  Figure~\ref{fig_tangential_shear}.  Regardless of the choice of the
  center, the residual signal is consistent with zero.  We omit the
  null (45$\degr$ rotation) test to avoid cluttering the figure, but
  the level of scatter is comparable to the residual tangential shear
  values shown here.
\label{fig_residual_tangential_shear}}
\end{figure}

An alternative method to assess how well the 2D halo model describes
the data is to examine the tangential shear values constructed from
the residual ellipticity
catalog. Figure~\ref{fig_residual_tangential_shear} displays such test
results azimuthally averaged around three different centers: the NW
subcluster, the SE subcluster, and the global center. The residual
signal is consistent with zero without any apparent large scale
correlations.

\begin{figure*}
\begin{center}
\includegraphics[width=8cm]{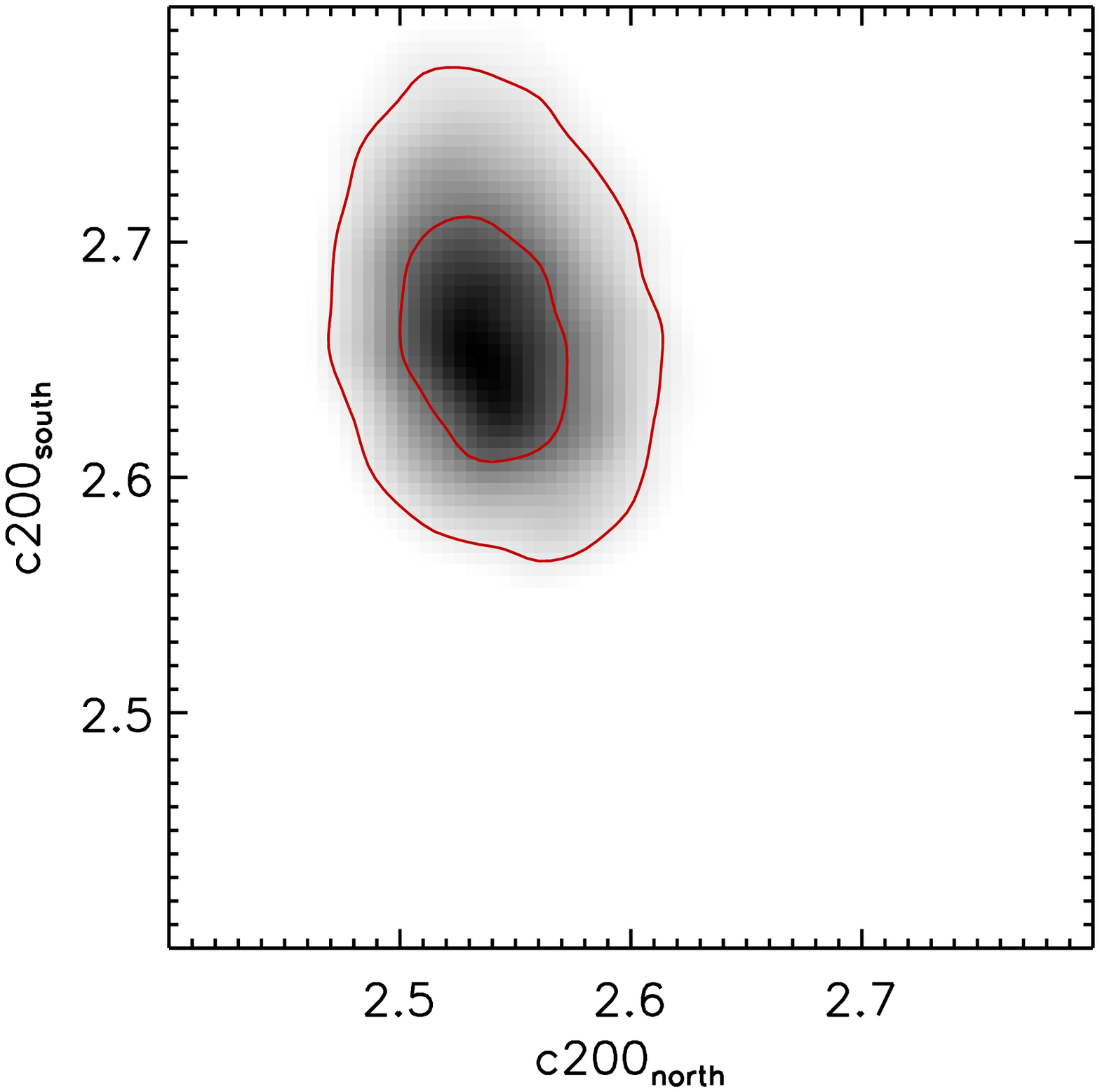}
\includegraphics[width=8cm]{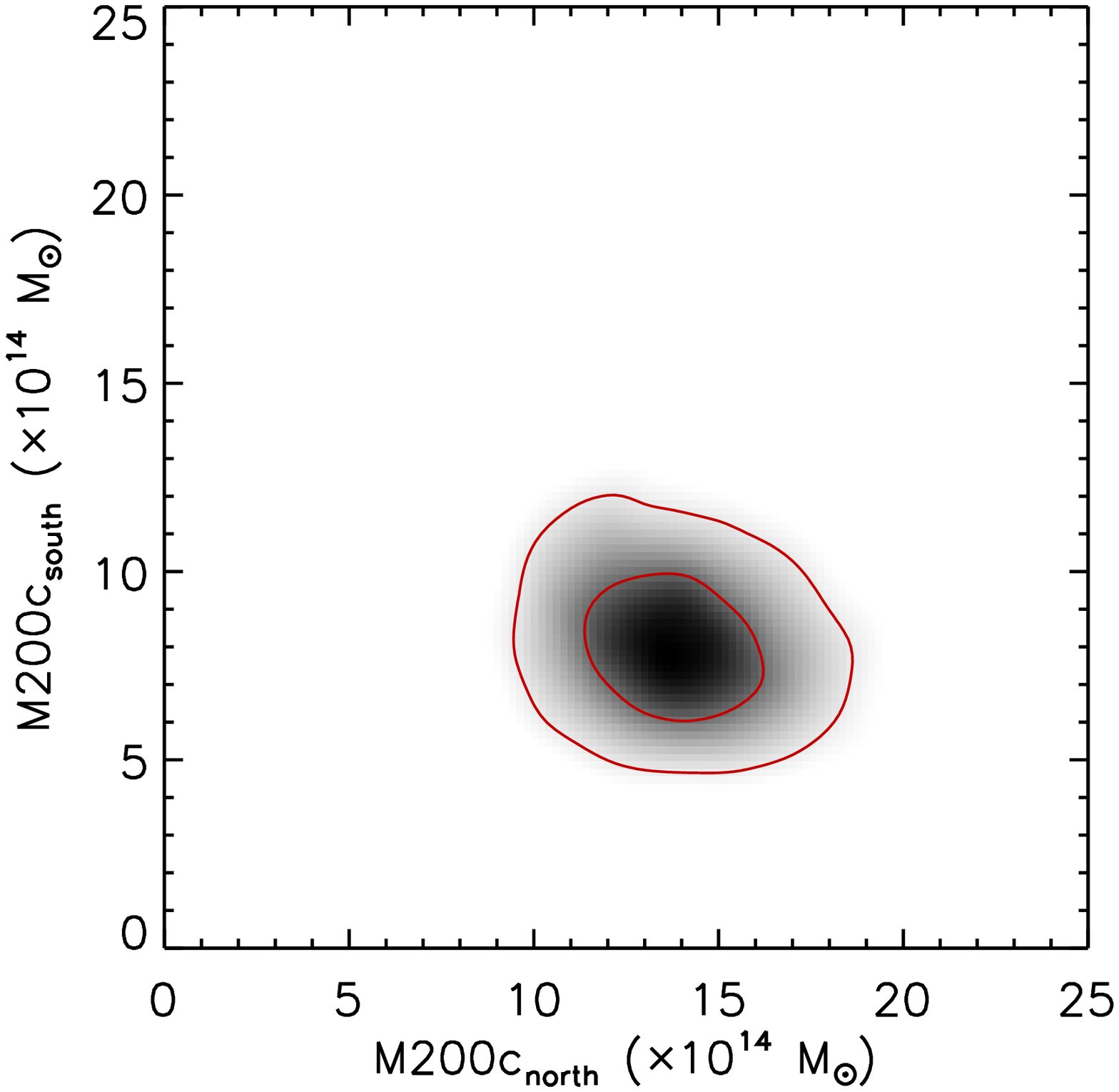}
\end{center}
\caption{Concentration and Virial Mass of \elgordo. The results are
  obtained from our 80,000 MCMC samples by treating the cluster as a
  sum of two NFW halos.  We show 1- and 2-$\sigma$ contours.  The
  mass-concentration relation of Duffy et al.\ (2008) is used, and
  thus if we know the result in one panel, the one in the other is
  determined.
\label{fig_c_and_m}}
\end{figure*}

\begin{figure*}
\begin{center}
\includegraphics[width=8cm]{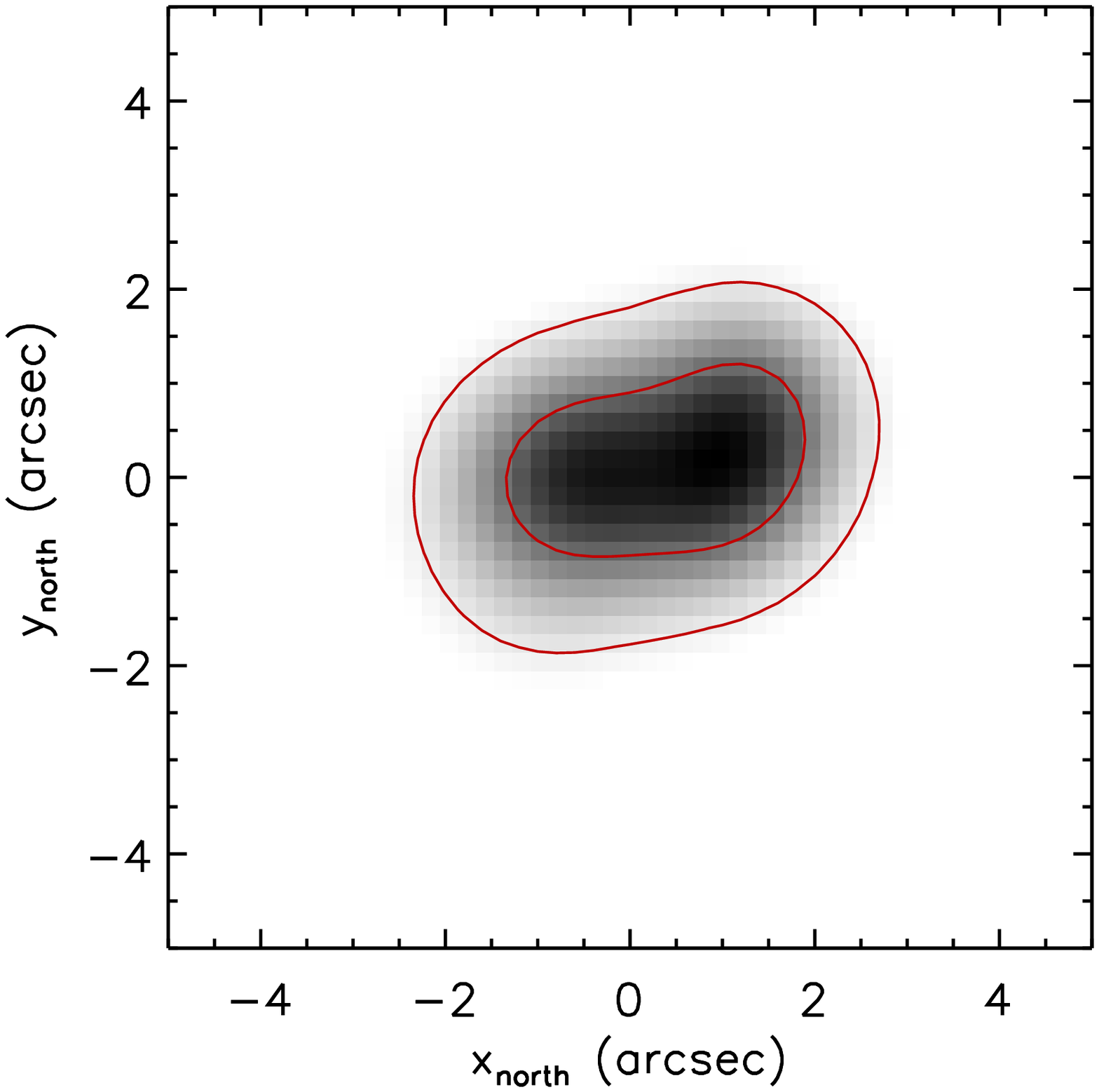}
\includegraphics[width=8cm]{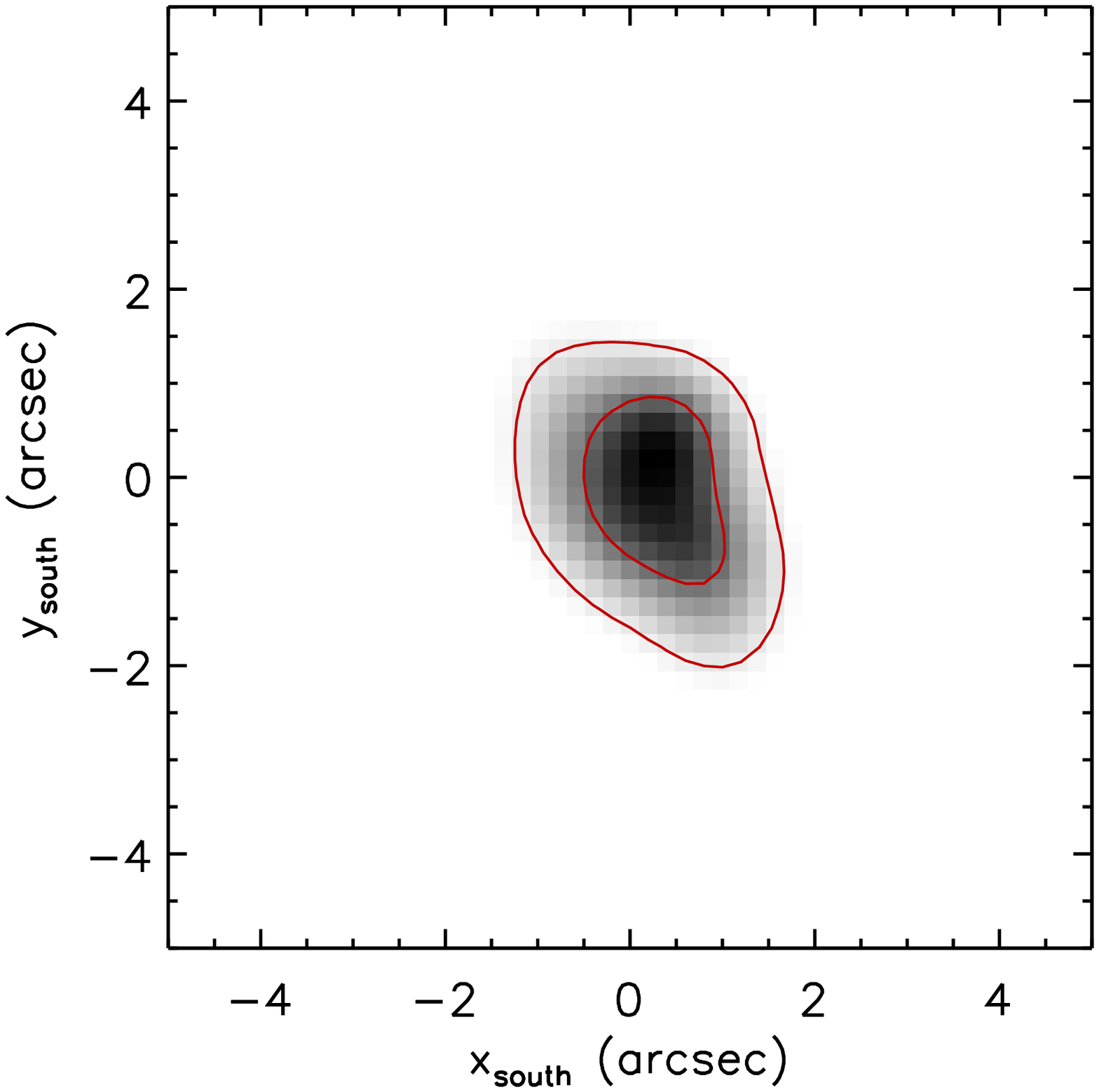}
\end{center}
\caption{Centroid distribution obtained from our 80,000 MCMC
  samples. We show 1- and 2- $\sigma$ contours.  We model \elgordo\ as
  a superposition of two NFW halos. Flat priors are assumed for the
  centroids with no boundary constraint.  The results are displayed in
  the observed orientation (i.e., the same as in
  Figure~\ref{fig_whisker_n_mass}).
\label{fig_centroid}}
\end{figure*}

It is important to examine how the two cluster masses are correlated
in this full 2D analysis. A potential problem is that the two cluster
masses might be highly anti-correlated. In other words, it may be
possible that the fit would allow the two halos to trade mass with
each other nonnegligibly while the $\chi^2$ value remains relatively
unchanged.  Our MCMC analysis shows that the degeneracy among fitted
values, if any, is very weak as shown in Figure~\ref{fig_c_and_m}.  We
suspect that we do not suffer from this potential degeneracy because
the mass estimates are mostly influenced by the {\it strong}
weak-lensing signals at small radii ($r\lesssim70\arcsec$) from each
cluster's center. This also may explain why the masses obtained from
the previous 1D analysis are similar to the current full 2D results.
In Figure~\ref{fig_centroid}, we display the centroid distributions of
the two halos, which illustrate that the mass centroid is
well-constrained ($\lesssim3\arcsec$).

\begin{figure}
\begin{center}
\includegraphics[width=8cm]{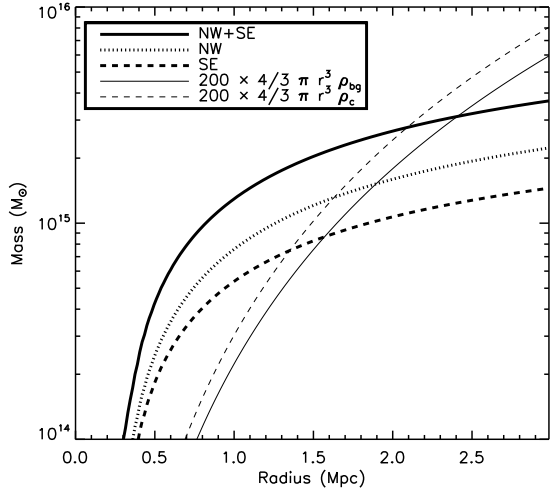}
\end{center}
\caption{Mass profile of \elgordo. We use the results from
  simultaneous 2D fits of two NFW profiles.  The thick dashed and
  dotted lines represent the masses within spherical volumes for
  individual halos whereas the thick solid line shows the sum of the
  two halos assuming that the projected distance is the actual
  physical separation (we choose the center of mass of the two halos
  for the origin).  The thin solid and dashed lines represent the
  total mass enclosed within the sphere at $r$ when the density inside
  becomes 200 times the mean and critical densities of the universe at
  the cluster redshift, respectively.  The $M_{200c}$ and $M_{200a}$
  values are easily determined by locating where the thin and thick
  lines intersect.  The statistical uncertainties of individual masses
  are about $\mytilde20$\% at $r=1.5\hubblem$ Mpc.
\label{fig_cluster_mass}}
\end{figure}

As mentioned above, one of the most interesting issues concerning
\elgordo\ is whether or not its total mass lies sufficiently above a
critical threshold that the cluster's mere existence gives rise to
some non-negligible tension with theoretical predictions. As we use
the mass function of Tinker et al.\ (2008) who choose $M_{200c}$ or
$M_{200a}$ as a reference, we need to combine the masses of the two
halos of \elgordo\ to estimate the equivalent quantity within a
spherical volume.  We implement the computation by populating a
three-dimensional ($1000^3$) grid with the sum of the density
distributions of the two halos. Figure~\ref{fig_cluster_mass}
displays, as a function of radius, the resulting mass profile of the
two individual halos of \elgordo\ (thick dotted and thick dashed
lines) and the combined total (thick solid line).  The thin solid and
dashed lines represent the integrated total mass within a spherical
volume of given radius for density values of 200 times the mean and
critical densities of the universe at the cluster redshift,
respectively. This allows the $M_{200c}$ and $M_{200a}$ values for the
cluster to be conveniently determined by locating where the thin and
thick lines intersect.  The mass uncertainties of individual halos are
about $\mytilde18$\% at $r=1.5\hubblem$ Mpc.
Figure~\ref{fig_cluster_mass} shows that extrapolation of the two NFW
halos to a radius of $r=2.4 \hubblem$ Mpc yields a combined mass of
$M_{200a}=(3.13\pm0.56)\times10^{15} \hubblem M_{\sun}$.

\begin{figure}
\begin{center}
\includegraphics[width=8cm]{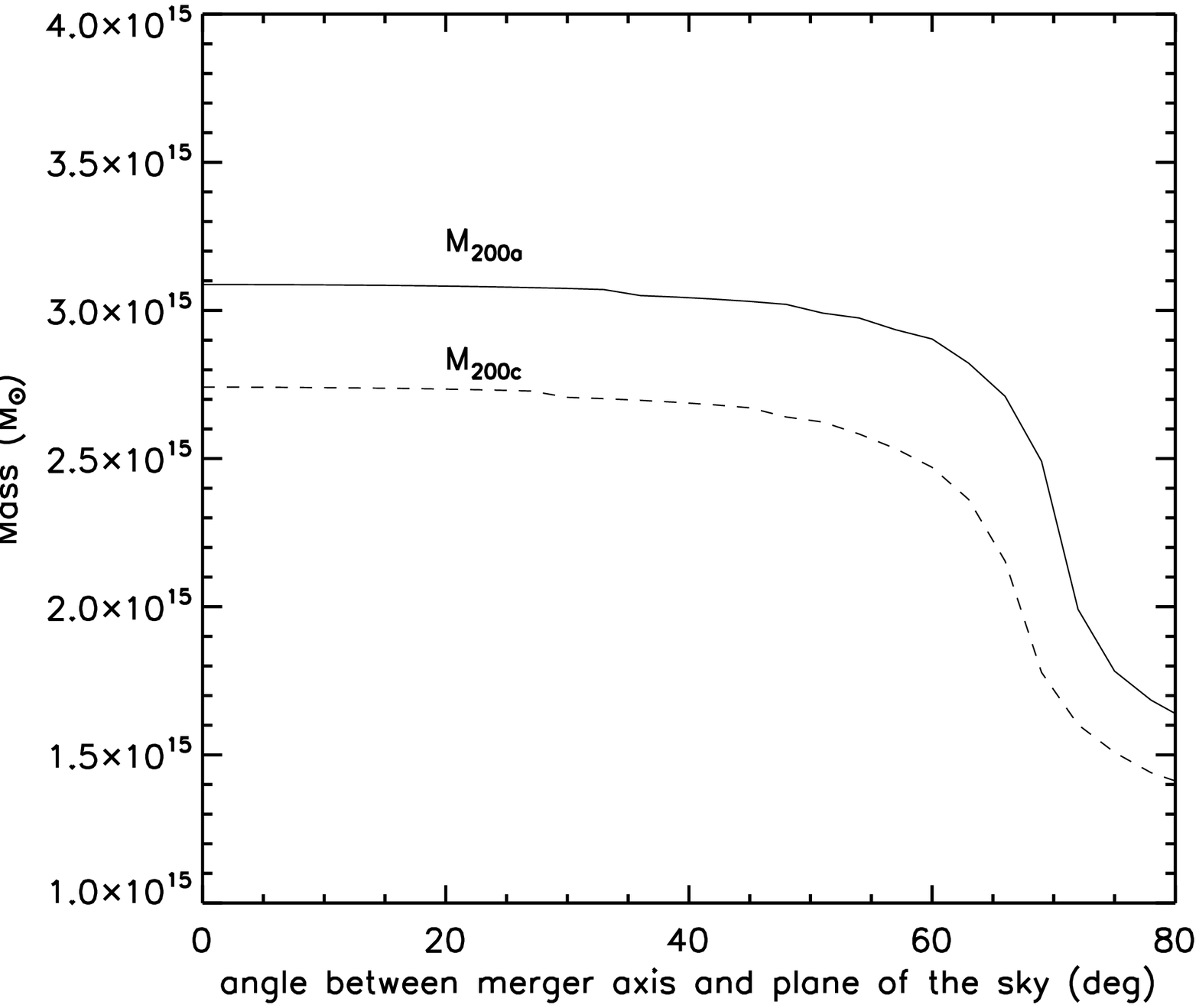}
\end{center}
\caption{Total cluster mass as a function of viewing angle. The sum of
  the two halos in \elgordo\ weakly depends on the viewing angle as
  shown here. The total cluster mass remains extremely high until we
  assume that angle between the merger axis and the plane of the sky
  becomes greater than $\mytilde65\degr$, in which case the separation
  between the two halos is more than $\mytilde2.3\hubblem$~Mpc.
\label{fig_angle_vs_mass}}
\end{figure}

The above mass estimation assumes that the two halos are at the same
distance from us. This, of course, leads to a maximum value for the
total mass.  Here we investigate how much the combined mass decreases
as we allow the two halos to have slightly different distances.  We
repeat the above total mass estimation, only now we populate the 3D
grid while varying the angle between the merger axis (the line
connecting the mass peaks) and the plane of the sky from zero to
80$\degr$. Figure~\ref{fig_angle_vs_mass} reveals that in fact the
total mass of the system is not sensitive to the viewing angle as long
as the angle is less than $\mytilde65\degr$, which would inflate the
projected distance of $\mytilde700\hubblem$ kpc to a physical
separation of $\mytilde2\hubblem$~Mpc.

M12 report that the line-of-sight velocity difference between the NW
and SE components is $586\pm96~\mbox{km}~\mbox{s}^{-1}$. This small
relative velocity favors a scenario wherein the merger is happening
nearly in the plane of the sky as long as the transverse velocity is
sufficiently large ($>$1000 $\mbox{km}~\mbox{s}^{-1}$. The presence of
double radio relics in \elgordo\ (Lindner et al. 2013) also suggests that the merger axis is
close to the plane of the sky.  Unfortunately, no study has yet
provided a secure constraint on the transverse velocity or three
dimensional geometry of the merger.

\subsection{Mass Estimation Uncertainties from Various Sources} \label{section_various_uncertainties}

The mass estimate uncertainties presented in
\textsection\ref{section_mass_determination} include only shape noise
arising from the finite number of source galaxies.  Additional sources
of uncertainty include the scatter in the mass-concentration
relation, possible triaxiality of the halo, and shear contribution from 
correlated/uncorrelated large-scale structures (LSSs).

{\bf Scatter in mass-concentration relation.} The Duffy et al.\ (2008)
mass-concentration relation possesses large scatter.  Because we
assume the mean relation to estimate the cluster mass, it is worth
examining how much the mass estimate changes when we consider this
scatter.  Unfortunately, there are no large ($M_{200c}> 10^{14}
M_{\sun}$) halos at $z\simeq1$ in the numerical simulation of Duffy et
al.\ (2008). Thus, the mass-concentration relation that we adopt is in
fact an extrapolation from the results obtained for the
$M_{200c}<10^{14} M_{\sun}$ halos.  In order to assess the impact of
the mass-concentration uncertainty, we fit the one-parameter NFW
profile while perturbing the concentration parameter around its mean
by $\delta c_{200}=0.5$ (the approximate scatter at
$M_{200c}\sim10^{14} M_{\sun}$ in the simulations).  We find that the
cluster mass changes by $\mytilde10$\% due to this effect.

{\bf Triaxiality, Large Scale Structure, and departure from NFW.}
Because our halo mass estimation inevitably assumes spherical
symmetry, the deprojected mass estimate can be over/under-estimated
depending on the actual degree and orientation of the triaxiality of
\elgordo. The masses of halos whose major axes are aligned along the
line of sight are overestimated whereas underestimation will occur
when the major axes are perpendicular to the line of sight.  Adding to
this complexity are departure of the cluster mass profile from the NFW
assumption and the LSSs along the line of sight.  It is
straightforward to quantify the scatter due to these sources using
numerical simulations (e.g., Meneghetti et al.\ 2010; Becker \&
Kravtsov 2011; Oguri \& Hamana 2011).  According to Becker \& Kravtsov
(2011) who investigate the weak-lensing mass scatter from fitting an
NFW profile to the shear profile, the intrinsic scatter is
$\mytilde20$\% for massive halos and increases to $\mytilde30$\% for
groups. Because the contribution from uncorrelated LSSs is about the
same regardless of the halo mass, more massive clusters are less
affected by LSSs.  As each component of \elgordo\ is a massive system,
it is legitimate to assume that the corresponding scatter (for each
halo) may be $\mytilde20$\%.  While fitting an NFW model, Beck \&
Kravtsov (2011) varied both total mass and concentration
independently. Therefore, the aforementioned scatter in the
mass-concentration relation is already included in this estimation of
the scatter.

\subsection{Luminosity Estimation and Mass-to-Light Ratio}
\begin{deluxetable*}{lcc}
\tablewidth{0pt}
\tabletypesize{\scriptsize}

\tablecaption{Luminosity and mass-to-light ratio  of \elgordo. \label{tab_ml}}
\tablehead{  \colhead{ } & \colhead{NW subcluster} & \colhead{SE subcluster} }
 \startdata
$B$-band Luminosity    &    $3.22\times10^{12}\hubblel L_{B\sun}$  &  $3.40\times10^{12}\hubblel L_{B\sun}$ \\
Column Mass  ($r<386 \hubblem$~kpc)                &  $4.20\pm0.34~(3.92\pm0.36) \times10^{14} \hubblem M_{\sun}$  & $2.98\pm0.25~ (3.42\pm0.31)\times10^{14} \hubblem M_{\sun}$  \\
Mass-to-light Ratio                                    &  $130\pm10 ~(122\pm11)  \hubbleml M_{\sun}/L_{B\sun}$  &  $88\pm7~(101\pm9) \hubbleml M_{\sun}/L_{B\sun}$ 
\enddata
\tablenotetext{a}{The numbers in parentheses are derived from aperture
  mass densitometry.}
\end{deluxetable*}

High mass-to-light ratio ($M/L$) values for galaxy clusters have been
firmly established today and provide critical evidence for the
dominance of dark matter in the large-scale structure of the
universe. Studies of cluster $M/L$ values in a broad range of
environments provide key information for us to understand the
formation of clusters and their member galaxies.
\elgordo\ distinguishes itself from other clusters by its extremely
high mass despite being at high redshift. Thus, an investigation of
the $M/L$ values of this unusual cluster is an important extension of
the environmental baseline in cluster $M/L$ studies.

We define two maximally non-overlapping circular apertures each with a
radius of $r=386\hubblem$ kpc (i.e., half the projected distance between the
NW and SE clumps) centered on the luminosity peaks. The
cluster galaxies are identified by combining our spectroscopic survey
data (Sif{\'o}n et al.\ 2013) and photometric redshift catalog
(M12). The rest-frame $B$-band at $z=0.87$ overlaps both the F775W and
F850LP filter throughput curves. We choose to use the {\it HST}/ACS
photometry to estimate the rest-frame $B$-band luminosity of
\elgordo\ via the following relation:
\begin{equation}
B_{rest}= -0.565 (F775W-F850LP) + 1.38 - DM,
\end{equation}
where DM is the distance modulus at the cluster redshift. The above
relation is derived from our synthetic photometry using the Kinney et
al.\ (1996) Spectral Energy Distribution (SED) template.

The column mass within the aperture is computed using both the
best-fit NFW parameters and aperture-mass densitometry (Fahlman et
al.\ 1994). Readers are referred to Bartelmann (1996) for the useful
equations for the computation of the NFW profile column density.  The
best-fit NFW parameters (from the 2D fit method) give
$(4.20\pm0.34)\times10^{14} \hubblem M_{\sun}$ and
$(2.98\pm0.25)\times10^{14} \hubblem M_{\sun}$ for the NW and SE
clusters, respectively, within the $r=386\hubblem$ kpc aperture.  The
NFW-fitting results for both NW and SE are consistent with the values
from the aperture densitometry, which gives
$(3.92\pm0.36)\times10^{14} \hubblem M_{\sun}$ and
$(3.42\pm0.31)\times10^{14} \hubblem M_{\sun}$, respectively.
Aperture densitometry requires us to define a control annulus, and we
use the $r=1-1.5\hubblem$ Mpc region for each cluster. We derive the
mean column density within the annulus using the best-fit NFW
parameters, taking into account the two overlapping profiles. The mean
surface mass densities for the NW and SE annuli are estimated to be
$\bar{\kappa}=0.067$ and $0.065$, respectively.  Note that since these
values are non-negligibly higher than zero, our estimates based on the
aperture mass densitometry are not entirely independent of our
parametric estimation.

We summarize the $M/L$ values of \elgordo\ in Table 3. Our analysis
shows that the $M/L$ value of the NW subcluster ($\mytilde130 \hubbleml M_{\sun}/L_{B\sun}$) is higher 
than that ($\mytilde88 \hubbleml M_{\sun}/L_{B\sun}$) of
the SE subcluster (25--50\% depending on the method).  
Comparison of
this $M/L$ value with those of other clusters requires caution because
the radial variation of the $M/L$ ratio is cluster-dependent.  For
example, the cumulative $M/L$ profile of Cl0152$-$1357 at $z=0.83$
rises steeply at small radii, peaking at $\sim270\hubblem$ kpc with
$M/L\mytilde150~ \hubbleml M_{\sun}/L_{B\sun}$. At larger radii, the
$M/L$ value gradually decreases and reaches $\sim95~ \hubbleml
M_{\sun}/L_{B\sun}$ at $r=1\hubblem$ Mpc (Jee et al.\ 2005a).  On the other
hand, the cumulative $M/L$ profile of MS1054$-$0321 at a similar
redshift of $z=0.83$ continues to increase with radius, reaching
$\sim120~ \hubbleml M_{\sun}/L_{B\sun}$ at $r=1\hubblem$ Mpc (Jee et
al.\ 2005b).  Within the $r=386\hubblem$ kpc aperture the difference
between the two $z\sim0.83$ clusters' $M/L$ values is non-negligible
($\mytilde140 ~\hubbleml M_{\sun}/L_{B\sun}$ and $\mytilde90~
\hubbleml M_{\sun}/L_{B\sun} $ for Cl0152$-$1357 and MS1054$-$0321,
respectively).  With this caveat in mind, we conclude that the $M/L$
value of \elgordo\ is well-bracketed by the $M/L$ values of these two
clusters.

\section{DISCUSSION} \label{section_discussion}

\subsection{Comparison with Other Observations}

\elgordo\ was first discovered by ACT as the cluster with their
survey's most significant SZ decrement and an X-ray luminosity
comparable to that of the Bullet Cluster (Marriage et al.\ 2011).
Using their own mass scaling relation, Williamson et al.\ (2011)
estimated $M_{200a} = (1.89\pm0.45) \times10^{15} \hubblem M_{\sun}$
from the South Pole Telescope (SPT) data.

More robust mass estimates of \elgordo\ were presented by M12, who
performed a joint analysis of the {\it Chandra} X-ray, ACT SZ, and
dynamical data and quoted $M_{200a}=(2.16\pm0.32)\times10^{15}\hubblem
M_{\sun}$. This value is statistically consistent with the result of
Williamson et al.\ (2011).

However, it is noteworthy that the M12 mass estimates based on
individual scaling relations show large variations.  For example, the
$M_{500c}-M_{gas}$ scaling relation predicts
$M_{200a}\sim3.1\times10^{15} \hubblem M_{\sun}$ whereas the dynamical
relation ($M_{200a}-\sigma$) gives $M_{200a}\sim1.86\times10^{15}
\hubblem M_{\sun}$, a factor of 1.7 times smaller than the former.

Our total mass estimate from the current weak-lensing study $M_{200a}
=(3.13\pm0.56)\times10^{15} \hubblem M_{\sun}$ is fully
consistent with the X-ray-based mass estimates of M12, including their
value based on $Y_X$ which is
$M_{200a}=2.88_{-0.55}^{+0.78}\times10^{15} \hubblem M_{\sun}$. We
find that the discrepancy between the cluster's dynamical mass and the
current weak-lensing mass can be reconciled if we treat the
\elgordo\ system as two components.  For example, the dynamical masses
of the two individual subclusters
($M_{200a}=1.76_{-0.58}^{+0.62}\times10^{15} \hubblem M_{\sun}$ and
$1.06_{-0.59} ^{+0.64}\times10^{15} \hubblem M_{\sun}$ for the NW and
SE subclusters, respectively) estimated by M12 are consistent with our
lensing results (also, remember that the lensing-based velocity
dispersion estimates are in good agreement with the spectroscopic
measurements).  M12 quote a total mass of the system
$M_{200a}=1.86_{-0.49}^{+0.54}\times10^{15} \hubblem M_{\sun}$ based
on the combined total velocity dispersion of
$\mytilde1321~\mbox{km}~\mbox{s}^{-1}$, which is only $\mytilde6$\%
higher than their mass of the NW cluster.  Instead, if we regard
\elgordo\ as two components with each following the Duffy et
al.\ (2008) mass-concentration relation, the combination of the two
dynamical masses of M12 will yield
$M_{200a}=(3.8\pm0.8)\times10^{15}\hubblem M_{\sun}$ consistent with
our lensing estimate.  Note that combining the two halos increases
$r_{200a}$, which makes the total $M_{200a} $ greater than the sum of
the two individual components' $M_{200a}$ values.  We further note
that the dynamical mass estimates are based on measurements of the
radial velocities of cluster member galaxies.  If the axis of the two
infalling  subclusters
is close to the plane of the sky, then there would be a
significant component of kinetic energy that is not yet virialized and
which would not be included in the total combined velocity dispersion
measurement.

\begin{figure}
\includegraphics[width=8cm]{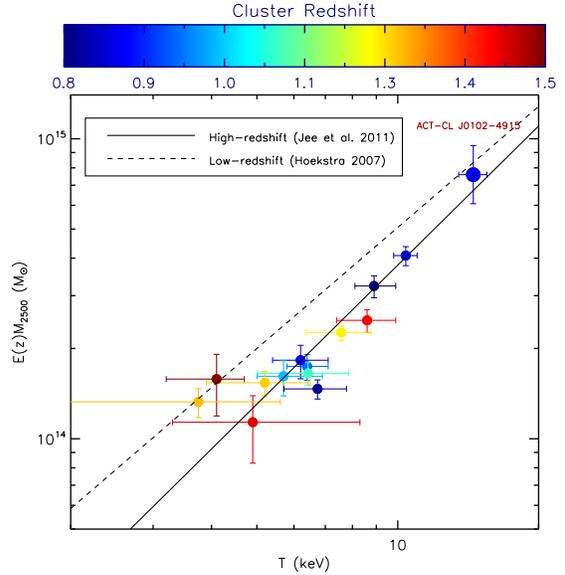}
\caption{Mass versus temperature relation of \elgordo. The data points
  except for \elgordo\ display the weak-lensing masses and X-ray
  temperatures of the 13 high-redshift ($z\gtrsim0.8$) clusters
  presented in Jee et al.\ (2011). The solid line shows a best-fit
  result to these 13 data points ($M\propto T^{1.54})$. The dashed
  line ($M\propto T^{1.34})$ represents the low-redshift result from
  Hoekstra (2007).  The two slopes are statistically consistent
  whereas the normalization (i.e., translation) is different at the
  $>3\sigma$ level.  Jee et al.\ (2011) interpret the difference as
  indicating a possible evolution of the normalization in the $M- T_X$
  relation. We compute $M_{2500}$ of \elgordo\ by defining a
  fictitious single NFW halo that yields
  $M_{200c}=2.75_{-0.59}^{+0.74}\times10^{15}\hubblem M_{\sun}$.
\label{fig_mass_temp}}
\end{figure}

Having shown consistency between the weak-lensing and X-ray--derived
masses, we now compare the X-ray temperature versus weak-lensing mass
relation of \elgordo\ with those of other high-redshift clusters.
Previous mass-temperature relation studies quote cluster masses in
$M_{2500c}$, which refers to the mass within a  smaller radius
$r_{2500c}$ than the typical virial radius $r_{200a}$. Thus, unlike
$M_{200a}$, $M_{2500c}$ is sensitive to the choice of the cluster
center in \elgordo\ because the shape of the combined halos are
``peanut-like".  Therefore, we define a single cluster whose virial
mass matches the total cluster mass $M_{200c}=(2.76\pm0.51)
\times10^{15}\hubblem M_{\sun}$ (Table 2).  The corresponding
NFW parameters are obtained from the Duffy et al.\ (2008) relation.
As shown in Figure~\ref{fig_mass_temp}, \elgordo\ follows the $M-T_X$
relation of other galaxy clusters.

The weak-lensing, X-ray, and dynamical studies consistently support
the extreme mass of \elgordo, projecting the system to be the most
massive cluster at $z>0.6$.  However, we find that the SZ mass
$1.64_{-0.42}^{+0.62}\times10^{15}\hubblem M_{\sun}$ of M12 is
significantly lower than the other measurements, indicating that
perhaps more efforts are needed to calibrate the SZ mass proxy
$yT_{CMB}$.

Zitrin et al.\ (2013) present a strong-lensing analysis of
\elgordo\ and provide a rough mass estimate of $M_{200c}
\simeq2.3\times10^{15} \hubblem M_{\sun}$ for the entire system. This
total mass estimate roughly corresponds to the 1-$\sigma$ lower limit
of the current weak-lensing result.  Since the strong-lensing analysis
of Zitrin et al.\ (2013) is based on poorly constrained three band
{\it HST}/ACS photometric redshifts, no significant constraint on the
slope of the mass profile is provided. Therefore, a detailed
discussion on the discrepancy in the total mass estimation by weak and
strong lensing should be deferred until secure spectroscopic redshifts
are obtained for the multiply-lensed background galaxies.  However, it
is still worthwhile to mention one discrepancy in the mass ratio of
the two subclusters between the current weak-lensing study and the
strong-lensing analysis.  The Zitrin et al.\ (2013) strong-lensing
model suggests that the SE clump is more massive than the NW one by a
factor of 1.5 whereas our weak-lensing analysis indicates that the NW
cluster is nearly twice as massive as the SE cluster, which is
supported by the peak intensity in mass reconstruction, the 2D
ellipticity distribution (i.e., whisker plot), and the 1D tangential
shear profiles.

This issue of the mass ratio between the two subcluster components is
important when one attempts to model the system with numerical
simulations because the mass ratio is pivotal in determining many of
the observed features including the likelihood of the survival of gas
peaks, the temperature structure, the morphology/direction of the
wake, and so on.

\subsection{Is \elgordo\ a High-redshift Analog of the Bullet Cluster?}

Comparison of \elgordo\ to the Bullet Cluster is first mentioned in
Marriage et al.\ (2011), however, only in the context of discussing
the significance of the overall SZ decrement.  M12 suggest that
\elgordo\ might be a high-redshift analog of the Bullet Cluster also
as regards its stage of merger.  They find that 1) the peak of the
X-ray emission is the coolest region, 2) the Fe abundance of the X-ray
peak is significantly enhanced, 3) the temperature structure is
indicative of shock heating, 4) the X-ray surface brightness falls off
steeply beyond the SE subcluster, 5) there is a wake toward the NW
subcluster, 6) the radio data indicate the possible presence of an
intense double radio relic (now confirmed by Lindner et al. 2013), and 7) the SE galaxy distribution is
offset to the southeast with respect to the X-ray peak.  M12 interpret
these lines of evidence as indicating that the cool X-ray peak is
moving southeast after passing through the core of the other
subcluster.
 
However, there is one major difference in the X-ray morphology between
the Bullet Cluster and \elgordo. In the Bullet Cluster, there are two
distinct X-ray surface brightness peaks that can be associated with
the main cluster and the subcluster. In contrast, the {\it Chandra}
X-ray map of \elgordo\ shows only a single X-ray peak located close to
the SE mass peak. Although the X-ray emission is extended from this
peak to the NW component, we cannot identify any distinct second X-ray
peak in this region. This difference could be attributed to the
different mass ratio of its components ($\mytilde2:1$) compared to
that of the Bullet Cluster ($\mytilde10:1$).  It looks like the
interaction in \elgordo\ has mostly disrupted the central gas core
initially present in the NW component. This scenario is supported by
the presence of a strong wake in the SE-NW direction to the NW of the
X-ray peak.
 
As in the case of the Bullet Cluster, there is an offset
between the gas and weak-lensing mass peaks in \elgordo, although the
separation in the Bullet Cluster is larger ($d\sim150\hubblem$ kpc)
than in \elgordo\ ($d\sim100\hubblem$ kpc).
The statistical significance of the position offset between the gas and weak-lensing 
mass peaks is a strong lower limit on the significance of the position offset 
between the gas and dark matter peaks.  Since a non-negligible fraction of 
the total mass at the location of the X-ray peak is contributed by the gas itself, 
if we were to remove the baryonic mass from the weak-lensing mass distribution, 
the pure dark matter peak would move to the west in \elgordo, further from the gas 
peak. To make this correction and remove the cluster gas mass from the 
weak-lensing mass reconstruction requires a careful analysis of the X-ray data, which 
is underway and will be the focus of an upcoming study

In the Bullet Cluster the
X-ray peaks are trailing the corresponding mass peaks if we assume
that the two components are moving apart from each other.  The
direction of the offset is consistent with the hypothesis that dark
matter particles are collisionless whereas the cluster plasma is
subject to ram pressure.  However, in \elgordo\ the gas peak is
leading the mass peak if we assume a similar stage of the merger.
We have considered two possible scenarios for the merger in \elgordo:
1) we are viewing after core passage, but before first turn around,
and the merger speed is low or 2) the merger speed is high, but we are
viewing after the first turnaround as the two components come together
for a second core passage.  In the first case the $\rho v^2$ gas
pressure that tends to cause the dark matter-gas disassociation in the
Bullet Cluster is negligible (or much less) in \elgordo\ and we would
expect that the gas and dark matter should not be significantly
disassociated.  For the second case we view the dark matter as moving
more freely than the gas, preceding it after first core passage (as
seen in the Bullet Cluster) and then, being less encumbered than the
gas, returning more quickly back toward a second encounter with the NW
component.  Whether or not the dark matter can overtake and pass the
gas component on its second inward core passage (as we seem to require
in \elgordo) remains to be studied with $n$-body and hydrodynamical
simulations of the merger.
  
\subsection{How Rare is a Massive Cluster like \elgordo?}

\begin{figure}
\begin{center}
\includegraphics[width=8cm]{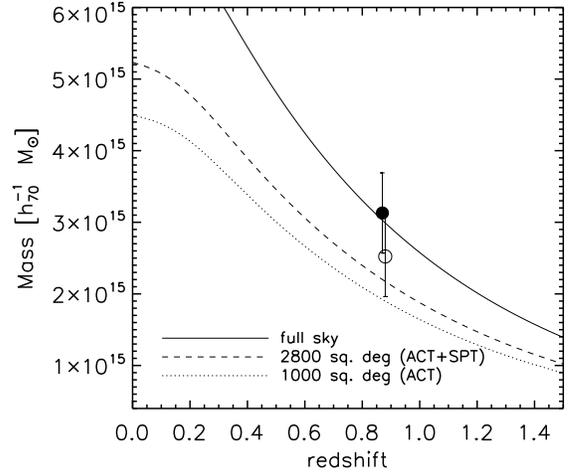}
\end{center}
\caption{Exclusion curves and weak-lensing mass estimate of
  \elgordo. We display 95\% probability curves, which exclude both
  cosmological parameter and sample variances at the 95\% level.  The
  open and filled circles show our weak-lensing mass estimates with
  and without the application of the Eddington bias correction,
  respectively.
\label{fig_exclusion}}
\end{figure}

Recent discoveries of extremely massive high-redshift clusters (e.g.,
Brodwin et al.\ 2012, Stanford et al.\ 2012, Jee et al.\ 2009, Foley
et al.\ 2011, Menanteau et al.\ 2012, Planck Collaboration et
al.\ 2011) kindled the debate over the consistency of the existence of
these massive clusters with the current $ \Lambda$CDM paradigm (e.g.,
Cayon et al.\ 2011, Jimenez \& Verde 2009, Gonzalez et al.\ 2012,
Hoyle et al.\ 2011, Meneghetti et al.\ 2011, Mortonson et al.\ 2011,
Waizmann et al.\ 2012, Harrison \& Coles 2012, Hotchkiss 2011).  At
the heart of this debate is the proper estimation of the survey
volumes and their selection functions that led to those discoveries. 

Conservative studies tend to
use the entire existing cluster surveys or the full sky area as a
reference to estimate the probabilities whereas others on the opposite
extreme consider only the selection function of the parent survey for
their abundance estimation. Moreover, some conservative studies
claim that the minimum redshift that one uses for the survey volume estimation
should be significantly lower than the redshift of the cluster in question (e.g., Hotchkiss 2011).
 
M12 concluded that although \elgordo\ is
an extremely rare system, the presence of the cluster is still
consistent with the standard $\Lambda$CDM cosmology in the lower part
of its allowed mass range.  In this paper, we revisit the question
with the current weak-lensing mass estimate.

Because \elgordo\ consists of two components, an important question is
whether or not it is legitimate to treat it as a single system when
its rarity is investigated.  A critical parameter to be considered is
the viewing angle with respect to the merger axis. In \textsection
\ref{section_mass_determination}, we demonstrated that the total
cluster mass remains high as long as the angle between the merger axis
and the plane of the sky is less than $\mytilde70\degr$. However, at
this angle, the physical separation of the two clusters becomes about
$2\hubblem$ Mpc, in which case cluster finding algorithms in $N$-body
data (most popular mass functions are calibrated with $N$-body data)
may identify them as two distinct clusters.  Addressing exactly where
the division happens is beyond the scope of the current study. In the
following, our analysis assumes that the Tinker et al.\ (2008) mass
function is still valid in estimating the abundance of a massive
binary system such as \elgordo.

A traditional method to test the possible tension between the
existence of an extremely massive cluster and the given cosmology is
to estimate the probability of the cluster discovery within the
assumed cosmology.  In a survey of the entire sky, the number of
clusters with mass and redshift greater than $M_{min}$ and $z_{min}$,
respectively is given by
\begin{equation}
N(M,z) = \int_{z_{min}}^{z_{max}}  \frac{dV(z)}{dz} dz \int_{M_{min}}^{M_{max}}  \frac{dn}{dM}  
dM 
\label{eqn_abundance}
\end{equation}
\noindent
where $dV/dz$ is the volume element and $dn/dM$ is the mass
function. For massive high-redshift clusters, the result is sensitive
to the mass function $dn/dM$ near $z_{min}$ and $M_{min}$.

The choice of the threshold mass $M_{min}$ is somewhat
subjective. Mortonson et al.\ (2011) argue that an Eddington bias
(Eddington 1913) is a non-negligible factor when the cosmological
significance using most massive clusters is discussed.  Although we
include this factor when we perform the Mortonson et al.\ (2011)
exclusion curve test below, our analysis with the traditional method
is carried out without the consideration of the Eddington bias.
Therefore, we use the statistical 1-$\sigma$ lower limit of
$M_{200a}=(3.13\pm0.56) \times 10^{15} \hubblem M_{\sun}$.

Similarly, the choice of the minimum redshift $z_{min}$ is also a subject for debate.
Equation~\ref{eqn_abundance} clearly ignores the case where a similarly rare
cluster exists at a lower redshift with a higher mass (Hotchkiss 2011).
In some surveys where one is preferentially looking for 
high-redshift clusters (e.g., archival data search), the issue is
not critical. However, in relatively complete surveys, the negligence can lead to
substantial bias in rarity estimation (Hotchkiss 2011).

Hotchkiss (2011) also discussed the case, where a similarly rare cluster exists
at higher redshift but with a {\it lower mass} not included in Equation~\ref{eqn_abundance}.
However, considering the typical detection limit of the existing surveys, this bias should be small
when the cluster redshift is sufficiently high ($z\sim1$).

Assuming the cosmological parameters favored by the Wilkinson
Microwave Anisotropy Probe 9-year results (Bennett et al.\ 2012), we
estimate that the expected number of clusters like \elgordo\ is
$\mytilde0.01$ over the full sky.  Because the existing SZ surveys
(SPT+ACT) cover only $\mytilde7$\% of the entire sky, this full-sky
assumption is conservative. Recent {\it Planck} cosmological
parameter estimation studies (Planck collaboration et al.\ 2013b)
favor slightly higher values in both matter density ($\Omega_M$) and
fluctuation on a 8 Mpc scale ($\sigma_8$), which increases the
expected abundance of clusters like \elgordo\ nearly by a factor of
five (i.e., probability of $\mytilde0.05$). This illustrates how sensitive the mass function of 
high-redshift clusters at the high end is to these two parameters.

Apart from the aforementioned ambiguity in choosing the lower limits of the integration, 
the above approach is
inefficient for quantifying the impact of both cosmological
parameter uncertainties and sample variance at the same time. In other
words, the method does not address the case where the
abundance of a massive cluster is not extremely low when the assumed
cosmology is, for example, at the 2-$\sigma$ tail of the fiducial
cosmology.

Mortonson et al.\ (2011) defined an ``exclusion curve'' that overcomes
this pitfall. M12 compared the exclusion curve with their mass of
\elgordo\ and showed that the central value of the cluster mass lies
on the 95\% exclusion curve for the combined ACT+SPT survey area (2800
deg$^2$).  Nevertheless, because of their mass uncertainties, they
concluded that the presence of \elgordo\ does not give rise to any
significant tension with the standard $\Lambda$CDM.  We display the
same Mortonson et al.\ (2011) 95\% exclusion curves with the current
weak-lensing mass estimate of \elgordo\ in Figure~\ref{fig_exclusion}.
The central value of \elgordo\ lies slightly above the 95\% exclusion
curve computed for the full sky, and its 1-$\sigma$ lower limit is
above the existing SZ survey area.
 
As mentioned above, a more conservative approach uses Eddington bias
to account for the asymmetric scatter probability for steep mass
functions: namely that the chance of up-scatter is higher than that of
down-scatter. We estimate the reduced mass $M^{\prime}$ using the
Mortonson et al.\ (2011) prescription: $M^{\prime}=\exp(1/2 \gamma
\sigma_{\ln M} )M$, where $\gamma$ is the local power-law slope
($dn/d\ln M \sim M^{\gamma}$), and $\sigma_{\ln M}$ is the 1-$\sigma$
uncertainty of $\ln M$ (log-normal distribution is assumed for mass
errors). We also display this Eddington-bias-corrected mass value
(open circle) in Figure~\ref{fig_exclusion}, which shows that the
1-$\sigma$ lower limit of this reduced mass is slightly below the
exclusion curve computed for the total ACT survey area (1000 deg$^2$).

The uncertainties in Figure~\ref{fig_exclusion} show only the
statistical uncertainties. As discussed in
\textsection\ref{section_various_uncertainties}, the total mass
uncertainty can be as large as $\mytilde20$\% to $\mytilde30$\% when
we include the major systematic uncertainties such as triaxiality.

Harrison \& Hotchkiss (2013) suggest some modifications to 
the Mortonson et al. (2011) exclusion curve in such a way that
the exclusion curve includes the cases where similarly rare clusters
might exist at lower (higher) redshift with higher (lower) masses.
This modified exclusion curve gives a threshold mass approximately 
a factor of two higher than that of Mortonson et al. (2011) at the cluster
redshift.

Therefore, despite the extremely high mass of \elgordo\ when the universe is half
its current age, the above rarity test shows that the cluster by
itself will not pose a significant challenge to the current
$\Lambda$CDM model. 

It is important to remember that our weak-lensing mass estimate
$M_{200a}$ is obtained by extrapolation. The current ACS field covers
approximately a $\mytilde2.75\hubblem~\mbox{Mpc} \times
\mytilde2.75\hubblem~ \mbox{Mpc}$ square region whereas $r_{200a}$ for the
entire system is $\mytilde2.4\hubblem$ Mpc. Although currently there
is no indication that the cluster mass density may fall off steeply
outside the ACS field of view, a more robust $M_{200a}$ mass
estimation requires a much larger weak-lensing field.

In addition, our theoretical understanding of the mass function at the
extreme end is incomplete.  Extreme clusters such as \elgordo\ are
rare even in the current largest cosmological simulations.  Therefore,
the use of the Tinker et al.\ (2008) mass function in the current
study is an extrapolation.

\section{CONCLUSIONS}  \label{section_conclusion}
We have presented a detailed weak-lensing analysis of \elgordo\ at
$z=0.87$, the most significant SZ decrement in the ACT and SPT
surveys.  Our analysis confirms that \elgordo\ consists of two massive
components with a projected separation of $
\mytilde700\hubblem$~kpc. Our mass determination requires care because
of the limited field of view in weak-lensing data and also this binary
structure. We estimate the mass of \elgordo\ by simultaneously fitting
two axisymmetric Navarro-Frenk-White (NFW) profiles allowing their
centers to vary. The masses of the northwestern (NW) and the
southeastern (SE) components are
$M_{200c}=(1.38\pm0.22)\times10^{15} \hubblem M_{\sun}$ and
$(0.78\pm0.20)\times10^{15} \hubblem M_{\sun}$,
respectively. These two lensing masses are consistent with the results
from dynamical studies.

The small line-of-sight velocity difference between the NW and SE
components ($\mytilde600~\mbox{km}~\mbox{s}^{-1}$) and the presence of
double radio relics suggests that the merger is proceeding nearly in
the plane of the sky.  With this plane-of-the-sky-merger hypothesis,
extrapolation of the two NFW halos to a radius $r=2.4\hubblem$~Mpc
yields a combined mass of $M_{200a}=(3.13\pm0.56)\times10^{15}
\hubblem M_{\sun}$, which is consistent with our two-component
dynamical analysis and previous X-ray measurements.

At face value, the existence of such an extreme cluster may be viewed
as a challenge to the current $\Lambda$CDM paradigm. However, such a
claim awaits further studies addressing the validity of the
extrapolation of the cluster mass profile beyond the current field
size, as well as the extrapolation of the current empirical mass
function calibrated with relatively low-volume $N$-body simulations.

\elgordo\ resembles the Bullet Cluster when it comes to its extreme
mass and binary distribution of galaxies and dark matter.  However,
the X-ray map shows only a single distinct gas peak located close to
the less massive SE subcluster.  Although we detect an offset between
the SE mass center and the X-ray peak, the mass peak does not seem to
be leading the gas peak if we are viewing the cluster soon after first
core passage during a high speed merger as in the Bullet Cluster.

\acknowledgements

JPH acknowledges Glennys Farrar and Craig Lage for useful
conversations on the dynamical state of El Gordo and Rachel Somerville
and Tomas Dahlen for help obtaining the photometric redshift catalog
for the GOODS fields.  Support for Program number HST-GO-12755.01-A
was provided by NASA through a grant from the Space Telescope Science
Institute, which is operated by the Association of Universities for
Research in Astronomy, Incorporated, under NASA contract NAS5-26555.
CS acknowledges support from the European Research Council under FP7
grant number 279396.

\clearpage
\appendix
\section{IMPACTS OF CTI ON WEAK-LENSING ANALYSIS OF \elgordo} \label{section_cti}
 
 \subsection{CTI Measurement}
The current {\it HST}/ACS data  of \elgordo\ was taken
on 2012 September and October.  
This is more than three years since Servicing
Mission 4 (SM4) in 2009 May and more than ten years since the camera was
installed in 2002 March. Cumulative damage on the CCDs due to space
radiation is severe, and the resulting charge transfer inefficiency
(CTI) is of great concern for the use of the ACS instrument in
weak-lensing studies. Here we present our investigation
of CTI and its impact on our weak-lensing using the PROP 12755 data. 
  
Many techniques have been suggested to characterize CTI effects
including differential aperture photometry, monitoring warm pixels and
their trails, etc. In this paper, we measure the ACS CTI utilizing
sub-seeing features (SSFs). We refer to any group of connected pixels
whose size is less than that of the PSF, but whose collective
significance is well above the sky rms as SSFs. Most SSFs are
cosmic-rays or uncorrected hot/warm pixels. Because they are not
affected by the PSF, their collective shapes are useful indicators of
CTI trails.

\begin{figure}
\begin{center}
\includegraphics[width=8cm]{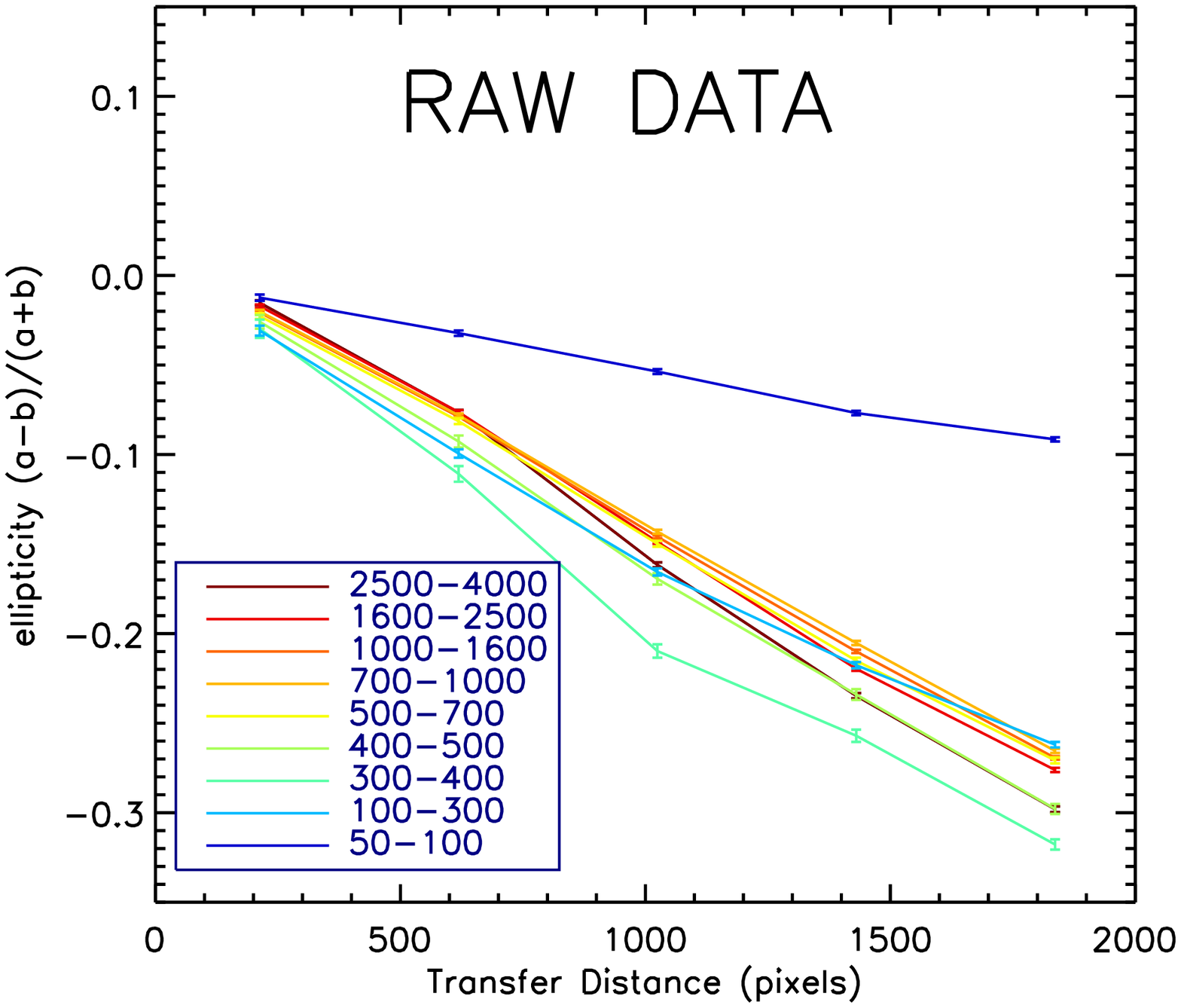}
\includegraphics[width=8cm]{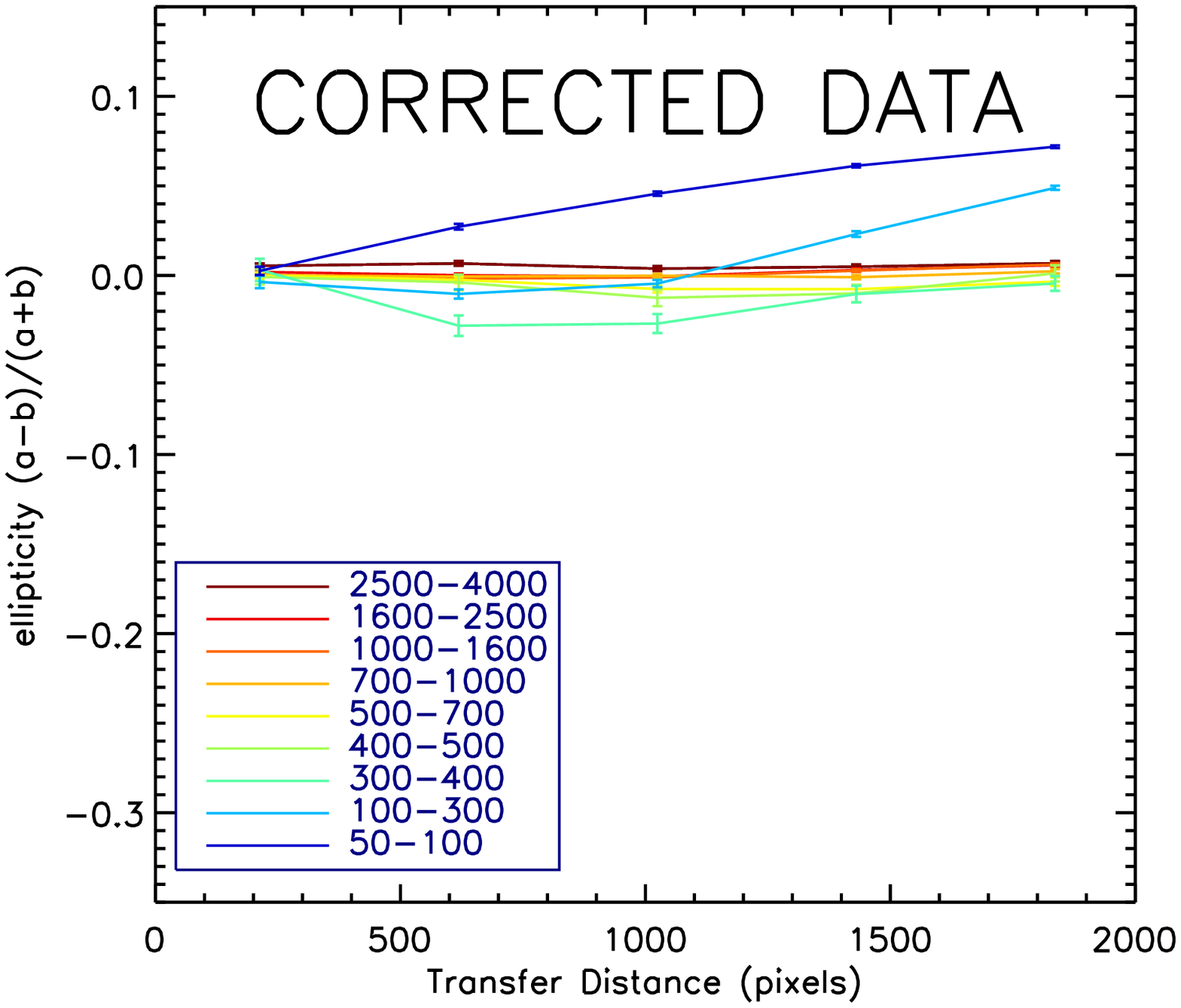}
\end{center}
\caption{CTI effects on the ellipticity of sub-seeing features
  (SSFs). We measure the ellipticity of the SSFs (e.g., cosmic-rays)
  to characterize the CTI effect. We define the $x$-axis orthogonal to
  the readout direction, and thus the CTI-induced elongation gives a
  negative ellipticity.  The curves of different color represent the
  SSFs of different flux ranges.  The left panel shows the results
  obtained from the raw images of \elgordo\ whereas the right panel
  displays the measurements from images where the CTI effect has been
  corrected for by the STScI pipeline using the Ubeda \& Anderson
  (2012) algorithm.  The correction reduces the CTI effect
  substantially, although the residual errors are still non-negligible
  at the faint limit.  Note the CTI over-correction in the
  $50-100\mbox{e}^{-}$ regime.
\label{fig_cti_correction}}
\end{figure}

In the left panel of Figure~\ref{fig_cti_correction}, we display the
ellipticity of the SSFs detected on the current \elgordo\ {\it HST}
images as a function of transfer distance. Because the analysis is
performed on the {\tt FLT} images (i.e., prior to geometric distortion
correction), the parallel CTI happens purely along the readout
direction. As observed, our definition of the $y$-axis being this
readout direction makes the CTI-induced ellipticity negative. Two
features of the ACS CTI are noteworthy.  First, the CTI increases
linearly with transfer distance, which is consistent with theoretical
expectations and previous results.  Second, no strong flux-dependence
is seen except for the data with the lowest counts (50-100 $\mbox{e}
^{-1}$).  This behavior is different from what we obtain with earlier
ACS images. For example, Figure~30 of Jee et al.\ (2011) shows that
the CTI slopes measured from the year 2006 data set depend sensitively
on flux.

The right panel of Figure~\ref{fig_cti_correction} is the same as the
left panel except that we repeat the measurement using the {\tt FLC}
images. These {\tt FLC} images provided by the STScI pipeline are
corrected for the CTI with the pixel-based method (Ubeda \& Anderson
2012), and therefore the result reveals residual CTI effects on SSFs
after the correction. The performance is remarkable for the bright
($300-4000\mbox{e} ^{-1}$) SSFs, as indicated by their residual
ellipticities close to zero. For the faint SSFs ($50-300\mbox{e}
^{-1}$), we find that the pixel-based algorithm overcorrects the
CTI. This overcorrection provides evidence for the ACS CTI mitigation
in the low flux regime first reported by Jee et al.\ (2009).

\subsection{Impacts on Galaxy Shapes}

Having examined the CTI using the SSFs before and after the
correction, we now turn to the question: whether or not the current
pixel-based correction is sufficient for our weak-lensing study of
\elgordo. In other words, we need to investigate if there should be
additional correction to our galaxy shape measurements in order to
remove the residual CTI effects shown in the right panel of
Figure~\ref{fig_cti_correction}. To answer this question, we must
quantify the impacts of the residual CTI effects on our source galaxy
ellipticity and examine the level of the statistical noise with
respect to that of the systematic noise caused by the residual CTI.

\begin{figure}
\begin{center}
\includegraphics[width=8cm]{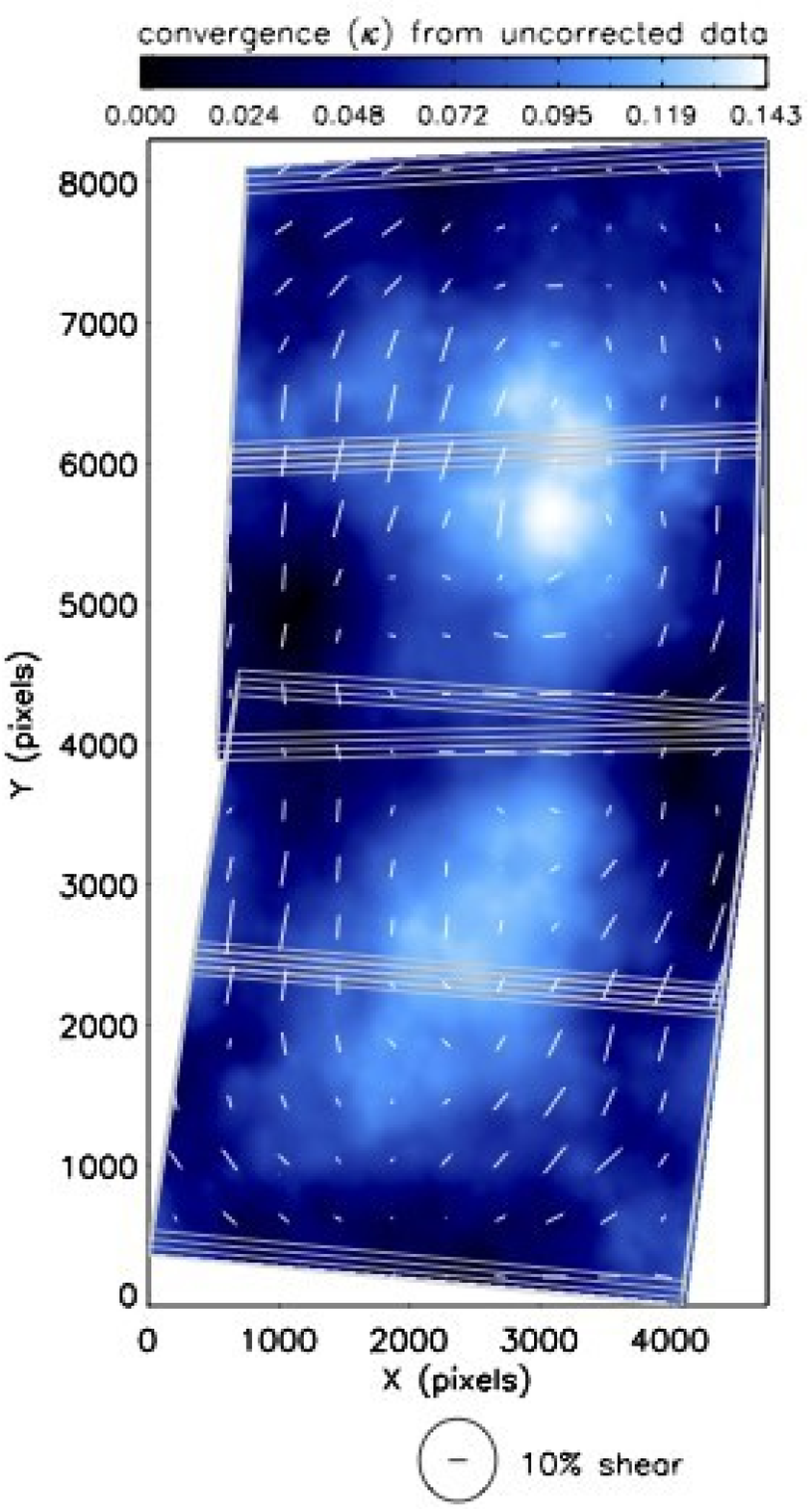}
\includegraphics[width=8cm]{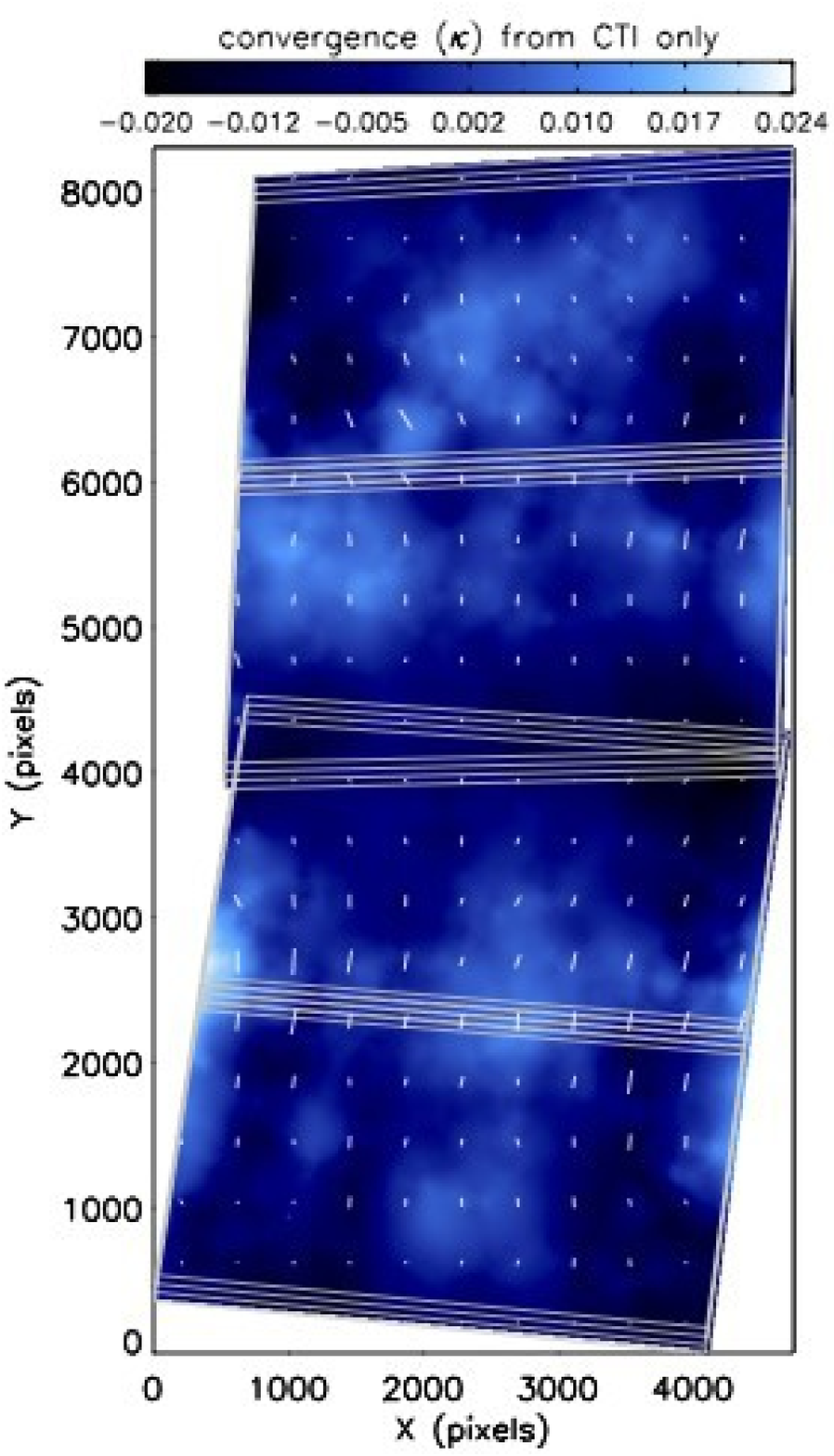}
\end{center}
\caption{CTI effects on the weak-lensing analysis. The left panel
  shows the shears (sticks) and the corresponding mass reconstruction
  (blue) from the raw \elgordo\ images.  Comparison of this result
  with the one shown in the left panel of
  Figure~\ref{fig_whisker_n_mass} illustrates that the global features
  of \elgordo\ is still seen in the CTI-contaminated data, although
  the details differ. In the right panel, we subtract the
  CTI-corrected ellipticities of galaxies from the uncorrected
  ellipticities, and smooth the results with a FWHM=30$\arcsec$
  Gaussian kernel (as is done in Figure~\ref{fig_whisker_n_mass}).
  The spatial variation of the residual ellipticity is consistent with
  the theoretical expectation and also the results from the SSF test
  (Figure~\ref{fig_cti_correction}).  The largest CTI-induced
  systematics ($\gamma\sim0.1$) is seen near the boundaries of the two ACS
  CCDs (WFC1 and WFC2), where the distance to the readout register is
  longest.
\label{fig_cti_mass}}
\end{figure}

As a first step toward this goal, we perform a weak-lensing analysis
of \elgordo\ using the CTI-uncorrected (i.e., {\tt FLT}) images. The
basic image reduction, PSF correction, and shape measurement methods
are identical to those described in \textsection \ref{section_PSF} and
\textsection\ref{section_shear_measurement}.  The left panel of
Figure~\ref{fig_cti_mass} displays this result.  Despite the large
CTI-induced systematic errors, it is remarkable that the bimodal
distribution of the mass is clearly seen even in this weak-lensing
analysis with the uncorrected data. However, direct comparison with
Figure~\ref{fig_whisker_n_mass} (the result obtained from the {\tt
  FLC} images) reveals that the substructures are severely smeared and
distorted. In particular, the centroid of the SE cluster is no longer
prominently defined in Figure~\ref{fig_cti_mass}. The difference
between the two results is also obvious when we compare the whiskers
computed from the smoothed ellipticity of source galaxies. In the
right panel of Figure~\ref{fig_cti_mass}, we show the whiskers and
resulting mass reconstruction using the differential ellipticities of
the same source galaxies between the two catalogs (before and after
the CTI correction).  Both the direction and size distribution of the
whiskers are consistent with our expectation.  That is, the
orientation of the whiskers is parallel to the readout direction, and
the sizes of the whiskers are largest at the center of the two ACS
pointings. The difference in galaxy ellipticity provides the key
information for us to relate the CTI measured with SSFs to the
systematic shear in galaxies.  Because galaxies are much larger than
SSFs, the amount of ellipticity change in galaxies due to CTI is
smaller than that in SSFs.  The differential ellipticity
(Figure~\ref{fig_cti_mass}) shows that a mean galaxy ellipticity
change near the center of each ACS pointing is $\gamma\sim0.08$, which
is $\mytilde27$\% of the CTI-induced ellipticity in SSFs.  Without the
Ubeda \& Anderson (2012) pixel-based CTI correction, our shears near
the center of each ACS pointing would be biased by $\gamma\sim0.08$.

Now with this conversion factor at hand, we can estimate the residual
CTI effect on galaxy shears after the pixel-based correction is
carried out.  For the bright SSFs ($>300~\mbox{e}^{-1}$), the residual
is small, giving at most an ellipticity of $ \mytilde0.02$
(Figure~\ref{fig_cti_correction}).  Thus, we can neglect the impact of
the residual CTI for bright galaxies. For the faint SSFs
($<300~\mbox{e} ^{-1}$), the maximum residual ellipticity is
$\mytilde0.08$. Applying the conversion factor of $\mytilde27$\%, we
expect that the faint galaxies will be stretched by
$\delta\gamma\sim0.02$ near the ACS pointing boundaries. If we
conservatively assume that the magnitude range of the galaxies
affected by this CTI overcorrection is F775W$\gtrsim26.5$, about $
\mytilde30$\% of our source galaxies ($\mytilde800$ out of 2541) are
affected. Therefore, we estimate that the maximum shear systematic
error induced by the residual CTI is less than $ \gamma<0.01$.  One
caveat is that the faint (F775W$\gtrsim26.5$) galaxies are smaller,
and their conversion factor may become larger than our average
value. However, these faint galaxies are also downweighted in our
shear estimation. Our study suggests that the two effects nearly
cancel each other.  Because the above maximum shear systematic error
$\gamma<0.01$ is insignificant compared to the level of the
statistical noise and the large cluster lensing signal, we conclude
that the current pixel-based CTI correction is sufficient for our
weak-lensing analysis.

\section{IMAGE DRIZZLING METHOD AND WEAK-LENSING PERFORMANCE} 
\label{section_drizzle_comparison}

The {\it HST}/ACS PSFs are slightly undersampled, exhibiting
some non-negligble aliasing effects.  To reduce this artifact, some
studies suggest drizzling WFC images with an output pixel scale
smaller than the native pixel scale 0\farcs05. We investigate the
impact of the drizzling method on our weak-lensing results by
performing a separate data reduction and examining the difference in
the results.

 We note that although the analysis presented here
is performed using the PROP 12755 data set, the result holds
for the entire data set (PROP 12755 and 12477).
 We follow the COSMOS data reduction scheme (Koekemoer et
al.\ 2006). Namely, the output pixel scale, the {\tt pixfrac}
parameter, and the drizzling kernel are set to $0\farcs03$, 0.8, and
Gauss.
 
\begin{figure}
\begin{center}
\includegraphics[width=8cm]{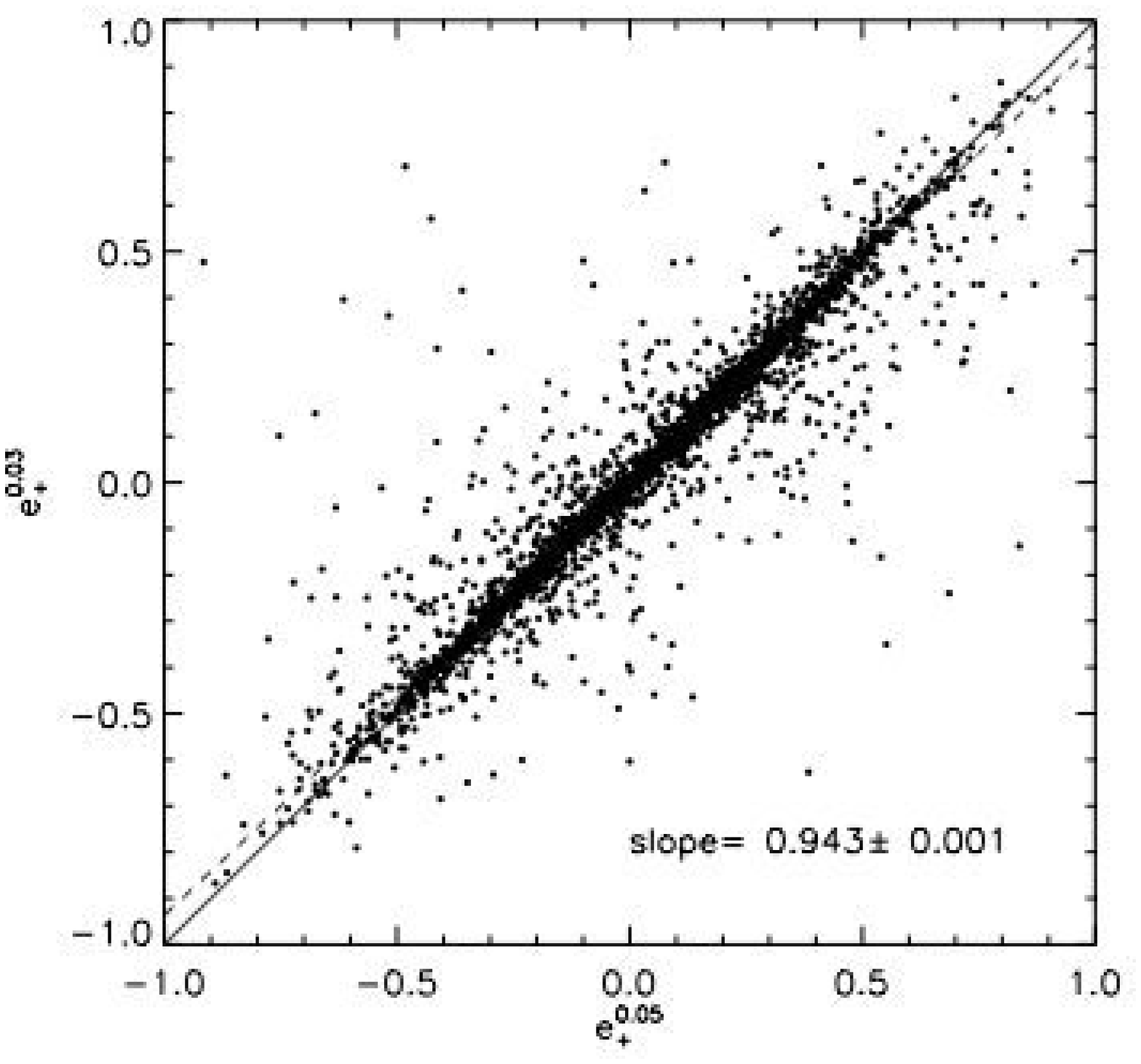}
\includegraphics[width=8cm]{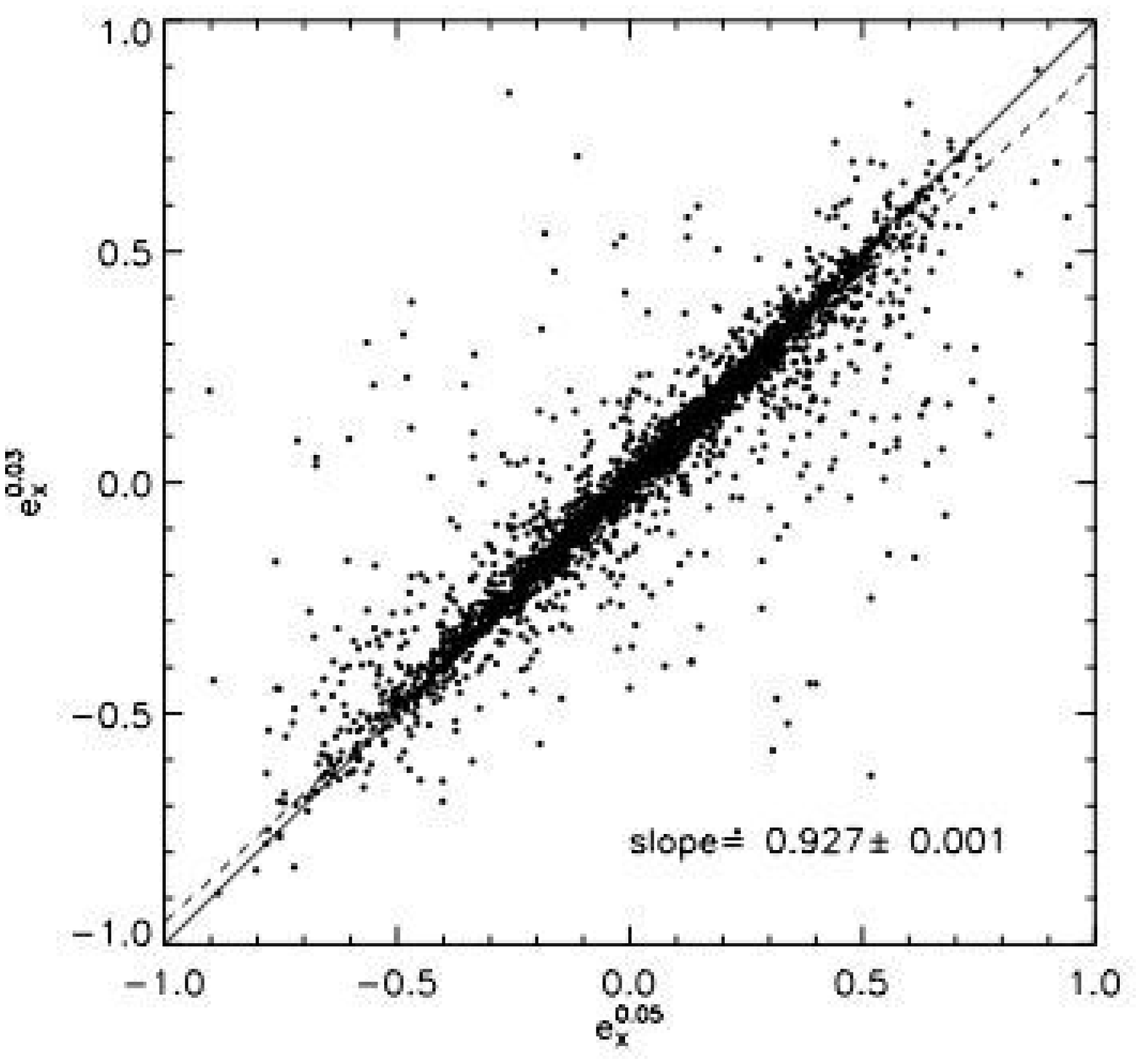}
\end{center}
\caption{Comparison of ellipticity components between the two data
  reductions. The superscript (0.05 or 0.03) represents the output
  pixel scale. The results from the two drizzling products are
  consistent, although there are indications that the ellipticity from
  the $0\farcs03$ output scale is systematically smaller (i.e., the
  slopes are less than unity). The solid line shows the $y=x$ equality
  while the dashed line is the fit to the data.
\label{fig_e_compare}}
\end{figure}

One of the most straightforward comparisons is to crosscheck the
ellipticity of the common objects as displayed in
Figure~\ref{fig_e_compare}. The results from the two versions of the
\elgordo\ images are highly consistent. The small scatter implies that
one would not reach a significantly different conclusion because of
the difference in the image output scale.  However, we find that the
ellipticity is slightly lower in the $0\farcs03$ output scale, as
indicated by the fitted slopes (dashed) being less than unity. This
means that for the conversion of ellipticity to shear we need to apply
a larger shear calibration factor. The exact cause for this difference
in shear calibration is not clear at the moment.

\begin{figure}
\begin{center}
\includegraphics[width=8cm]{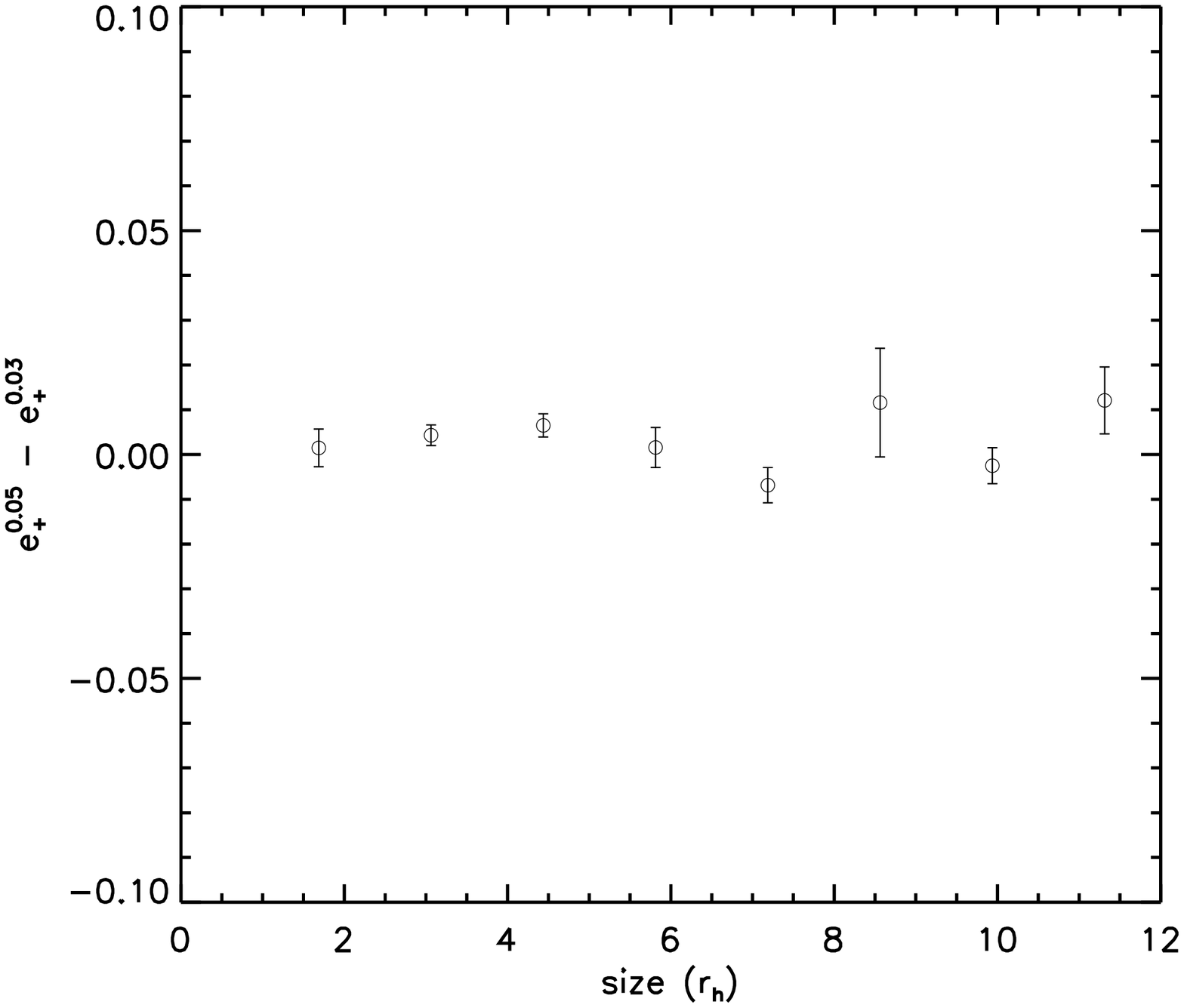}
\includegraphics[width=8cm]{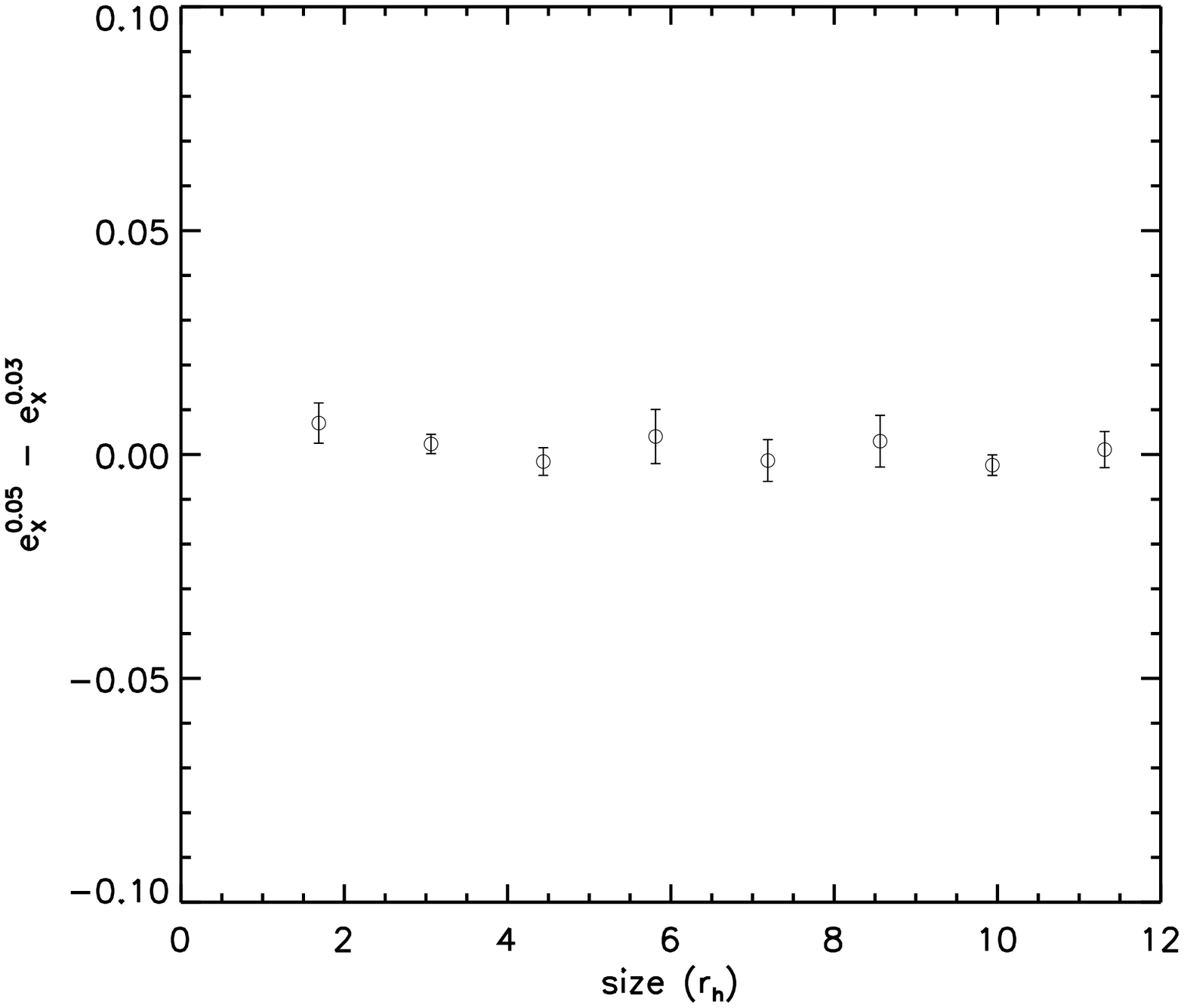}
\end{center}
\caption{Ellipticity difference as a function of object size. If the
  impact of the drizzling output scale on the mitigation of aliasing
  is large, we expect to observe a trend that depends on the object
  size.  However, we do not detect such a pattern.
\label{fig_de_snr}}
\end{figure}

Another useful diagnostic is the investigation of the ellipticity
difference between the two versions of the data reduction as a
function of object size.  If the aliasing arising from undersampling
is severe and can be relieved by the choice of the $0\farcs03$ output
scheme, we expect to detect some systematic difference in ellipticity
for small objects.  However, as shown in Figure~\ref{fig_de_snr}, no
obvious size-dependent pattern is present.

\begin{figure}
\begin{center}
\includegraphics[width=15cm]{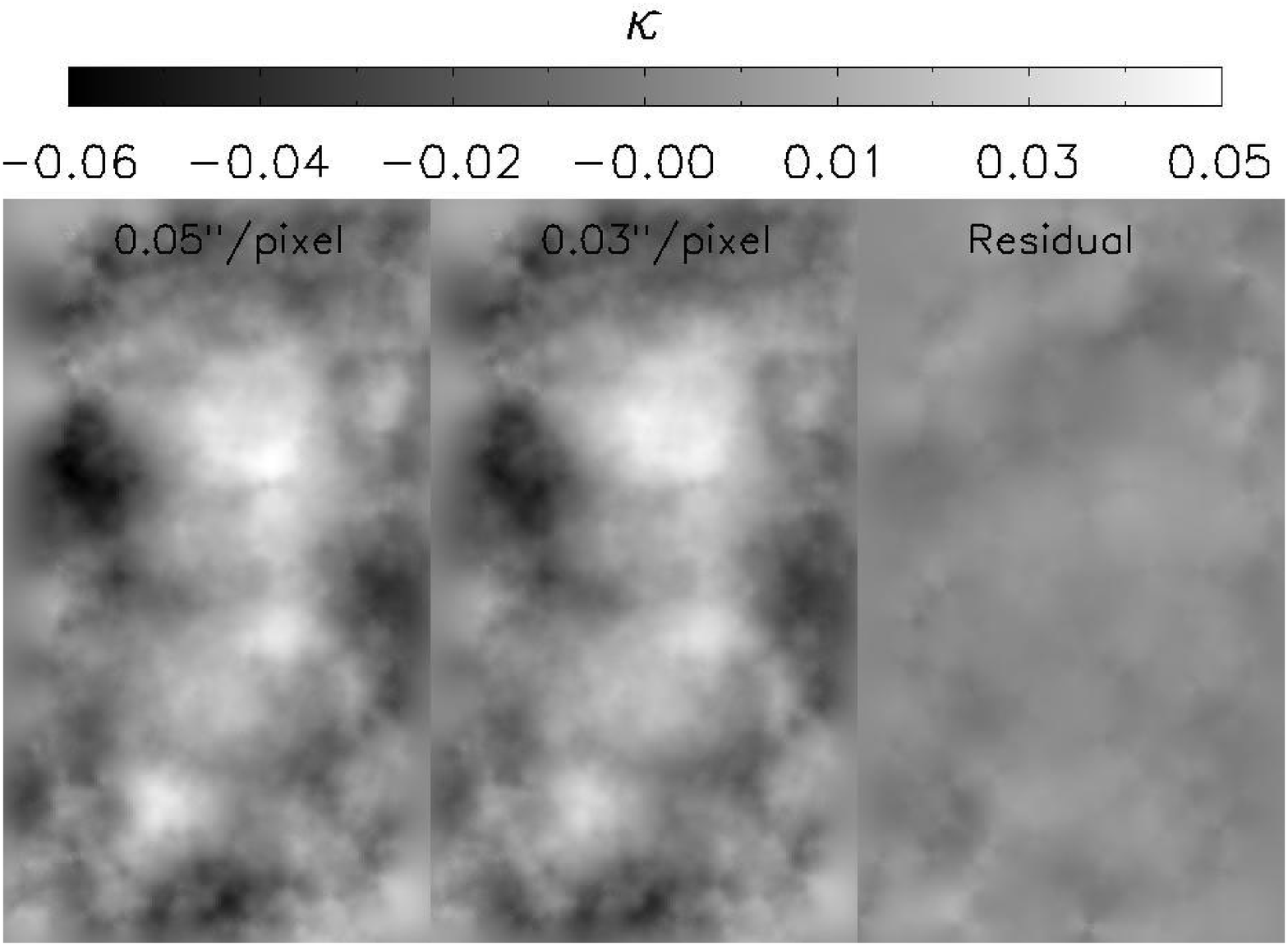}
\end{center}
\caption{Comparison of two-dimensional mass reconstructions from the two data reductions. 
We use {\tt 
FIATMAP} to produce the result.
The common source galaxies
are selected by looking for pairs within $\mytilde0\farcs15$ after the application of the criteria 
in \textsection
\ref{section_redshift}.
 The results are in excellent agreement. Because here we do not apply the corresponding 
shear 
calibrations, the agreement can become even better when this remaining correction is applied.
\label{fig_map_com}}
\end{figure}

Finally, we compare the scientific results from the two versions of
the data reduction. We find that the two weak-lensing masses agree
within $\mytilde2\%$ when we apply the corresponding shear calibration
factor to each shear catalog. The two-dimensional mass maps are
compared in Figure~\ref{fig_map_com}. We select common source galaxies
by looking for pairs within $\mytilde0\farcs15$ after applying the
criteria in \textsection\ref{section_redshift}.  The results are in
excellent agreement as indicated by the small residual across the
field of view (right).  Because here we do not apply the corresponding
shear calibration to each ellipticity catalog, the agreement can
be made even better by applying this remaining correction.

\section{MAGNITUDE-DEPENDENT LENSING SIGNAL} 
\label{section_mag_test}

We select source population in a broad magnitude range (e.g., $22<F775W<28$ where F775W is available). One may question whether these lower and upper bounds are legitimate choices. In case the faint end is too faint, our source selection is severely contaminated by too noisy (thus unusable) galaxies, which dilute the cluster lensing signal. Likewise, in case the bright end is too bright, the population at the bright end is mostly cluster galaxies or foreground objects containing no lensing signal. Therefore, here we demonstrate that our sources at both ends provide significant lensing signals and add to the overall S/N of our measurements.

\begin{figure}
\begin{center}
\includegraphics[width=15cm]{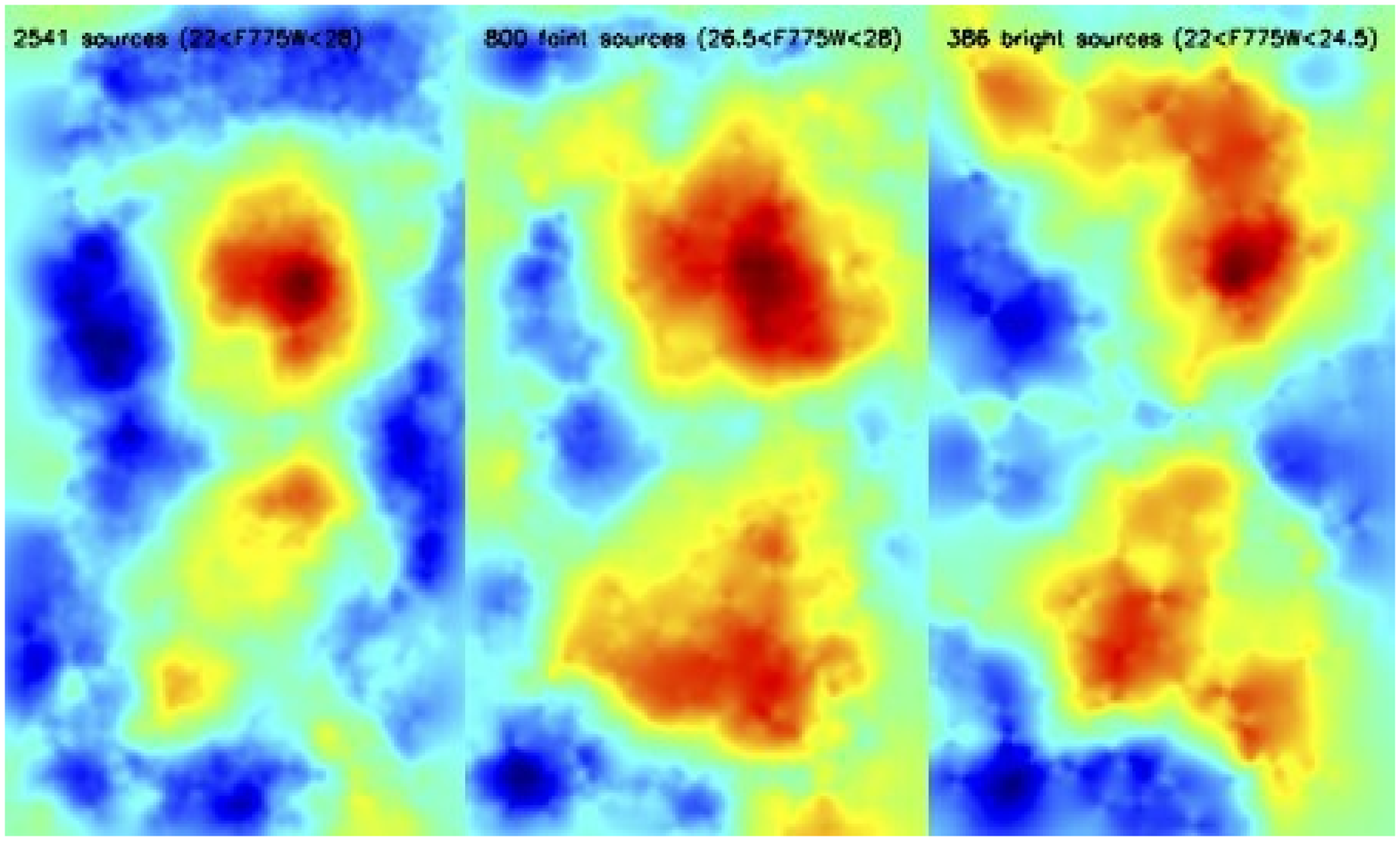}
\end{center}
\caption{Mass reconstruction with sources of different magnitudes. We define two sub-samples by applying different magnitude limits.  The mass reconstruction using the entire source population is displayed in the left panel. This mass map is created with {\tt FIATMAP} and is slightly different from the maximum entropy mass map presented in \textsection\ref{section_massmap}.  The middle and right panels show the results from the bright and faint sources, respectively. These two mass maps are much noisier because of the considerably smaller number of sources. However, it is easy to identify the bimodal mass distribution of \elgordo~ in both cases.
\label{fig_mag_dependence}}
\end{figure}

Within the area defined by the ACS pointings of the PROP 12755, the total number of sources is 2541. The mass reconstruction using the entire source population is displayed in the left panel of Figure~\ref{fig_mag_dependence}. This mass map is created with {\tt FIATMAP} without implementing the non-linear relation ($g=\gamma/(1-\kappa)$) between shear and ellipticity and is slightly different from the maximum entropy mass map presented in \textsection\ref{section_massmap}.
Then, we separate faint sources by selecting objects in the $26.5<F775W<28$ range. This sub-sample contains 800 galaxies, and the corresponding mass map is shown in the middle panel. In addition, we define a bright sample using the  $22<F775W<24.5$ magnitude range. The total number of sources in the bright sample is 386.
The resulting mass map is shown in the right panel. The two mass maps created from these two sub-samples are much noisier because of the considerably smaller source densities. However, it is easy to identify the bimodal mass distribution of \elgordo~in both cases. Thus, both sub-samples contain significant lensing signals, which serve as justification for our choice of the magnitude limits.

\end{document}